\documentclass[a4paper]{article}
\pdfoutput=1

\usepackage{jcappub} 

\usepackage[T1]{fontenc} 
\usepackage[utf8]{inputenc}
\usepackage{pythonhighlight} 
\usepackage{textcomp,gensymb}
\usepackage{autobreak}

\usepackage{placeins}

\usepackage{autobreak}
\usepackage{physics}
\usepackage{feynmp-auto}
\usepackage{slashed}
\usepackage{lipsum}
\usepackage{adjustbox}

\usepackage{ulem}
\usepackage{amsmath}
\usepackage{amssymb}
\usepackage{bm}
\usepackage{bbold}
\usepackage{multirow}
\usepackage{makecell}
\usepackage[dvipsnames]{xcolor}

\usepackage{graphicx}
\usepackage{hyperref}
\usepackage{fontawesome}
\usepackage[american]{babel}
\usepackage{enumerate}
\usepackage{mathtools}
\usepackage{support-caption}
\usepackage{subcaption}
\usepackage{soul}
\setstcolor{red}
\usepackage{makebox}
\usepackage{comment}
\usepackage{booktabs}

\renewcommand{\[}{\left[}

\usepackage{pifont}

\newcommand\varpm{\mathbin{\vcenter{\hbox{%
  \oalign{\hfil$\scriptstyle\hspace{-0.2ex}+\hspace{-0.2ex}$\hfil\cr
          \noalign{\kern-.5ex}
          $\scriptscriptstyle({-})$\cr}%
}}}}

\newcommand\varmp{\mathbin{\vcenter{\hbox{%
  \oalign{\hfil$\scriptstyle\hspace{-0.2ex}-\hspace{-0.2ex}$\hfil\cr
          \noalign{\kern-.5ex}
          $\scriptscriptstyle({+})$\cr}%
}}}}

\DeclareMathSymbol{\comma}{\mathpunct}{letters}{"3B} 

\definecolor{ForestGreen}{rgb}{0.13, 0.55, 0.13}

\usepackage{xspace}
\newcommand*{\GW}{\ensuremath{\mathrm{GW}}\xspace}

\usepackage[capitalise]{cleveref}

\crefname{section}{Section}{Sections}
\crefname{table}{Table}{Tables}
\crefname{figure}{Fig.}{Figs.}
\crefname{equation}{Eq.}{Eqs.}
\crefname{appendix}{Appendix}{Appendices}

\graphicspath{{figures/}}

\allowdisplaybreaks

\title{{\it Prominence}: A discriminator of gravitational wave signals}

\author[a,b]{Jo\~ao~Gon\c{c}alves,}
\author[c]{Danny Marfatia,}
\author[b,d]{and Ant\'onio~P.~Morais}

\affiliation[a]{Departamento de F\'{i}sica da Universidade de Aveiro, Campus de Santiago, 3810-183 Aveiro, Portugal.}
\affiliation[b]{Laborat\'{o}rio de Instrumenta\c{c}\~{a}o e F\'{i}sica Experimental de Part\'{i}culas (LIP), Universidade do Minho, 4710-057 Braga, Portugal}
\affiliation[c]{Department of Physics and Astronomy, University of Hawaii at Manoa, Honolulu, HI 96822, USA}
\affiliation[d]{Departamento de F\'{i}sica, Escola de Ci\^{e}ncias, Universidade do Minho, 4710-057 Braga, Portugal}

\emailAdd{jpedropino@ua.pt}
\emailAdd{dmarf8@hawaii.edu}
\emailAdd{amorais@fisica.uminho.pt}

\abstract{The concept of {\it prominence} is familiar to signal engineers, topographers and mountaineers. We introduce Prominence $\cal P$ as a discriminator of gravitational wave (GW) signals. We treat black hole and neutron star binaries as astrophysical background sources, and show how $\cal P$ can be used to distinguish between GW spectra produced by first-order phase transitions, domain walls and cosmic strings, and combinations thereof. Prominence can also be used to discriminate between these and off-piste sources of GWs. 
The uncertainty in the measured energy density in GWs at Pulsar Timing Arrays needs to be at the sub-percent to percent  level for 
$\cal{P}$ to achieve discrimination at 3$\sigma$. 
LISA and ET data are expected to have sufficiently small uncertainties that Prominence can play a central role in their analysis. 
We define and apply a $\chi^2$ statistic based on an inner product in signal space to show that sources indistinguishable at $2\sigma$ using the signal-to-noise ratio can be distinguished using $\cal P$ with significances above and in some cases far above $3\sigma$. \href{https://github.com/Mrazi09/Prominence-GW-discriminator}{\faGithub}

}

\begin{document}

\maketitle
\flushbottom

\section{Introduction}\label{sec:intro}

Gravitational waves (GWs) constitute a powerful tool for probing both the dynamics of the early Universe and various astrophysical phenomena. In addition to transient signals, such as those originating from binary black hole mergers \cite{LIGOScientific:2016aoc}, a stochastic gravitational wave background (SGWB) from the superposition of numerous unresolved sources may be detectable~\cite{Allen:1996vm}. Beyond the expected astrophysical contributions, a cosmological component may also be present. If detected, such a component could serve as evidence for new physics beyond the Standard Model~\cite{Caprini:2018mtu}.

Cosmological backgrounds can originate from processes prior to the era of Big Bang Nucleosynthesis at energy scales that may far exceed the electroweak scale. Among the possible sources, first-order phase transitions (FOPTs), proceeding via bubble nucleation, generate GWs through sound waves in the plasma, bubble collisions, and turbulence \cite{Weir:2017wfa}. Topological defects, such as cosmic strings (CSs), line-like false vacuum remnants formed after the spontaneous breaking of a global or gauge symmetry, may also produce GWs via the oscillation and decay of string loops~\cite{Vilenkin:1981bx}. If instead a discrete symmetry is broken, domain walls (DWs) are formed and their decay can similarly give rise to a GW signal~\cite{Preskill:1991kd}. Another possible source arises from standard inflation during which quantum fluctuations of the metric generate tensor perturbations. These perturbations are stretched to superhorizon scales and, upon reentering the horizon, manifest as GWs.

Extensions of the Standard Model featuring new gauge symmetries and extended scalar sectors predict such cosmological GW sources. The same underlying symmetry-breaking mechanism responsible for a phase transition could simultaneously generate CS or DW relics~\cite{Caprini:2018mtu}. Thus, a multi-peak structure in the SGWB spectrum is likely, underscoring the importance of observables capable of discriminating between the components contributing to these peaks. In this context, we introduce the concept of \textit{prominence}, a measure employed by topographers and mountaineers to quantify the height of a peak relative to its surrounding peaks. Unlike the absolute peak amplitude, Prominence is sensitive to the shapes of both the background and SGWB peaks, offering the ability to distinguish between different SGWB sources, which is not achievable with only the Signal-to-Noise-Ratio (SNR).

The structure of this article is as follows. In \cref{sec:prom_stats}, we introduce the concept of prominence. In \cref{sec:GWs_backs}, we discuss the SGWB spectra from CSs, DWs and FOPTs, scalar-induced GWs (SIGWs), and the expected astrophysical backgrounds.  In \cref{sec:results}, we describe the statistical methods used to quantify the discriminating power of Prominence and then present several illustrative examples. We summarize in \cref{sec:summary}.

\section{Prominence}\label{sec:prom_stats}

\begin{figure*}[t]
	\centering
    \includegraphics[width=\textwidth]{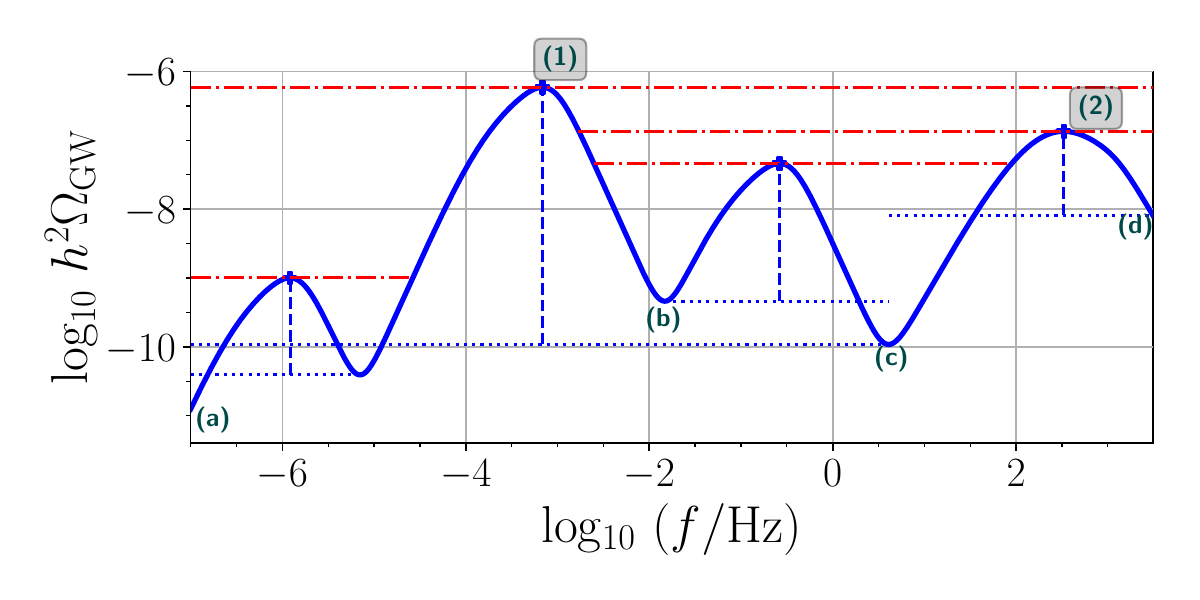} \\ 
    \includegraphics[width=\textwidth]{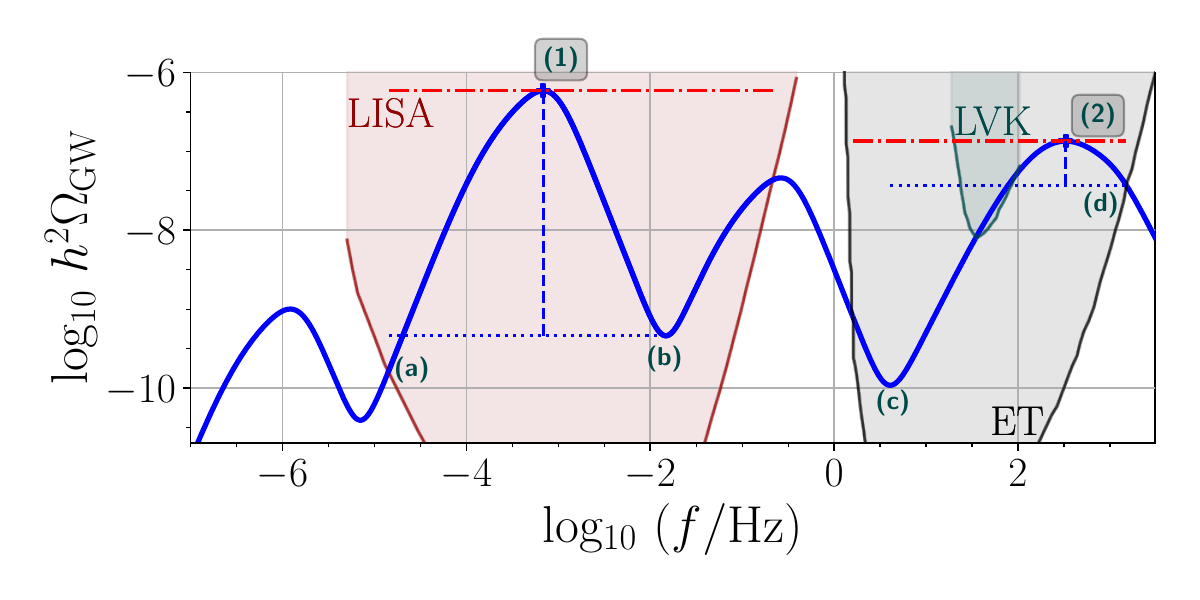} 
    \caption{ Illustration of Prominence in a multi-peaked SGWB spectrum. Top panel: by convention, the edges of the plot define the window border. Bottom panel: we define the window borders by the points at which the signal intersects the experimental sensitivity curves; in this case for LISA~\cite{LISA:2017pwj} and ET~\cite{Punturo:2010zz}. The blue dotted lines indicate the bases of the peaks and the vertical dashed lines define the Prominences. 
    The labels correspond to two examples described in the text. Also shown is the LVK bound~\cite{KAGRA:2021kbb}.  
    }
    \label{fig:prom_example}
\end{figure*}

Prominence, a concept originally designed for topographical applications, is a measure of how much a mountain or hill rises above the surrounding terrain. Formally, the Prominence of a peak is defined as the height of the peak's summit above the lowest contour line encircling it that contains no higher summit within it. In a broader context, Prominence also finds application in signal processing. To compute the prominence of each peak, we apply the following algorithm: 
\begin{itemize}
	\item In addition to the global maximum, identify all local maxima.
    \item By convention, the window border is defined by the edges of the plot. We modify the conventional definition to one that is more suitable for our application. We define the window border as the horizontal segment between the points at which the signal intersects the sensitivity curves of GW experiments in the frequency-amplitude 
    ($f$-$h^2\Omega_{\rm GW}$) plane. Select a peak and extend a horizontal line on both sides of the peak maximum until it reaches either the window border or intersects the signal at the slope of a higher peak.  These are shown as 
red dot-dashed lines in the top panel (conventional definition) and bottom panel (our definition) in Fig.~\ref{fig:prom_example}.
    An intersection with a degenerate peak height is ignored. 
	\item On either side of each peak, find the minimum signal in the interval defined by the corresponding red dot-dashed line. These points define the two \textit{valleys} of the peak and correspond to either a minimum or a signal endpoint on the border.
	 The {\it base} of the peak, indicated by the blue dotted lines in Fig.~\ref{fig:prom_example}, is defined by the level of the higher valley and is bounded by the two valleys. 
    The {\it Prominence} of the peak is the height of the summit measured from the base, or equivalently from the higher of the two valleys. 
	\item The Prominence of peak $i$ is computed as
	\begin{equation}
		\mathcal{P}_i = \log_{10}\left[h^2\Omega_\mathrm{GW}^\mathrm{peak}\right]_i - \log_{10}\left[h^2\Omega_\mathrm{GW}^\mathrm{base}\right]_i\,,
	\end{equation}
\end{itemize}
which is defined to be $\mathcal{O}(1)$.
Note that this procedure ignores the signal beyond the endpoints, regardless of its magnitude. In other words, the endpoints are always treated as valleys relative to the neighboring peak. This is reasonable given that an experiment is sensitive in a fixed frequency range.

To make our discussion clear, consider two concrete examples: peaks~(1) and~(2) in \cref{fig:prom_example}. We begin with peak~(2). Starting from the summit, extend a horizontal line across the entire frequency range until it intersects either the slope of a higher peak or a window border. This corresponds to the second highest red dot-dashed line in the figure. In the top panel, the window border is set by the edges of the plot, so the line intersects the right edge of the frequency window and defines a valley at~(d). On the left of the summit, the line intersects the right slope of peak~(1). In the bottom panel, the window is defined by the segment between the points at which the signal intersects the ET sensitivity curve.  Then, the red dot-dashed line ends almost on the ET sensitivity curve on the left, and close to this curve on the right. The intersection of the signal and the ET sensitivity curve on the right defines valley~(d). Within the frequency range covered by the red dot-dashed line, in the top panel we identify two minima on the left of the peak, labeled (b) and (c), while in the bottom panel we identify only one minimum, labeled (c). In the top panel, since (c) is lower than (b), it is the second valley. In the bottom panel (c) is the second valley because it is unique. The Prominence of peak~(2) is the vertical distance between the higher valley~(d) and the summit.  The base is marked by the blue dotted line that passes through the higher 
valley~(d) and extends to the location of 
valley~(c) in both panels.

For peak (1), begin by extending a horizontal line from the summit. Because peak~(1) is the highest peak, in the top panel the (highest) red dash-dotted line intersects the window border on both sides. 
In the lower panel, the red dot-dashed line  ends close to (but not on) the LISA sensitivity curve on both sides. The two valleys are the lowest minima on either side of the peak. In the top panel, these correspond to~(a) and~(c). In the bottom panel, (b) is the valley on the right of the peak, and valley~(a) on the left is the point of intersection of the signal and the LISA sensitivity curve. In the top panel, since (c) is the higher valley, it determines the base (blue dotted line) between~(a) and~(c). In the bottom panel, since (b) is the higher valley, it determines the base between~(a) and~(b).
 The Prominence is the height of the peak from the base, or equivalently the height of the peak from the higher valley~(c), in the top panel, and the higher valley~(b), in the bottom panel.
Note that the Prominence of peaks~(1) and (2) are different in the two panels because of the difference in definition of the window border.

Prominence is sensitive to both the background structure and the shape of the signal, providing a more nuanced measure than SNR.

\section{Analytical templates for cosmological GW signals and backgrounds}\label{sec:GWs_backs}

\subsection{Cosmic strings}\label{sec:CS}

Cosmic strings are linear accumulations of energy formed during the spontaneous breaking of a continuous symmetry in a phase transition \cite{Kibble:1976sj}. For gauge symmetries, CSs mainly decay via gravitational radiation, generating a SGWB. Their gravitational interactions depend on a single parameter, the string tension or energy per unit length $\mu$, set by the symmetry breaking scale~\cite{Blanco-Pillado:2024aca}:
\begin{equation}\label{eq:GMU}
    G\mu \approx 10^{-6} \left(\frac{v}{10^{16}~\mathrm{GeV}}\right)^2\,,
\end{equation}
where $v$ is the vacuum expectation value of the broken phase, and the gravitational constant is $G = 1/M_p^2$, with $M_p = 1.22 \times 10^{19}~\mathrm{GeV}$ the Planck mass.

The SGWB spectrum depends on the properties of the CS network. The contribution of long strings stretching across the Universe is generally smaller than that from loops~\cite{CamargoNevesdaCunha:2022mvg}. For oscillating loops, the spectral energy density is a weighted sum over harmonic modes $k$:
\begin{equation}\label{eq:Om_GW_CS}
    \Omega_{\mathrm{GW}}(f) = \frac{8\pi}{3H_0^2} (G\mu)^2 f \sum_{k=1}^\infty P_k C_k(f)\,,
\end{equation}
where $H_0 = 100h~\mathrm{km/s/Mpc}$ is Hubble constant today. The power emitted in mode $k$ is \cite{Blanco-Pillado:2024aca}
\begin{equation}\label{eq:power_k}
    P_k = \frac{\Gamma k^{-q}}{\zeta(q)}\,,
\end{equation}
with total power $\Gamma \approx 50$ in units of $G\mu^2$~\cite{Blanco-Pillado:2017oxo}, and $\zeta(q)$ the Riemann zeta function. The index $q$ depends on the emission mechanism: $q=5/3$ for kinks, $q=4/3$ for cusps, and $q=2$ for kink-kink collisions. We only consider emission from cusps because it is dominant.
The weight function $C_k(f)$ is defined as
\begin{equation}\label{eq:CK_weight}
    C_k(f) = \frac{2k}{f^2} \int_0^\infty \frac{dz}{H(z)(1+z)^6} \, n\left(\frac{2k}{(1+z)f}, t(z)\right),
\end{equation}
where $z$ is the redshift, $n(l,t)$ is the loop number density and $t(z)$ is cosmic time. Assuming a flat Friedmann-Robertson-Walker cosmology, the Hubble parameter is
\begin{equation}\label{eq:Hubble}
    H(z) = H_0 \sqrt{1 - \Omega_M - \Omega_R + \Omega_M(1+z)^3 + \Omega_R \mathcal{C}(z)(1+z)^4}\,,
\end{equation}
with $\Omega_M = 0.3081$ and $\Omega_R = 1.291 \times 10^{-5}$~\cite{Planck:2018vyg}. The correction factor,
\begin{equation}\label{eq:CZ_factor}
    \mathcal{C}(z) = \frac{g_*(z)}{g_*(0)} \left(\frac{h_*(z)}{h_*(0)}\right)^{-4/3}\,
\end{equation}
arises from the varying relativistic degrees of freedom in the early Universe. Here, $g_*$ and $h_*$ are the energy and entropy degrees of freedom, respectively. Although $\mathcal{C}(z)$ varies with redshift,  fixing $\mathcal{C} = 0.8$ is a good approximation in the frequency range $[10^{-3},10]\,\mathrm{Hz}$~\cite{Marfatia:2023fvh}.

Analytical approximations for the SGWB spectrum in different frequency ranges are provided in Ref.~\cite{Marfatia:2023fvh}. In the nHz range, relevant for Pulsar Timing Arrays (PTAs), the energy density spectrum is approximated by
\begin{equation} \label{eq:Omega_hsq_PTA_approx}
	h^2 \Omega_{\rm GW}(f)\big|_{\rm PTA} \approx 
    4.2 \times 10^{-9} \left( \frac{f}{f_{\text{yr}}} \right)^{3/2} \frac{\Gamma}{50} \left( \frac{G\mu}{10^{-11}} \right)^2\times
	\sum_{k=1}^{\infty} \frac{k^{-17/6}}{(1 + 2.075\, u_k)^{1.945} \, \zeta(17/6)}\,,
\end{equation}
where $f_{\mathrm{yr}} \simeq 32~\mathrm{nHz}$ and
\begin{equation}\label{eq:uk_fun}
    u_k = \frac{2.89}{2k} \frac{f}{f_{\mathrm{yr}}} \frac{\Gamma}{50} \left(\frac{G\mu}{10^{-11}}\right)\,.
\end{equation}
Above a mHz, relevant for space-based and ground-based detectors, the spectrum is approximately flat and given by
\begin{equation}\label{eq:Omega_hsq_laser}
    h^2 \Omega_{\rm GW}(f) \big|_{\rm laser} \simeq 4.78 \times 10^{-5} \, \mathcal{C} \sqrt{G\mu}\,.
\end{equation}

\subsection{Domain walls}\label{sec:DW}

If a phase transition spontaneously breaks a discrete symmetry, domain walls form~\cite{Kibble:1976sj}. 
Cosmologically stable DWs overclose the early Universe because their energy density falls slower ($\sim 1/t$), than radiation in the radiation-dominated era ($\sim 1/t^2$) and matter in the matter-dominated era ($\sim 1/t^2$).
However, if the discrete symmetry is softly broken, the DWs become unstable and collapse. These effects can be parameterized by an explicit symmetry-breaking term that introduces a bias in the potential, $V_\mathrm{bias}$, and lifts the degeneracy in the 
vacua. The energy released during DW collapse is a source of gravitational radiation. 

\begin{figure}
    \centering
    \includegraphics[width=\textwidth]{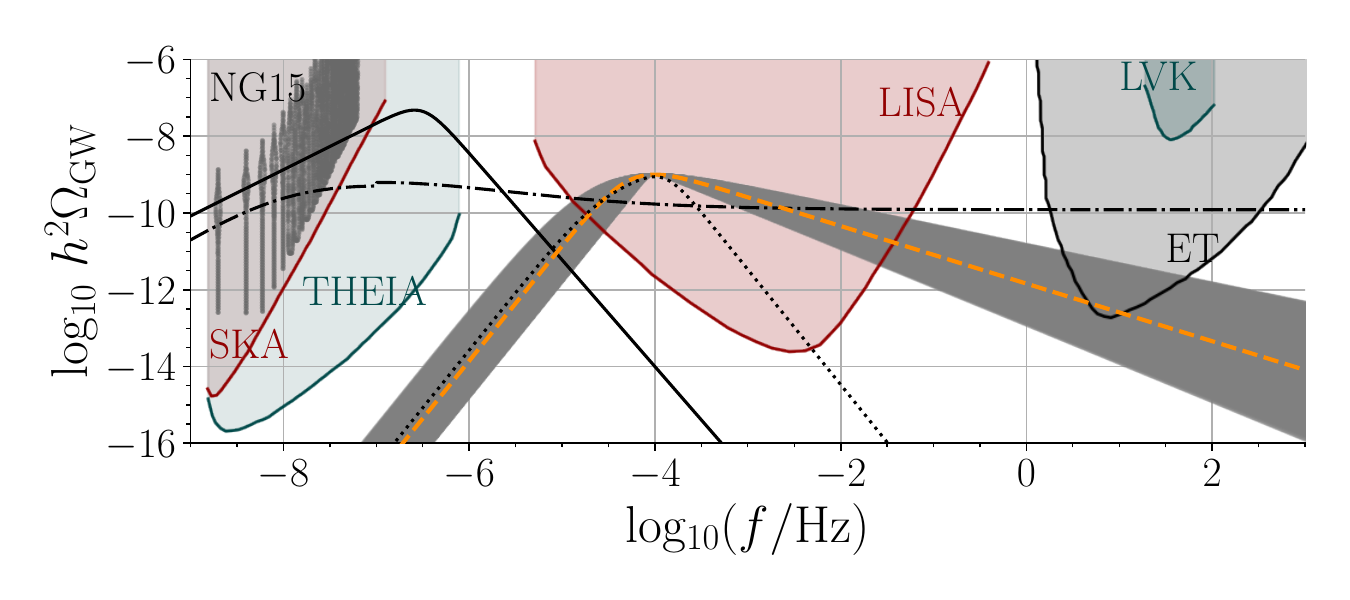}
    \caption{
    The orange dashed curve is the SGWB spectrum for DWs for spectral parameters $b = 0.75$ and $c = 1.65$, and the gray band corresponds to the DW spectrum for $b \in [0.5, 1]$ and $c \in [0.3, 3]$. The DW parameters are $\mathrm{log_{10}} [\mathcal{E}/\mathrm{GeV}^3] = 27.36$ and $\mathrm{log_{10}} [V_\mathrm{bias}/\mathrm{GeV}^4] = 19.15$. The black dash-dotted curve is the GW signal from CSs with $G\mu = 10^{-11}$.
    The black dotted curve is the SGWB for an FOPT with $\beta/H(T_*) = 50$, $T_* = 37.81~\mathrm{GeV}$, $v_w = 1$ and $\alpha = 1.64$. The black solid curve is the SIGW signal with $\mathrm{log_{10}}A = -7.32$, $\mathrm{log_{10}} (f_c/\mathrm{Hz}) = -6.59$, $\alpha_\mathrm{SIG} = 1.21$ and $\beta_\mathrm{SIG} = 2.78$. The 15-year NANOGrav dataset is shown as gray periodograms~\cite{NANOGrav:2023hvm}, and the LVK bound and sensitivities of various experiments are also shown. }
    \label{fig:DW_bc}
\end{figure}

Assuming DWs collapse approximately instantaneously during the radiation-dominated epoch, the peak frequency and amplitude of the SGWB today can be expressed as~\cite{Saikawa:2017hiv}
\begin{equation}\label{eq:ompeak_fpeak_DW}
    \begin{aligned}
    	& h^2\Omega^{\rm peak}_{\rm GW} \simeq 1.49 \times 10^{-10} \times \left(\frac{10.75}{g_\star}\right)^{1/3} \left(\frac{\mathcal{E}^{1/3}}{10^7\ \rm GeV}\right)^{12} \left(\frac{10^7\ \rm GeV^4}{V_{\rm bias}}\right)^{2}\,, 
        \\
    	& f_{\rm peak} \simeq 5.93 \times 10^{-6}\ {\rm Hz} \times \left(\frac{10^7\ \rm GeV}{\mathcal{E}^{1/3}}\right)^{3/2} \left(\frac{V_{\rm bias}}{10^7\ \rm GeV^4}\right)^{1/2}\,, 
    \end{aligned}
\end{equation}
where $\mathcal{E}$ is the surface energy density of the DW. Throughout, we fix the number of relativistic degrees of freedom to $g_* = 106.75$. We can express the GW spectrum as  
\begin{equation}
    h^2 \Omega_\mathrm{GW}(f) = h^2\Omega_\mathrm{GW}^\mathrm{peak} S(f/f_\mathrm{peak})\,,
\end{equation}
where $S$ is a spectral smoothing function given by \cite{Ferreira:2022zzo}
\begin{equation}
    S(f) = \frac{(a+b)^c}{(bf^{-a/c} + af^{b/c})^c}\,.
\end{equation}
Causality fixes $a = 3$, while numerical simulations allow $0.5 \leq b \leq 1$ and $0.3 \leq c \leq 3$~\cite{Ferreira:2022zzo}. The parameter $c$ controls the slope of the spectrum for frequencies above $f_{\mathrm{peak}}$, and $b$ governs the slope below $f_{\mathrm{peak}}$. In Fig.~\ref{fig:DW_bc}, the gray band shows the impact of varying $b \in [0.5,1]$ and $c \in [0.3,3]$ on the spectral shape, and the dashed orange curve corresponds to the central values $b = 0.75$ and $c = 1.65$, fixed in our numerical analysis. For comparison, we also show spectra for SIGW (solid black curve), CSs (dash-dotted black curve), and an FOPT (dotted black curve). 

\subsection{First-order phase transitions}\label{sec:PT}

First-order phase transitions, from which cosmic strings and domain walls may originate, produce GWs themselves. These transitions occur as the Universe cools and the scalar potential that governs the FOPT acquires a new global minimum. The true and false vacua are separated by a potential barrier, and transitions between them proceed via quantum tunneling, thermal fluctuations, or a combination thereof. During an FOPT, bubbles of the true vacuum nucleate and grow, driven by the latent heat released in the transition. These bubbles subsequently collide and merge, eventually filling the entire Universe. The SGWB from an FOPT is sourced by three distinct processes: 1) collisions of highly relativistic true-vacuum bubbles; 2) bulk fluid motion in the form of sound waves; and 3) magnetohydrodynamic turbulence. In our numerical analysis, we only consider the sound-wave contribution, which often dominates.

The dynamics of phase transitions can be described by four thermodynamic parameters: 1) the phase transition strength $\alpha$, which measures the amount of latent heat released; 2) the bubble wall velocity $v_w$; 3) the inverse time duration of the FOPT normalized to the Hubble rate, $\beta/H(T_*)$ and 4) the temperature at which the phase transition ends, $T_*$.

The sound-wave contribution to the SGWB spectrum is modeled by a double broken power law~\cite{Caprini:2024hue},
\begin{equation}\label{eq:DBPL}
\Omega^{\rm SW}_{\GW}(f, \Omega_2, f_1, f_2) = \Omega_\text{int} \times N \left( \frac{f}{f_1} \right)^{n_1} \left[ 1 + \left( \frac{f}{f_1} \right)^{a_1} \right]^{\frac{- n_1 + n_2}{a_1}} \left[1 + \left( \frac{f}{f_2} \right)^{a_2} \right]^{\frac{- n_2 + n_3}{a_2}} \,, 
\end{equation}
where the fixed fit parameters are $n_1 = 3$, $n_2 = 1$, $n_3 = -3$, $a_1 = 2$ and $a_2 = 4$ \cite{Caprini:2024hue}. $N$ is a normalization factor such that $\int^{+\infty}_{-\infty} S(f) d\ln f = 1$, where $S(f)$ is the RHS/$\Omega_\text{int}$. The geometric and thermodynamic parameters are related by
\begin{equation}
    \begin{aligned}
        & f_1 \simeq 0.2 \, H_{*,0} \, (H(T_*) R_*)^{-1} \,, \\
        & f_2 \simeq 0.5 \, H_{*,0} \, \Delta_w^{-1}~(H(T_*) R_*)^{-1} \,, \\
        & \Omega_\mathrm{int} = 0.11 F_{\GW,0} \, K^2 \left( H(T_*) \tau_{\rm SW} \right) \left( H(T_*)R_* \right) \,,
    \end{aligned}
\end{equation}
where $\Delta_w = |v_w - c_s|/\mathrm{max}(v_w,c_s)$. We assume a speed of sound $c_s = 1/\sqrt{3}$ and fix $v_w = 1$.  In terms of the inverse time duration, the average bubble size $R_*(T_*)$ is given by
\begin{equation}\label{eq:bubble_size}
  H(T_*)R_*(T_*) = (8\pi)^{1/3}\mathrm{max}(v_w,c_s) \left(\frac{\beta}{H(T_*)}\right)^{-1}\,.
\end{equation}
The lifetime of sound waves in units of Hubble time is $H(T_*) \tau_{\mathrm{SW}} = \mathrm{min} (2H(T_*) R_*(T_*)/\sqrt{3K}, 1)$ where $K = 0.6\alpha/(1+\alpha)$. The redshift factors are
\begin{equation}\label{eq:redshit_H_FGW0}
    \begin{aligned}
        &H_{*,0} \simeq 1.65\times 10^{-5}~\mathrm{Hz}\,\left(\frac{g_*(T_*)}{100}\right)^{1/6} \left(\frac{T_*}{\mathrm{GeV}}\right)\,, \\
        &h^2 F_{\mathrm{GW},0}  \simeq 1.65\times 10^{-5} \left(\frac{100}{g_*(T_*)}\right)^{1/3}\,.
    \end{aligned}
\end{equation}

\subsection{Scalar-induced GWs}

Scalar-induced GWs (SIGWs) are another possible source of cosmological GWs. At linear order in cosmological perturbation theory, scalar and tensor modes evolve independently. However, at second order, these modes become coupled with each other. Therefore, large first-order scalar perturbations can generate second-order tensor perturbations. These tensor perturbations appear as GWs when the scalar modes reenter the horizon at the end of inflation~\cite{Matarrese:1992rp}.
To maintain a model-independent approach, we consider the broken power-law spectrum~\cite{Cai:2023dls},
\begin{equation}
    h^2\Omega_{\mathrm{GW}}(f) =A \frac{\alpha_\mathrm{SIG}+\beta_\mathrm{SIG}}{\beta_\mathrm{SIG}\left(f / f_{c}\right)^{-\alpha_\mathrm{SIG}}+\alpha_\mathrm{SIG}\left(f / f_{c}\right)^{\beta_\mathrm{SIG}}}\,,
\end{equation}
where $A$ is the amplitude and $f_c$ is the peak frequency. The spectral indices $\alpha_\mathrm{SIG}$ and $\beta_\mathrm{SIG}$ control the slopes for $f < f_\mathrm{c}$ and $f > f_\mathrm{c}$, respectively.

\subsection{Backgrounds}\label{sec:backgrounds}

We describe the astrophysical backgrounds relevant to our analysis. Note that different backgrounds apply in different frequency ranges. 

\subsubsection*{PTA band ($f < 10^{-7}~\mathrm{Hz}$)}

Supermassive black hole binaries (SMBHBs) are believed to be formed from the merger of two galaxies. The black holes of each galaxy become gravitationally bound and form a binary system. Once they become sufficiently close, GW emission becomes efficient. Binaries with a combined mass in the range $[10^8, 10^{10}]~M_\odot$ are expected to emit GWs in the $[10^{-9},10^{-7}]~\mathrm{Hz}$ frequency range~\cite{Rajagopal:1994zj}. In this narrow band, the spectrum follows an approximate power law,
\begin{equation}\label{eq:smbhbs_pta_powerlaw}
    h^2 \Omega^\mathrm{PTA}_{\rm {GW}}(f)={\frac{2\pi^2 A_{\rm {BHB}}^2}{3H_0^2}}\bigg(\frac{f}{\rm{year}^{-1}}\bigg)^{5-\gamma_{\rm BHB}} {\rm {year}}^{-2}\,,
\end{equation}
where $A_\mathrm{BHB}$ and $\gamma_\mathrm{BHB}$ determine the amplitude and tilt of the spectrum, respectively. As in Ref.~\cite{NANOGrav:2023hvm}, we sample $(\mathrm{log_{10}} A_\mathrm{BHB},\gamma_\mathrm{BHB})$ with a bivariate normal distribution, whose mean and covariance matrix are
\begin{equation}\label{eq:musigma}
\bm{\mu}_{\scriptscriptstyle\rm BHB} = \begin{pmatrix} -15.6 \\ 4.7 \end{pmatrix} \,, \qquad \bm{\sigma}_{\scriptscriptstyle\rm BHB} = 10^{-1}\times\begin{pmatrix} 2.8 & -0.026 \\ -0.026 & 1.2 \end{pmatrix} \,.
\end{equation}

\subsubsection*{Interferometer band ($10^{-5}~\mathrm{Hz} \leq f < 10^3~\mathrm{Hz}$) }

\begin{figure*}[t]
	\centering
    \subfloat{\includegraphics[width=0.8\textwidth]{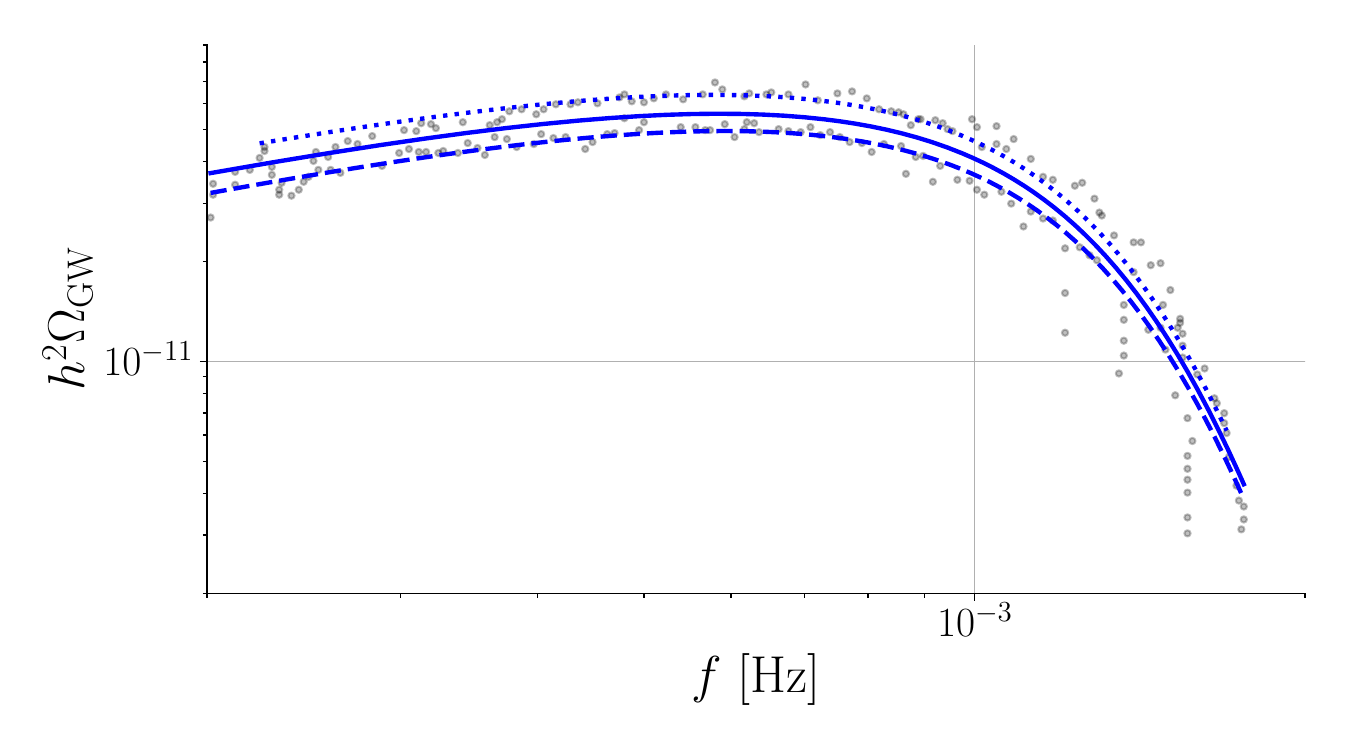}}
    \caption{SGWB spectrum from ultra-compact binaries in the Milky Way. The gray points are digitized from Fig.~3 of Ref.~\cite{Karnesis:2021tsh}. The solid curve corresponds to $n_s = -7/3$, the dashed curve to $n_s = -2.293$, and the dotted curve to $n_s = -2.327$. For all three curves, the other parameters are fixed to the best-fit point of Table~II of Ref.~\cite{Karnesis:2021tsh}, assuming an SNR threshold of 5 and median smoothing.}\label{fig:back_ns}
\end{figure*}

In this band, two contributions dominate: an extragalactic background from black hole and neutron star mergers outside the Milky Way, and a galactic background from ultra-compact binaries within our galaxy. The spectrum of the galactic background is modeled as \cite{Karnesis:2021tsh} 
\begin{equation}\label{eq:SGWB_gal}
h^2\Omega^{\textrm{Gal}}_{\rm GW}(f)= \frac{f^3}{2}
\left(\frac{f}{\textrm{Hz}}\right)^{n_s}   
\left[1+\tanh \left({\frac{f_{\textrm{knee}}-f}{f_2}} \right) \right] e^{-(f/f_1)^\nu} h^2 \Omega_{\rm Gal}  \,,
\end{equation}
where
\begin{equation}
\begin{aligned}
&\log_{10} \left( f_1 / {\rm Hz}\right) = a_1 \log_{10} \left( T_{\rm obs}/{\rm year} \right) + b_1\,, \\
&\log_{10} \left( f_{\textrm{knee}}/ {\rm Hz} \right) = a_k \log_{10} \left( T_{\rm obs}/{\rm year} \right) + b_k\,,
\end{aligned}
\end{equation}
with $a_1 = -0.15$, $b_1 = -2.78$, $a_k = -0.34$, $b_k = -2.55$, $\nu  = 1.66$ and $f_2 = 5.9 \times 10^{-4}$. These values correspond to the best-fit point for an SNR threshold of 5 with median smoothing (see Table~II of Ref.~\cite{Karnesis:2021tsh}). We vary the amplitude within 
$\mathrm{log_{10}}(h^2\Omega_\mathrm{Gal}) = -7.85 \pm 0.21$~\cite{Caprini:2024hue}, assuming a Gaussian prior.
We consider an observation time $T_\mathrm{obs} = 4~\mathrm{years}$. In Ref.~\cite{Karnesis:2021tsh}, the spectral index is fixed to $n_s = -7/3$, with the mention that allowing it to vary could lead to better fits. Keeping the other parameters fixed, we vary $n_s$ to determine the range of values for which the spectra fall within the gray band of Fig.~3 of Ref.~\cite{Karnesis:2021tsh}. Our analysis yields $n_s = -2.33^{+0.015}_{-0.055}$, as shown in \cref{fig:back_ns}.
We sample $n_s$ assuming a Gaussian prior. 

The extragalactic background follows a simple power law \cite{Caprini:2024hue},
\begin{equation} \label{eq:Ext_template}
    h^2 \Omega^{\textrm{Ext}}_{\rm GW}  = h^2 \Omega_{\rm Ext} \left(\frac{f}{f_\mathrm{ref}}\right)^{\alpha_\mathrm{Ext}}\,.
\end{equation}
Based on the results of Ref.~\cite{Babak:2023lro}, we sample $\mathrm{log_{10}} h^2 \Omega_{\rm Ext} = -12.28 \pm 0.17$ and $\alpha_\mathrm{Ext} = 0.66 \pm 0.34$ assuming Gaussian priors and reference frequency $f_\mathrm{ref} = 10^{-3}~\mathrm{Hz}$. For frequencies above $100~\mathrm{Hz}$, the supernova contribution impacts the shape of the spectrum and induces deviations from Eq.~\eqref{eq:Ext_template}~\cite{Zhou:2022nmt}. However, this contribution is subdominant, and we neglect it in our analysis.

\begin{figure}[htb!]
    \centering
    \includegraphics[width=\textwidth]{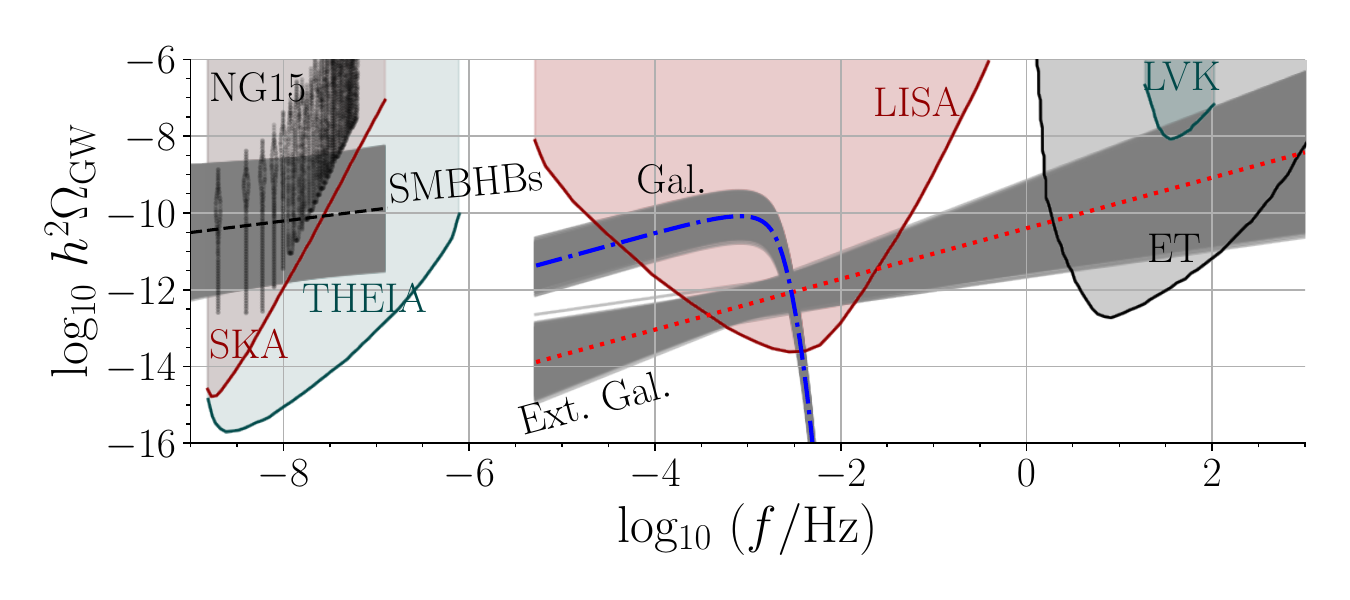} \\
    \caption{The main astrophysical backgrounds in the PTA and the interferometer bands. The dashed curve is the SMBHB background, the blue dash-dotted curve is the galactic background, and the red dotted curve is the extragalactic background. The bands show the $1\sigma$ uncertainties. 
}
    \label{fig:PTA_Bkg}
\end{figure}

In sum, the background spectrum for each contribution is shown in \cref{fig:PTA_Bkg}.
It is evident that the SMBHB template in the PTA band does not match the backgrounds in the interferometer band, despite the expectation that the SMBHB contribution should extend to higher frequencies, and eventually fall around 10~Hz; see Fig.~17 of Ref.~\cite{Ellis:2023oxs}. This discrepancy arises because, in constructing the interferometer background templates in Eqs.~\eqref{eq:SGWB_gal} and~\eqref{eq:Ext_template}, contributions from nearby, individually resolved SMBHB sources have been subtracted. Consequently, the expected background amplitude in the interferometer band is lower than in the PTA band. 
We present our numerical results separately for the PTA and interferometer bands.

\section{Results}\label{sec:results}

To quantify how Prominence can help discriminate among different SGWB sources, we simulate datasets with backgrounds varied within the previously mentioned ranges.  Each combination of background parameters produces a GW spectrum with a different shape and, consequently, different values of Prominence. For each signal hypothesis, we generate spectra by sampling the background parameters according to the distributions in Section~\ref{sec:GWs_backs}. For each individual peak, we construct a probability density function (PDF) of $\mathcal{P}$ corresponding to the sampled background spectra. This PDF characterizes the distribution of Prominence values over the possible background realizations. If the GW spectrum has multiple peaks, multiple values of $\mathcal{P}$ are computed for a given background, one for each peak. We separately construct a PDF for each individual peak and then sum the peak-specific PDFs with equal weights. The resulting distribution is normalized such that its integral over the full range of $\mathcal{P}$ is unity. This defines the PDF for a spectrum with multiple peaks. We compute $\mathcal{P}$ within the sensitivity regions of the LISA/ET/SKA experiments. The bin size is selected to match the expected uncertainties in the measurement of $h^2 \Omega_\mathrm{GW}$, since $\cal{P}$ is a relative measure of $h^2 \Omega_\mathrm{GW}$ and therefore sensitive to its precision. Current parameter reconstruction techniques utilized by LISA project the amplitude of a FOPT signal to be determined with an uncertainty at the level of $1\%$ to $2\%$~\cite{Caprini:2024hue}, which we take as a reference range in our numerical analysis. To determine $p$-values, we employ two different statistical tests: the two-sample Kolmogorov-Smirnov (KS) test~\cite{1958ArM3469H} and the Cramer-von Mises (CvM) test~\cite{10.1214/aoms/1177704477}. 

The KS test statistic to evaluate the similarity of two SGWB spectra measures the maximum difference between the cumulative distribution functions $\mathrm{CDF}_1$ and $\mathrm{CDF}_2$ of their $\cal{P}$ values: 
\begin{equation}\label{eq:KS_statistic}
    \mathrm{KS} = \max_{\cal{P}}\left|\mathrm{CDF}_{1}({\cal{P}})-\mathrm{CDF}_2(\cal{P})\right|\,.
\end{equation}
The CvM test statistic is defined as 
\begin{equation}\label{eq:CVM_statistic}
    \mathrm{CvM}=\int_{0}^{\infty}\left[\mathrm{CDF}_{1}({\cal{P}})-\mathrm{CDF}_2(\cal{P})\right]^{2} \mathrm{~d} \mathrm{CDF}_1(\cal{P})\,.
\end{equation}
For both test statistics, the null hypothesis is that the two distributions are identical, \textit{i.e.}, ${\rm{CDF}}_1({\cal{P}}) = {\rm{CDF}}_2(\cal{P})$ for all $\cal{P}$.
Accordingly, lower values of the test statistics (corresponding to higher $p$-values) indicate that the distributions are similar, while higher values of the test statistics (lower 
$p$-values) suggest large differences.

Throughout, we provide the SNR value of each spectrum, defined as
\begin{equation}\label{eq:SNR_calc}
    \mathrm{SNR} = \sqrt{ T_\mathrm{obs} \int_{f_\mathrm{min}}^{f_\mathrm{max}} df\, \frac{\Omega_{\mathrm{GW}}^2(f)}{ \Omega_{\mathrm{Sens}}^2(f)} } \,,
\end{equation}
where $\Omega_{\mathrm{GW}}(f)$ is the SGWB spectrum (comprised of signal and background) and $\Omega_{\mathrm{Sens}}(f)$ is the expected experimental sensitivity. The integration limits, $f_\mathrm{min}$ and $f_\mathrm{max}$, are set to the limits of the experimental sensitivity curve. Since the background is treated dynamically, SNR varies depending on the specific choice of background parameters. To adopt a conservative approach, we report the smallest SNR obtained. All signals are chosen such that $\mathrm{SNR} > 10$. 

\subsection{Impact of amplitude uncertainties}

Effects like noise fluctuations can lead to less accurate reconstruction of the underlying GW signal, which is reflected in a larger relative amplitude uncertainty $\Delta \Omega/\Omega$. We therefore start by investigating how the uncertainty in the GW amplitude affects the performance of our Prominence-based discrimination method. Consider the simplest scenario by comparing single-peaked GW signals arising from FOPTs and DWs. We focus on the most conservative case, where both the peak amplitude and frequency are identical for the FOPT and DW signals. 

\begin{figure*}[htb!]
	\centering
    \subfloat[]{\includegraphics[width=0.5\textwidth]{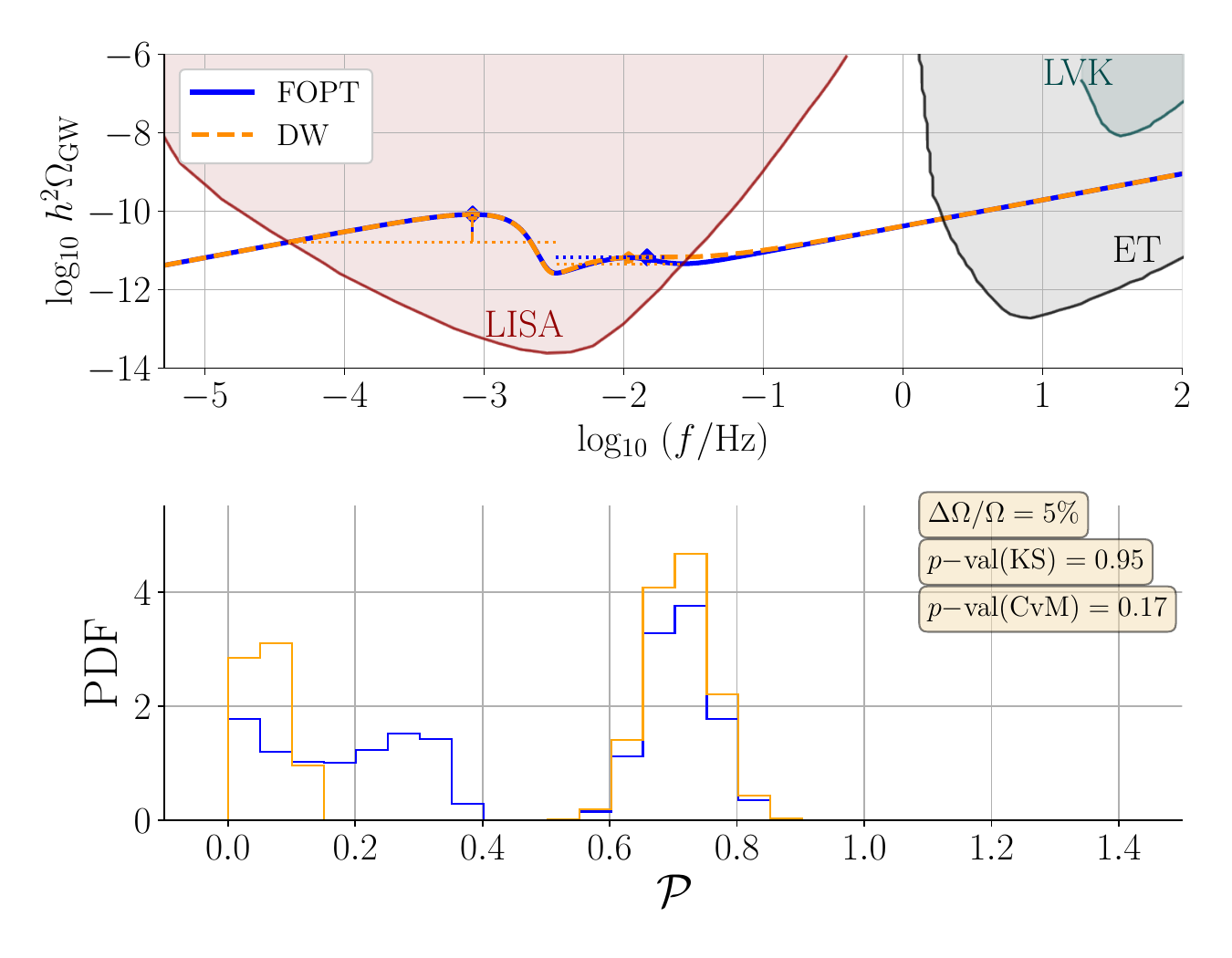}} 
    \subfloat[]{\includegraphics[width=0.5\textwidth]{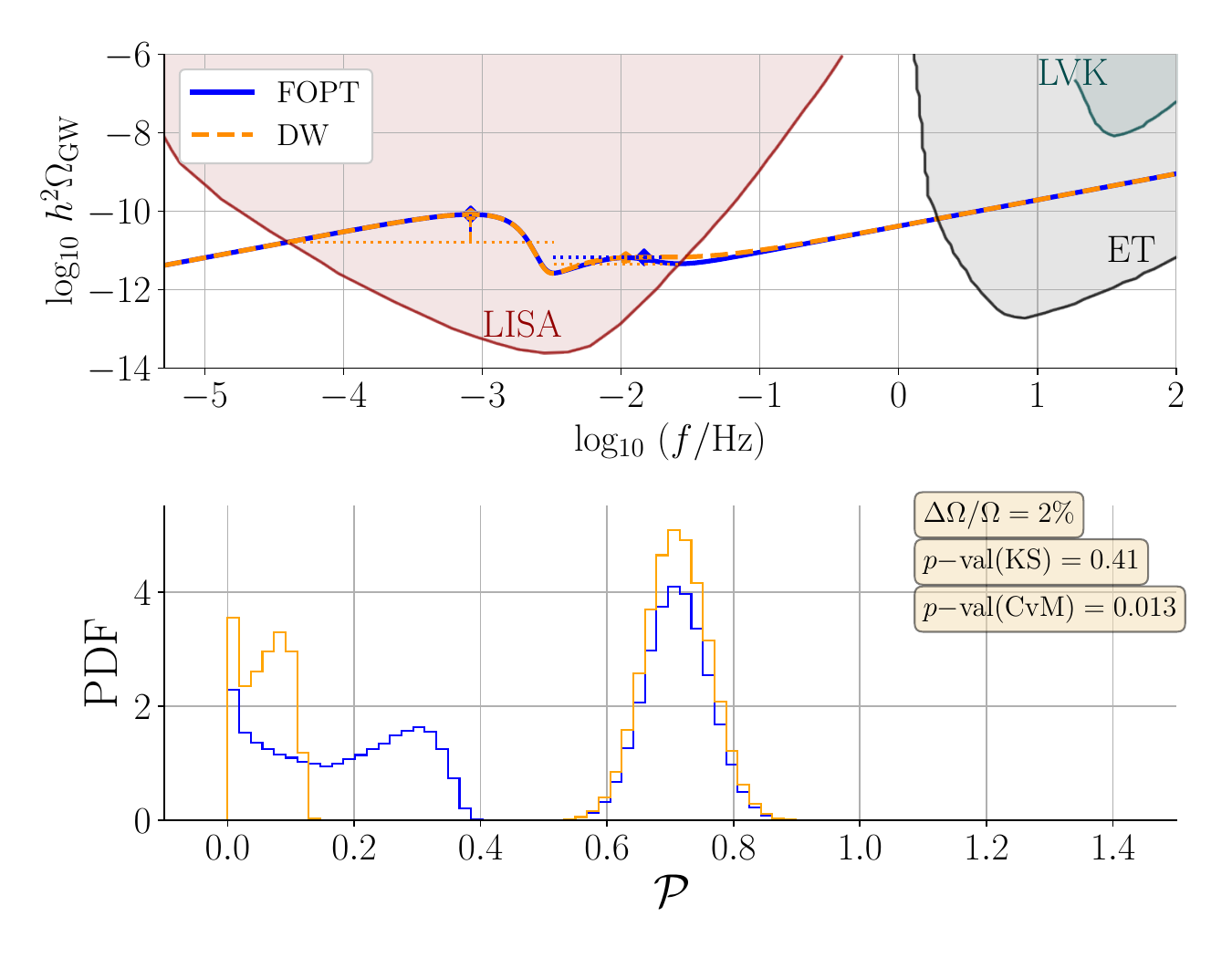}} \\
    \subfloat[]{\includegraphics[width=0.5\textwidth]{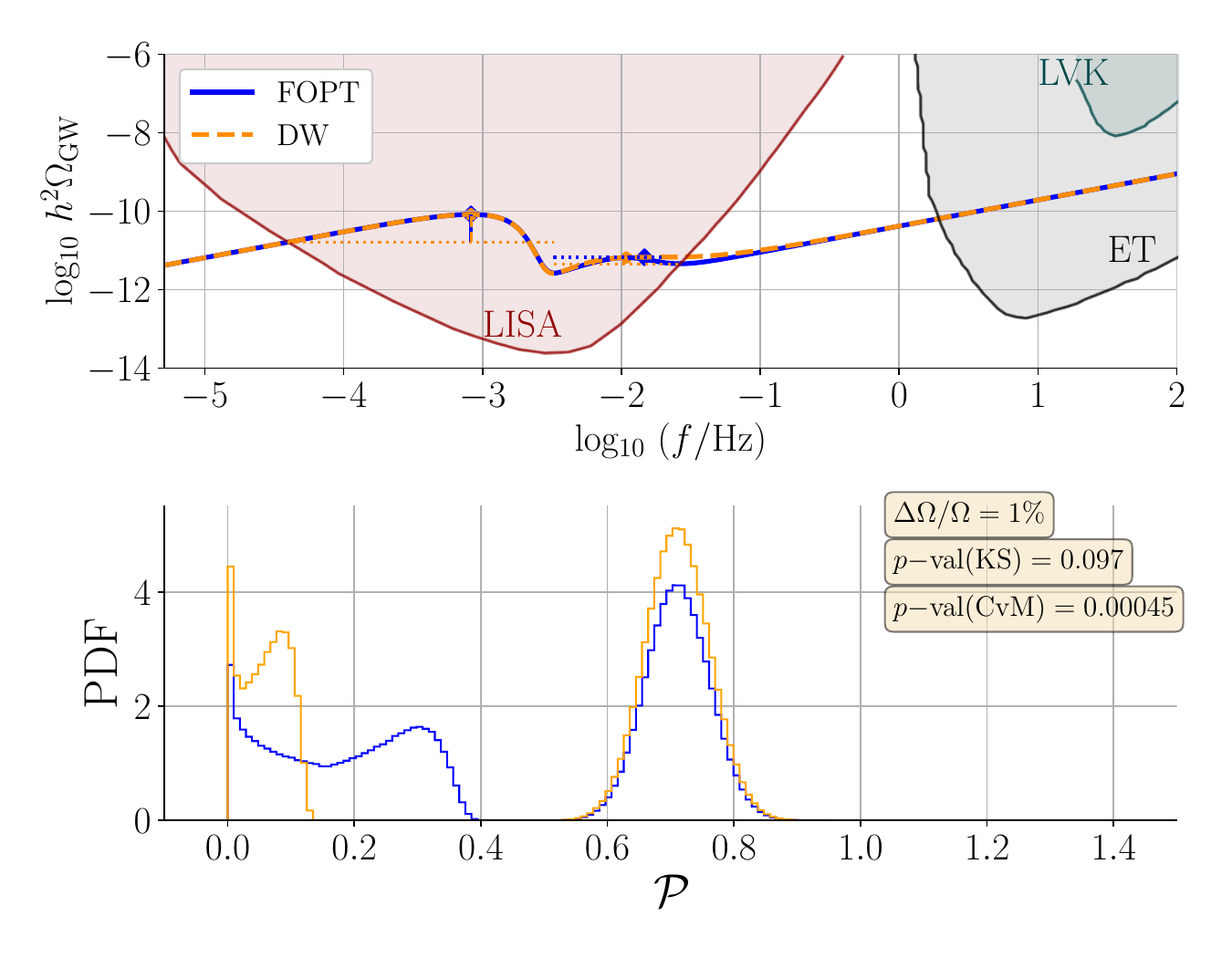}} 
    \subfloat[]{\includegraphics[width=0.5\textwidth]{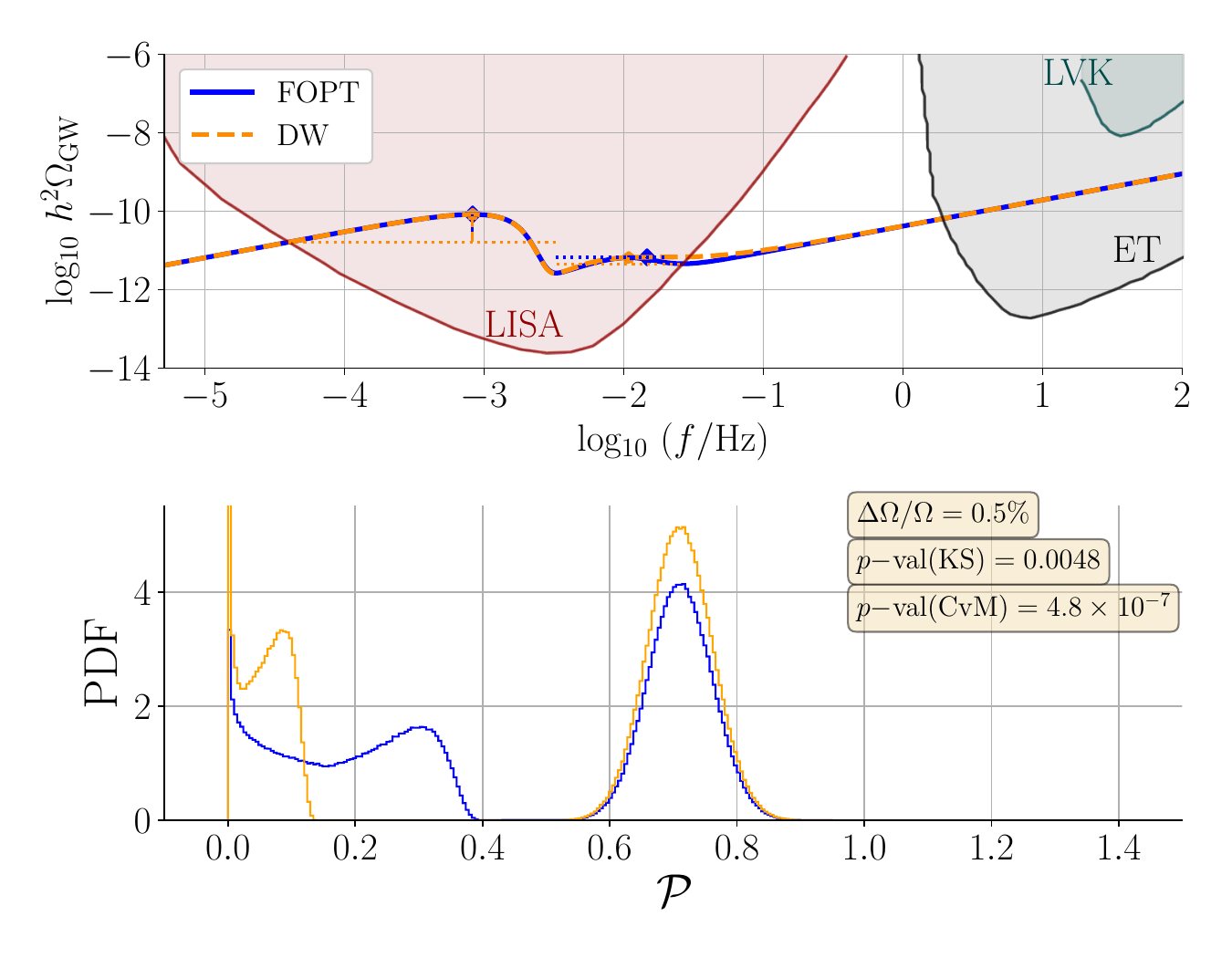}} \\
    \caption{SGWB spectra and the corresponding PDFs of $\cal{P}$. For the GW spectra, the background parameters are fixed to their central values. Blue (orange) curves correspond to the FOPT (DW) source. The legend displays the $p$-values from the KS and CvM tests, the measured uncertainty in $h^2\Omega_\mathrm{GW}$. The minimum SNR of the two signals at LISA and ET are 22.1 and 102.3, respectively. The vertical dashed lines indicate the Prominences of the peaks, and horizontal dotted lines mark their bases. Note how the discriminating power of $\mathcal{P}$ changes with the uncertainty. Even for largely overlapping signals, a discriminating power above 95\% CL can be achieved for $\Delta\Omega/\Omega = 1\%$.}
    \label{fig:prom_SNR_vary}
\end{figure*}

When constructing the Prominence PDFs, we determine the window border by the horizontal segment between the points where the signal intersects the LISA sensitivity curve.
In each subpanel of \cref{fig:prom_SNR_vary}, we present the PDFs of $\cal{P}$ below the corresponding SGWB spectra. For these SGWB spectra, the background parameters in Eqs.~\eqref{eq:SGWB_gal}--\eqref{eq:Ext_template} are fixed to their central values. The blue curves correspond to the FOPT, and the orange curves correspond to DWs. In all panels, the DW signal is computed for $\mathrm{log_{10}} [\mathcal{E}/\mathrm{GeV}^3] = 26.86$ and $\mathrm{log_{10}} [V_\mathrm{bias}/\mathrm{GeV}^4] = 19.32$. The FOPT signal is determined with $\beta/H(T_*) = 100$, $T_* = 1889.61~\mathrm{GeV}$ and $\alpha = 4.56\times 10^{-2}$. The $p$-values from the KS and CvM tests and the percent uncertainties in $h^2\Omega_\mathrm{GW}$ are displayed in each panel. The minimum SNR of the FOPT and DW signals at LISA and ET are 22.1 and 102.3, respectively.  The Prominences of the peaks are indicated by vertical dashed lines, and their bases are marked by horizontal dotted lines.

\begin{figure*}[!htbp]
	\centering
    \renewcommand{\thesubfigure}{\Alph{subfigure}}
    \subfloat[]{\includegraphics[width=0.5\textwidth]{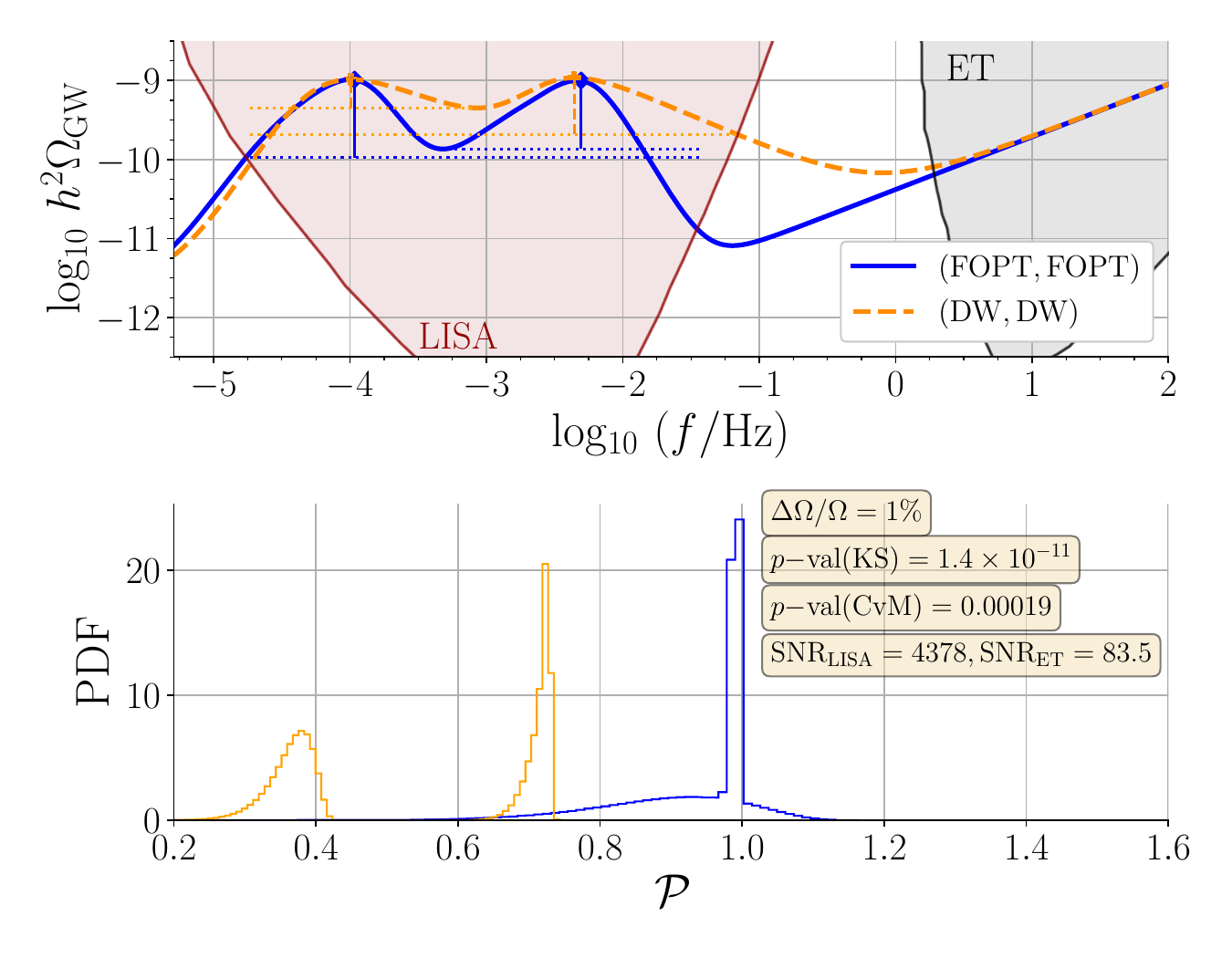}} 
    \subfloat[]{\includegraphics[width=0.5\textwidth]{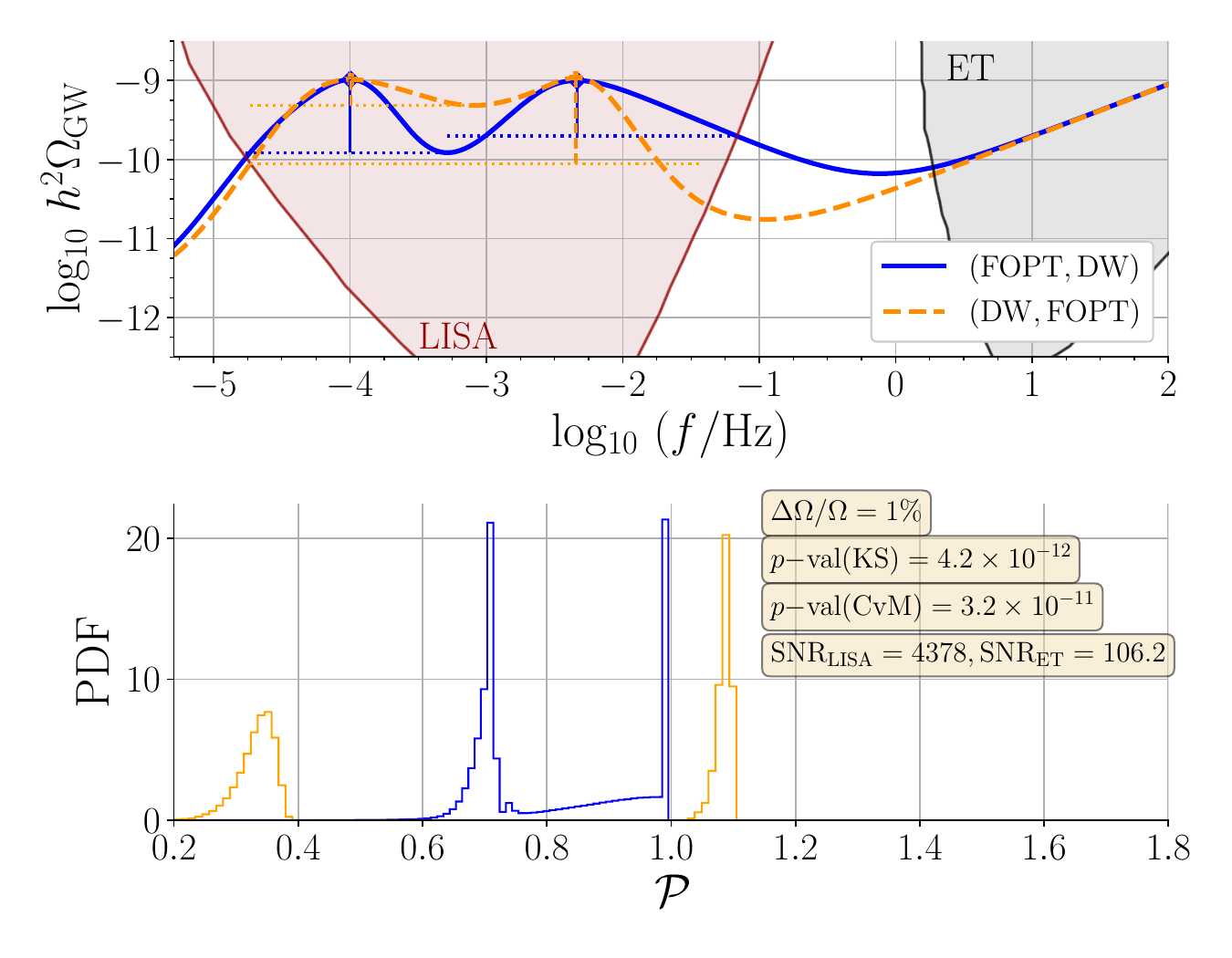}} \\
    \subfloat[]{\includegraphics[width=0.5\textwidth]{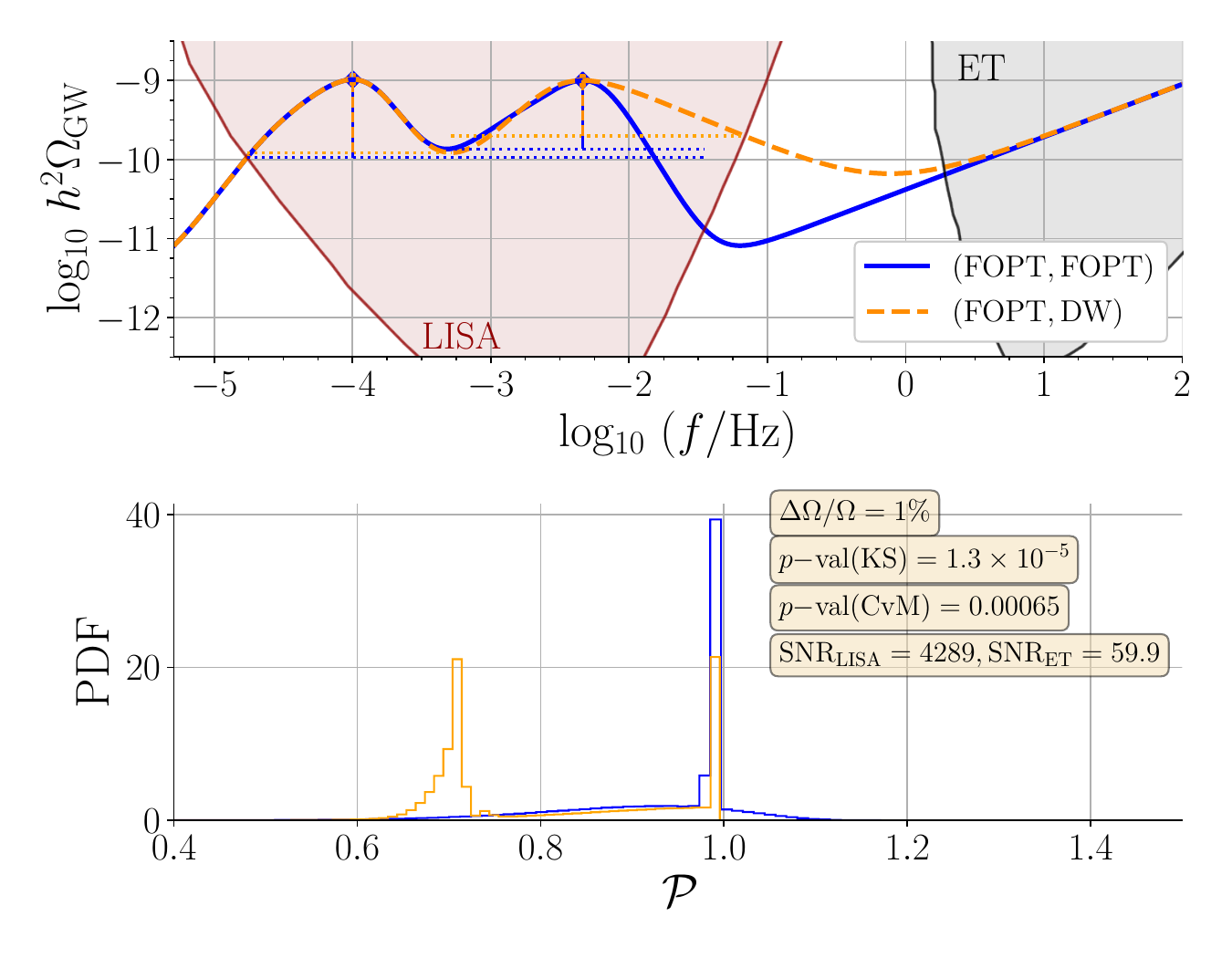}} 
    \subfloat[]{\includegraphics[width=0.5\textwidth]{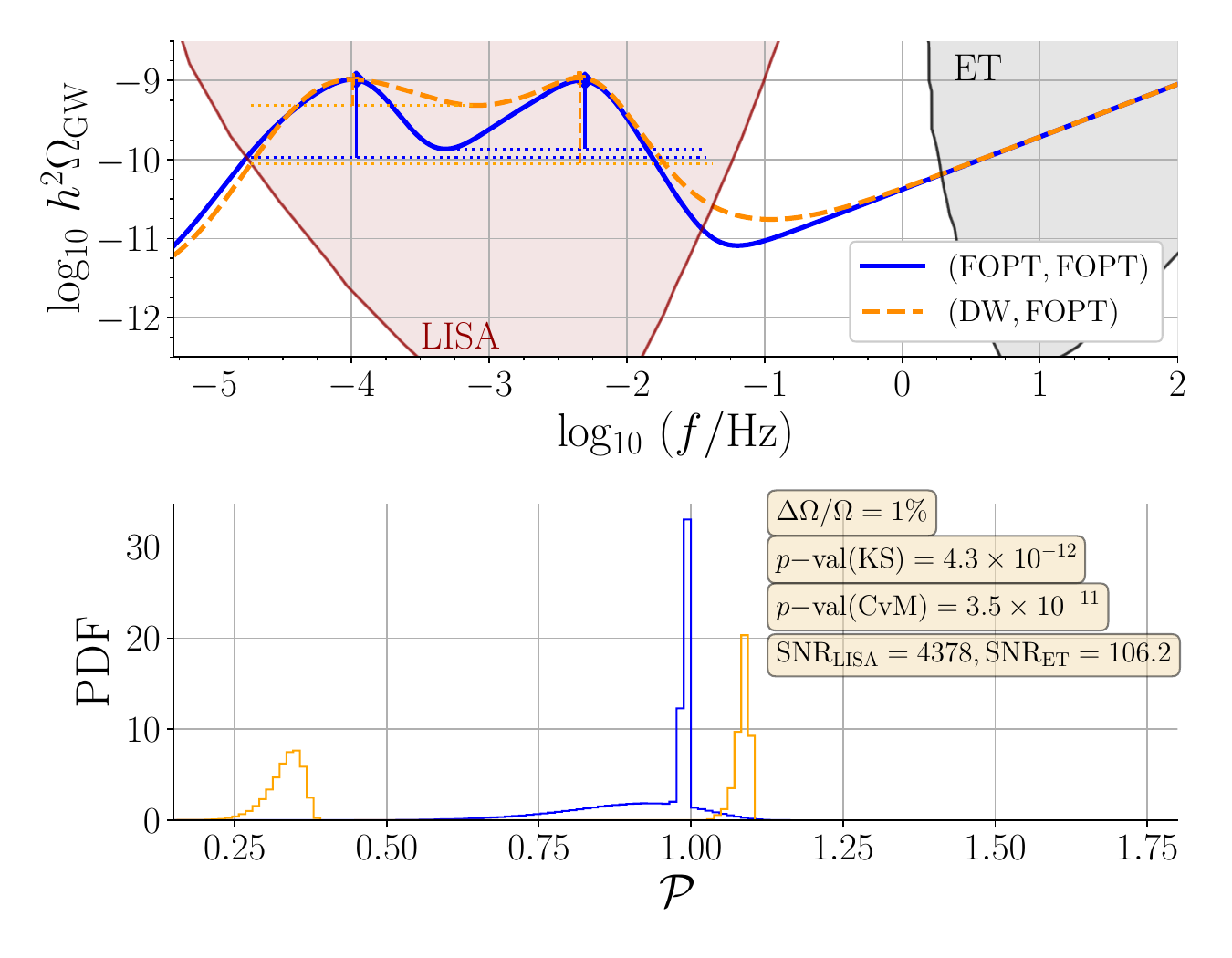}} \\
    \subfloat[]{\includegraphics[width=0.5\textwidth]{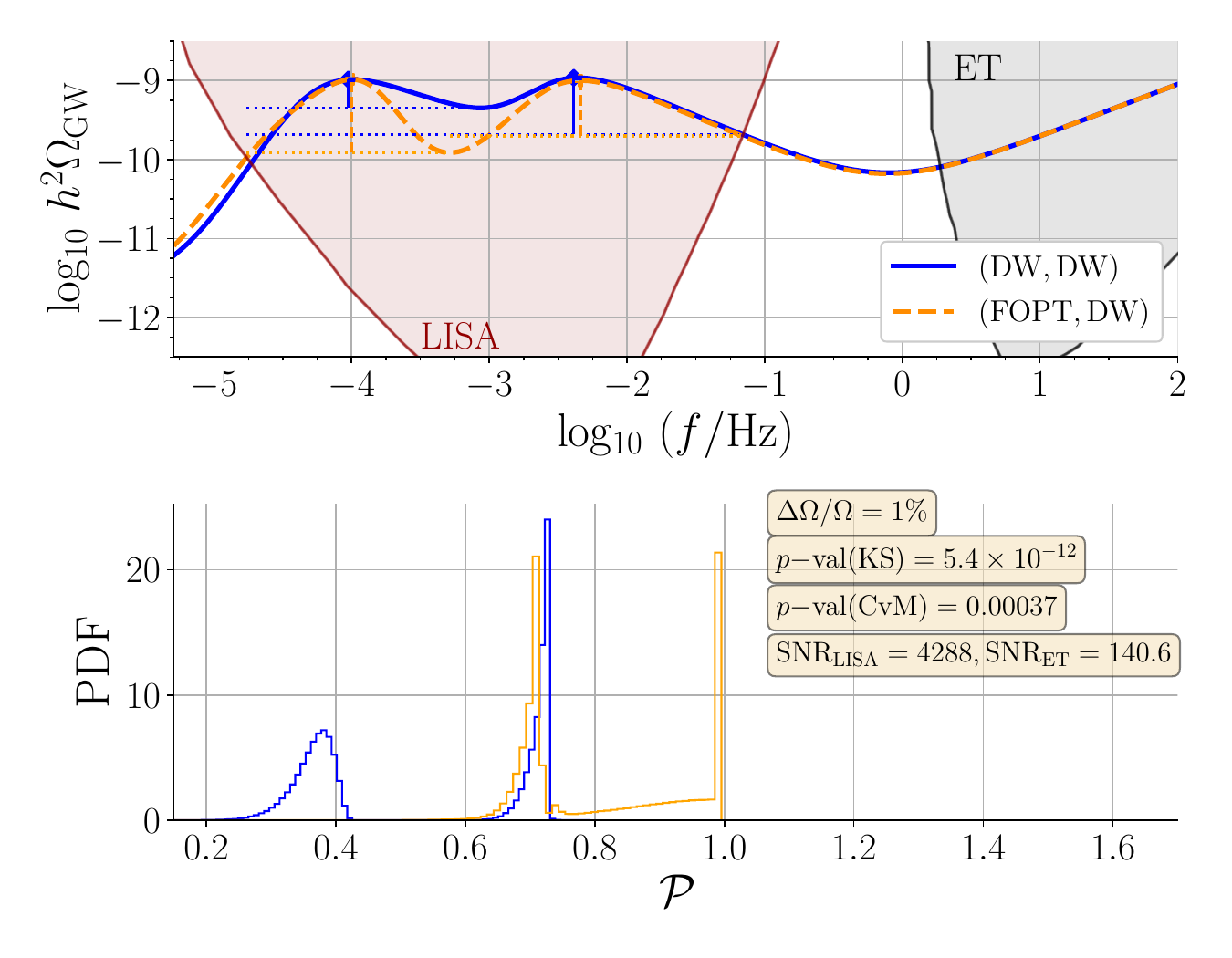}}
    \subfloat[]{\includegraphics[width=0.5\textwidth]{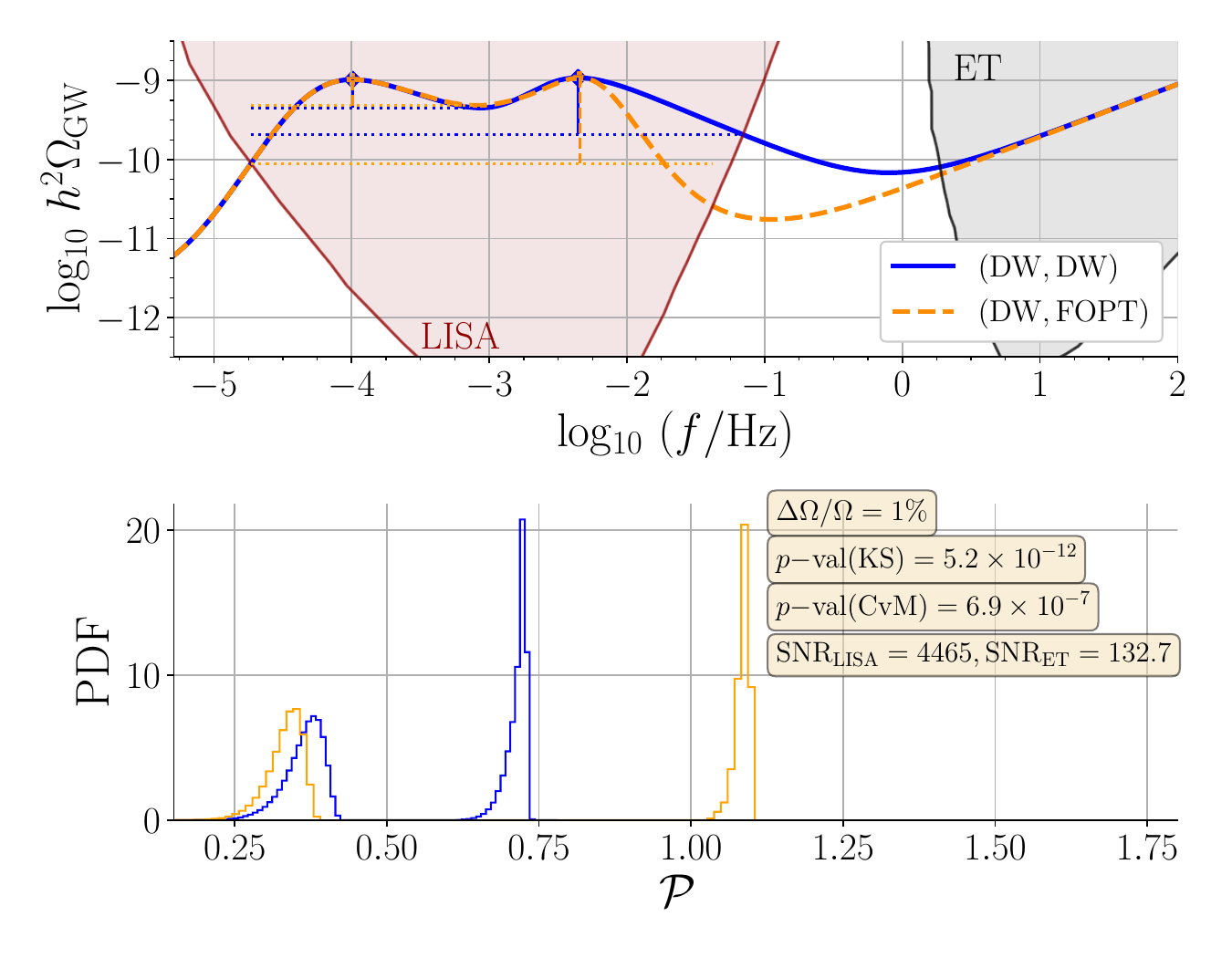}}
    \caption{Similar to Fig.~\ref{fig:prom_SNR_vary} for two-peaked GW spectra with both peaks in the LISA frequency band. Each panel is labeled in accordance with Table~\ref{tab:benchmark_cases}. We have fixed $\Delta\Omega/\Omega = 1\%$ based on the projected uncertainty achievable by LISA. Observe that $\mathcal{P}$ offers better discriminating power compared to the case of the single-peaked spectra in Fig.~\ref{fig:prom_SNR_vary}. 
    }
    \label{fig:prom_twopeak}
\end{figure*}

As a general trend, the $p$-values from the KS and CvM tests decrease as the uncertainties decrease. From panel~(a), which represents our most conservative case with a 5\% uncertainty in $h^2\Omega_\mathrm{GW}$, we find that both the KS and CvM tests yield large $p$-values (KS: 0.95, CvM: 0.17), which indicates no statistically significant difference between the FOPT and DW distributions. 
As panel~(b) shows, 98.7\% confidence level (CL) discrimination with the CvM statistic is possible with a 2\% uncertainty. For the projected uncertainty achievable with LISA, $\Delta \Omega/\Omega = 1\%$ (panel~(c)), the distinction between the FOPT and DW cases becomes more pronounced. Here, the CvM test returns a $p$-value of 0.00045, so the two sources can be distinguished at the 3.5$\sigma$~CL. 
Panel~(d) presents results for an even smaller value of $\Delta \Omega / \Omega$. The discriminating power is further enhanced as evidenced by a $p$-value $\lesssim 5\times 10^{-7}$, corresponding to a significance greater than $4\sigma$.
This demonstrates that Prominence provides robust discriminating power between the FOPT and DW sources, even if their SGWB peaks overlap. 

We now extend our analysis to scenarios with two-peaked GW spectra. As before, we focus on the case where the peak amplitudes and frequencies of the DW and FOPT contributions are identical. For a complete comparison, we examine all possible combinations of the two spectra as in \cref{tab:benchmark_cases}.
\begin{table}[t]
\centering
\begin{tabular}{|c|c|}
\hline
\textbf{Case} & \textbf{Comparison} \\ \hline
A & (FOPT, FOPT) vs. (DW, DW) \\ \hline
B & (FOPT, DW) vs. (DW, FOPT) \\ \hline
C & (FOPT, FOPT) vs. (FOPT, DW) \\ \hline
D & (FOPT, FOPT) vs. (DW, FOPT) \\ \hline
E & (DW, DW) vs. (FOPT, DW) \\ \hline
F & (DW, DW) vs. (DW, FOPT) \\ \hline
\end{tabular}
\caption{ Benchmark cases for the comparison of two-peaked GW spectra. For each pair $(-,-)$, the first (second) entry refers to the source of the lower (higher) frequency peak. }
\label{tab:benchmark_cases}
\end{table}
Here, for each pair $(-,-)$, the first (second) entry refers to the source of the lower (higher) frequency peak. In \cref{fig:prom_twopeak}, we present plots similar to those in \cref{fig:prom_SNR_vary}, for the two-peaked GW spectra. The signal parameters for the first peak are fixed as follows: for the DW spectrum, we set $\log_{10}[\mathcal{E}/\mathrm{GeV}^3] = 27.36$ and $\log_{10}[V_\mathrm{bias}/\mathrm{GeV}^4] = 19.15$; for the FOPT case, we set $\beta/H(T_*) = 50$, $T_* = 37.98~\mathrm{GeV}$ and $\alpha = 1.80$.  For the second peak, the DW parameters are set to $\log_{10}[\mathcal{E}/\mathrm{GeV}^3] = 27.36$ and $\log_{10}[V_\mathrm{bias}/\mathrm{GeV}^4]=19.15$, and for the FOPT spectrum, the only change in the parameters for the first peak is that $T_* = 1770.81~\mathrm{GeV}$. 

In general, we obtain the smallest $p$-values with the KS test, which range from $4.2\times 10^{-12}$ (Case~B) to $1.3 \times 10^{-5}$ (Case~C). Clearly, the KS test indicates that Prominence is a robust discriminator, effectively distinguishing between all combinations of the two-peaked (FOPT, DW) spectra. In contrast, the CvM test yields more conservative results, with the largest $p$-value of 0.00065 for Case~C. In all cases, the signals can be distinguished at greater than 3.4$\sigma$ with the CvM test. 
\begin{figure*}[t!]
	\centering
    \subfloat[$p$-$\mathrm{val(KS)} = 1.4\times 10^{-11}$, \quad $p$-$\mathrm{val(CvM)} = 0.00019$.]{\includegraphics[width=0.50\textwidth]{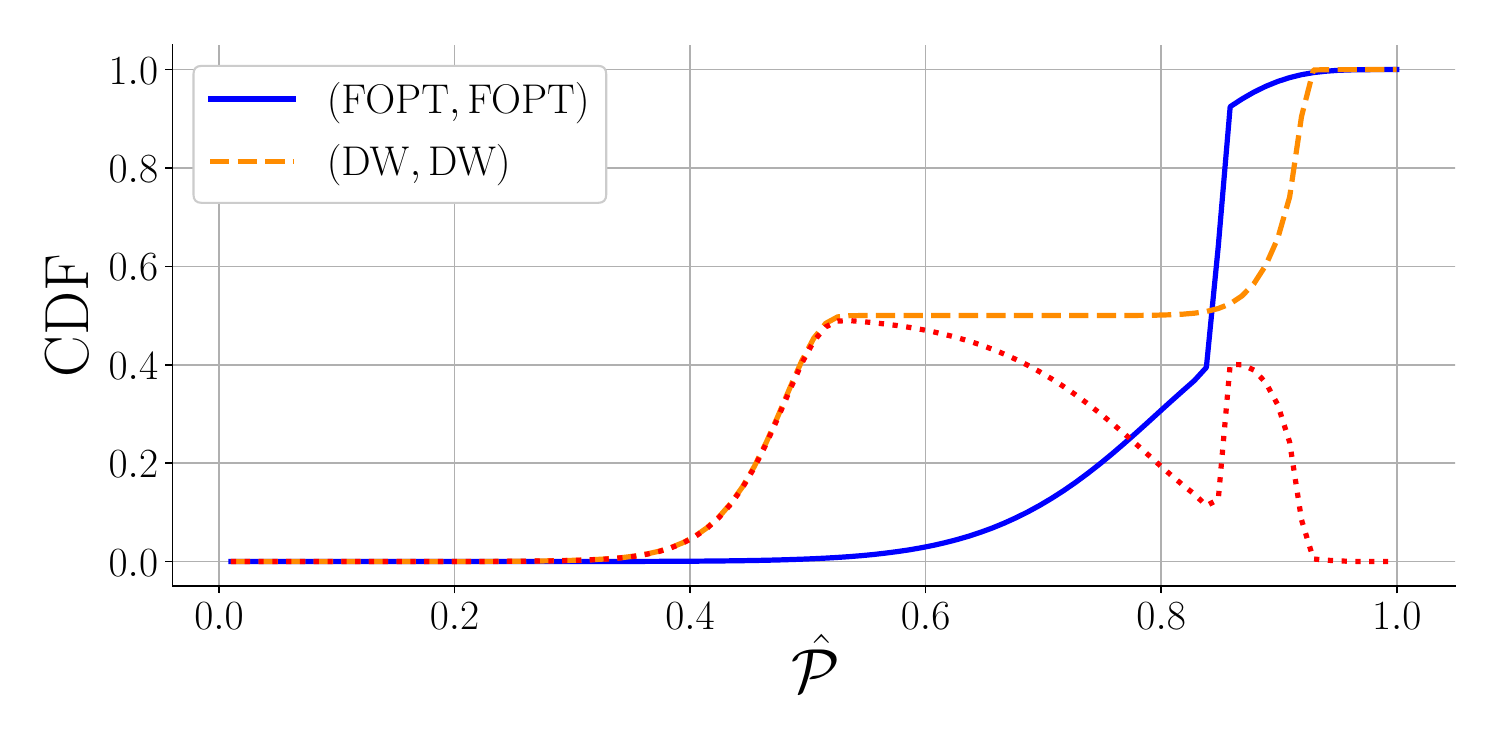}}
    \subfloat[$p$-$\mathrm{val(KS)} = 0.097$, \quad $p$-$\mathrm{val(CvM)} = 0.00045$.]{\includegraphics[width=0.50\textwidth]{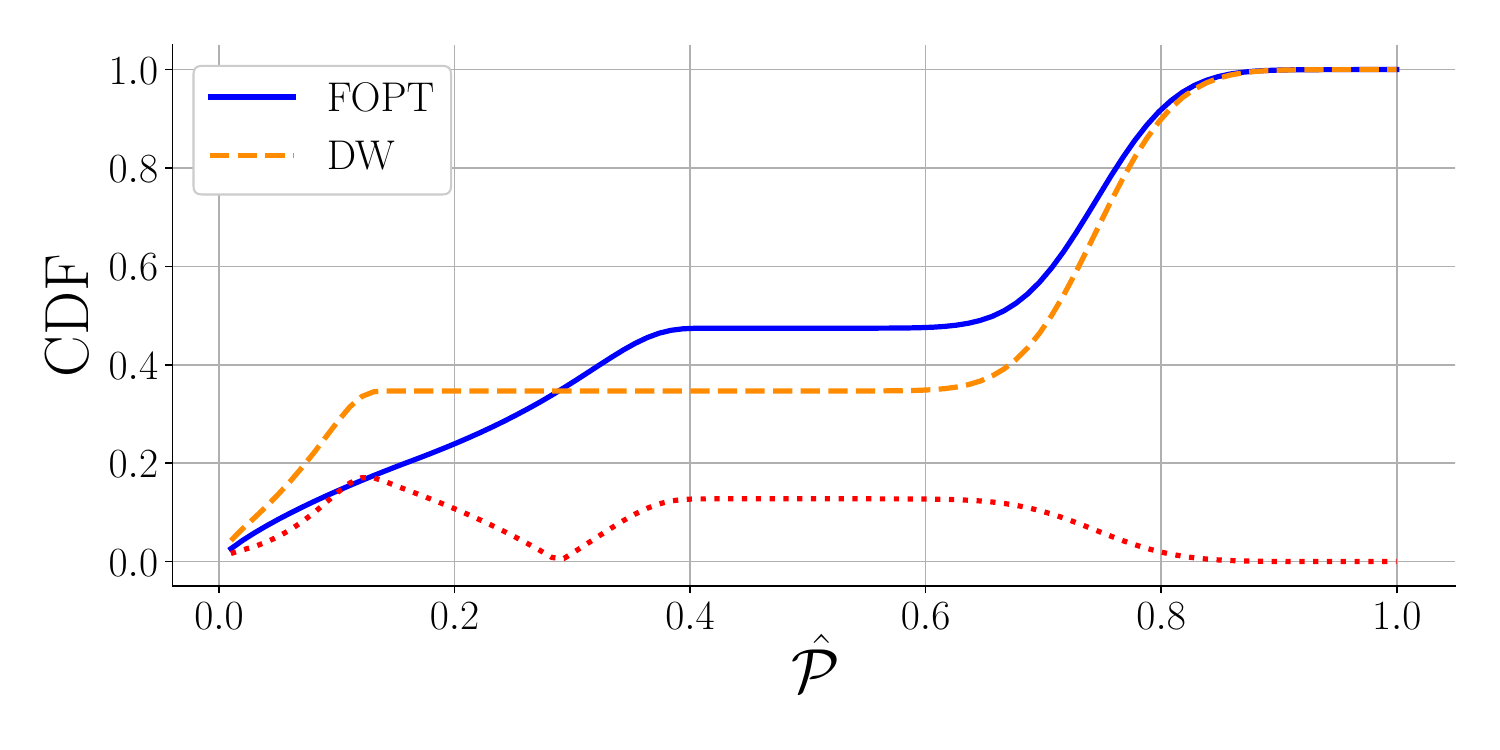}} \\
    \subfloat[$p$-$\mathrm{val(KS)} = 0.53$, \quad 
    $p$-$\mathrm{val(CvM)} = 0.0072$.]{\includegraphics[width=0.50\textwidth]{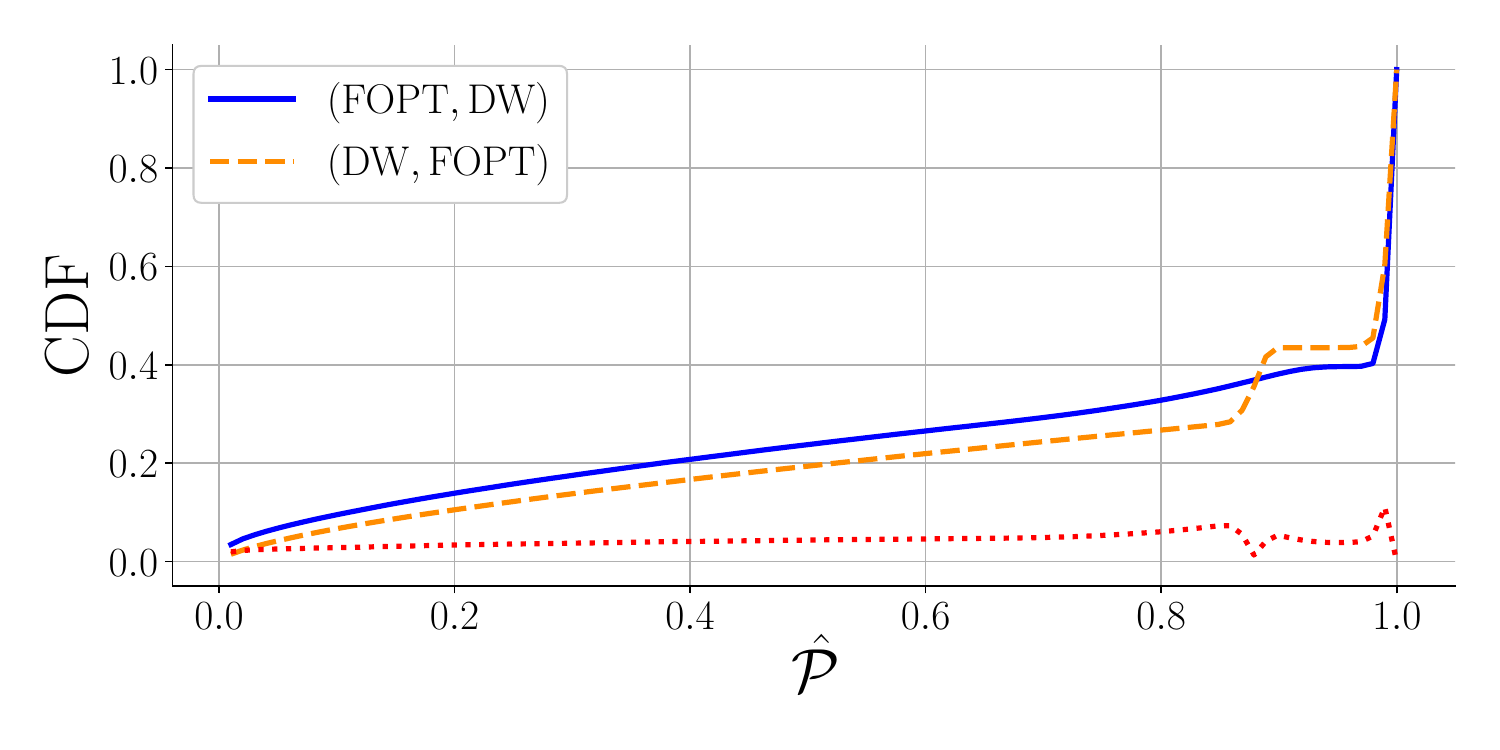}}
    \caption{CDFs as a function of the normalized Prominence $\hat{\mathcal{P}}$, which facilitates direct comparison between the different cases by making the domain of each curve the same. In each panel we provide the associated $p$-values for the KS and CvM tests. The dotted red curves are the magnitude of the difference between the solid blue and dashed orange curves. Panel~(a) corresponds to the scenario in Fig.~\ref{fig:prom_twopeak}(A), panel~(b) corresponds to the scenario in Fig.~\ref{fig:prom_SNR_vary}(c) and panel~(c) corresponds to the scenario in Fig.~\ref{fig:prom_twopeak_ET}(B). Although the CvM test is usually more conservative than the KS test, it can give smaller $p$-values than the KS test.
    Note how localized and global features in the CDFs impact the relative size of the $p$-values of the two tests.}
    \label{fig:prom_twopeak_CDFs}
\end{figure*}
This behavior can be understood from the definitions of the KS and CvM test statistics in Eqs.~\eqref{eq:KS_statistic} and \eqref{eq:CVM_statistic}. The KS test quantifies the maximum deviation between the CDFs of the samples, making it particularly sensitive to localized differences. In contrast, the CvM statistic is a measure of the aggregate difference in the CDFs over the entire $\cal{P}$ range, leading to more conservative $p$-values when there is substantial overlap in the distributions. 
To illustrate this, in Fig.~\ref{fig:prom_twopeak_CDFs} we show the CDFs for a few benchmark cases. Panel~(a) corresponds to the scenario in Fig.~\ref{fig:prom_twopeak}(A), panel~(b) to Fig.~\ref{fig:prom_SNR_vary}(c), and panel~(c) to \cref{fig:prom_twopeak_ET}(B). The Prominence axis is normalized 
(denoted $\hat{\mathcal{P}}$) so that the domain of each curve is the same for the different cases, facilitating direct comparison.
Since the KS statistic measures the maximum vertical distance between two CDFs, the resulting $p$-value is highly sensitive to sharp, localized variations in the CDFs. For example, in panel~(a), although the two distributions overlap below $\hat{\mathcal{P}}\sim 0.4$, they exhibit large differences above it, which leads to a strongly suppressed $p$-value. In contrast, the CvM test integrates the squared differences over the entire distribution, making it less sensitive to localized features and more indicative of global behavior. Consequently, in scenarios where sharp features are absent, the $p$-values from the CvM test tend to be smaller than from the KS test. This is true if the distributions exhibit a systematic shift or persistent differences, in which case the differences between the distributions span the entire range rather than being localized, as in panels~(b) and~(c).

To conclude this section, we present plots in \cref{fig:prom_twopeak_ET} similar to those in \cref{fig:prom_twopeak}, but with the higher frequency peak in the ET sensitivity band.
\begin{figure*}[htb!]
	\centering
    \renewcommand{\thesubfigure}{\Alph{subfigure}}
    \subfloat[]{\includegraphics[width=0.5\textwidth]{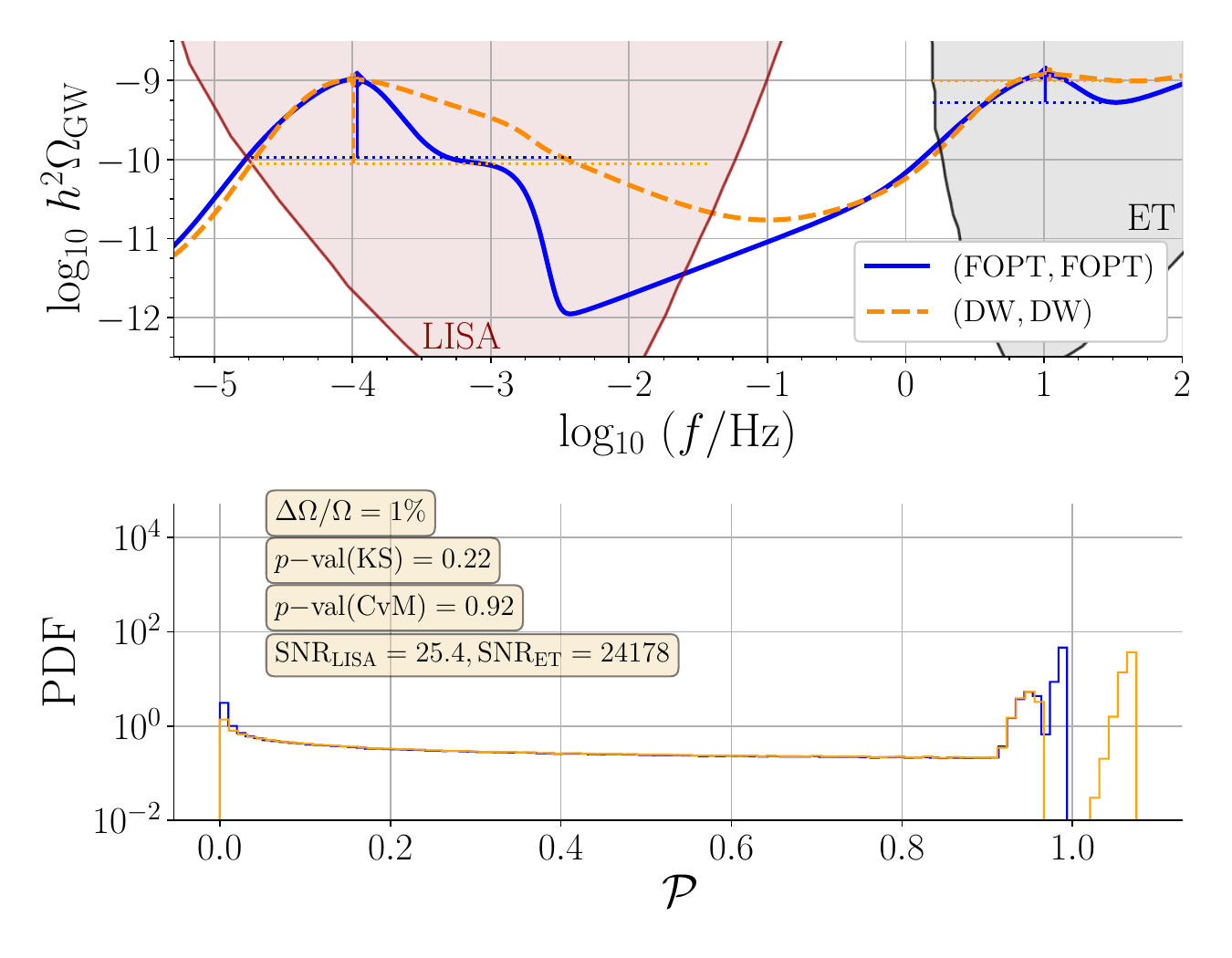}} 
    \subfloat[]{\includegraphics[width=0.5\textwidth]{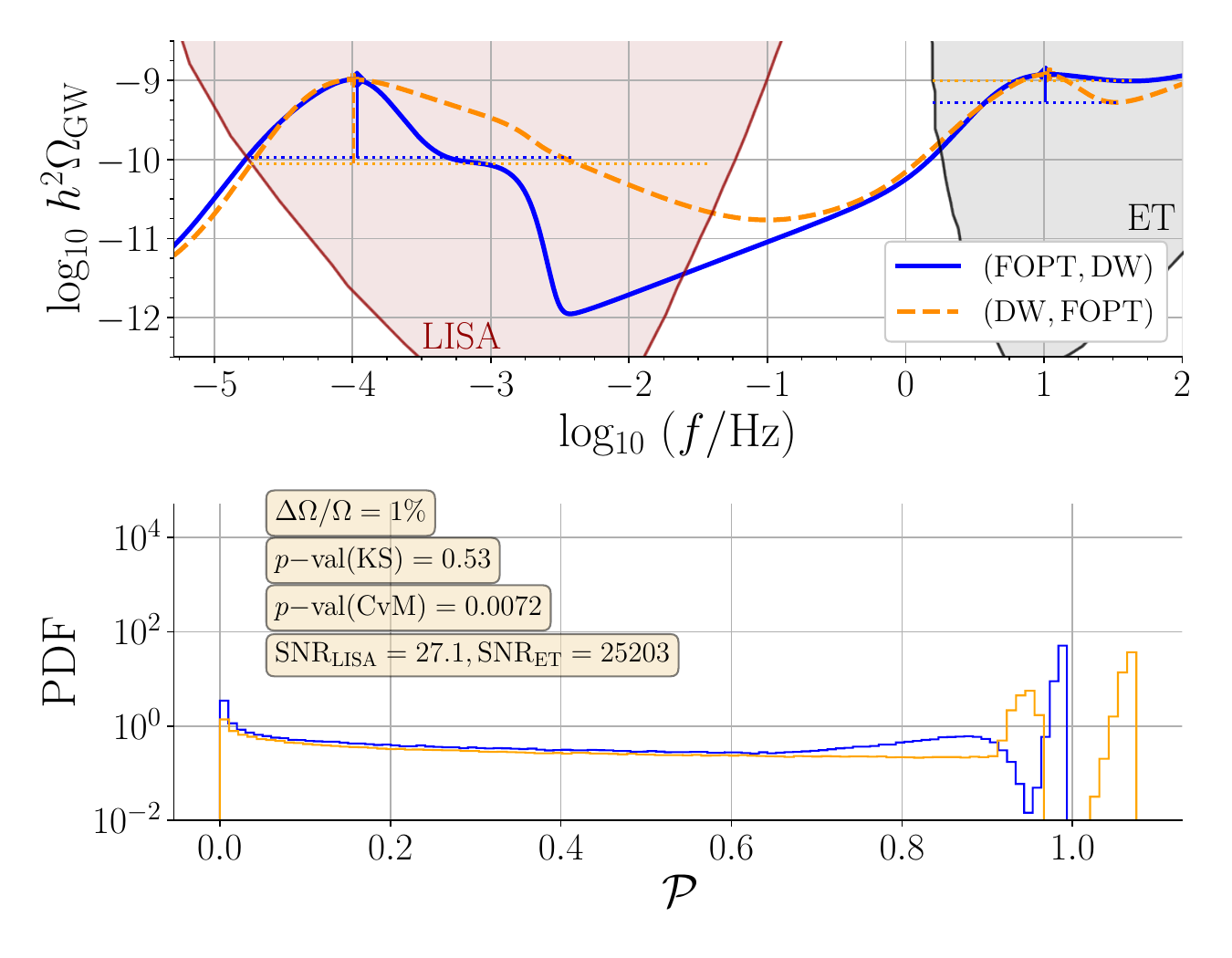}} \\
    \subfloat[]{\includegraphics[width=0.5\textwidth]{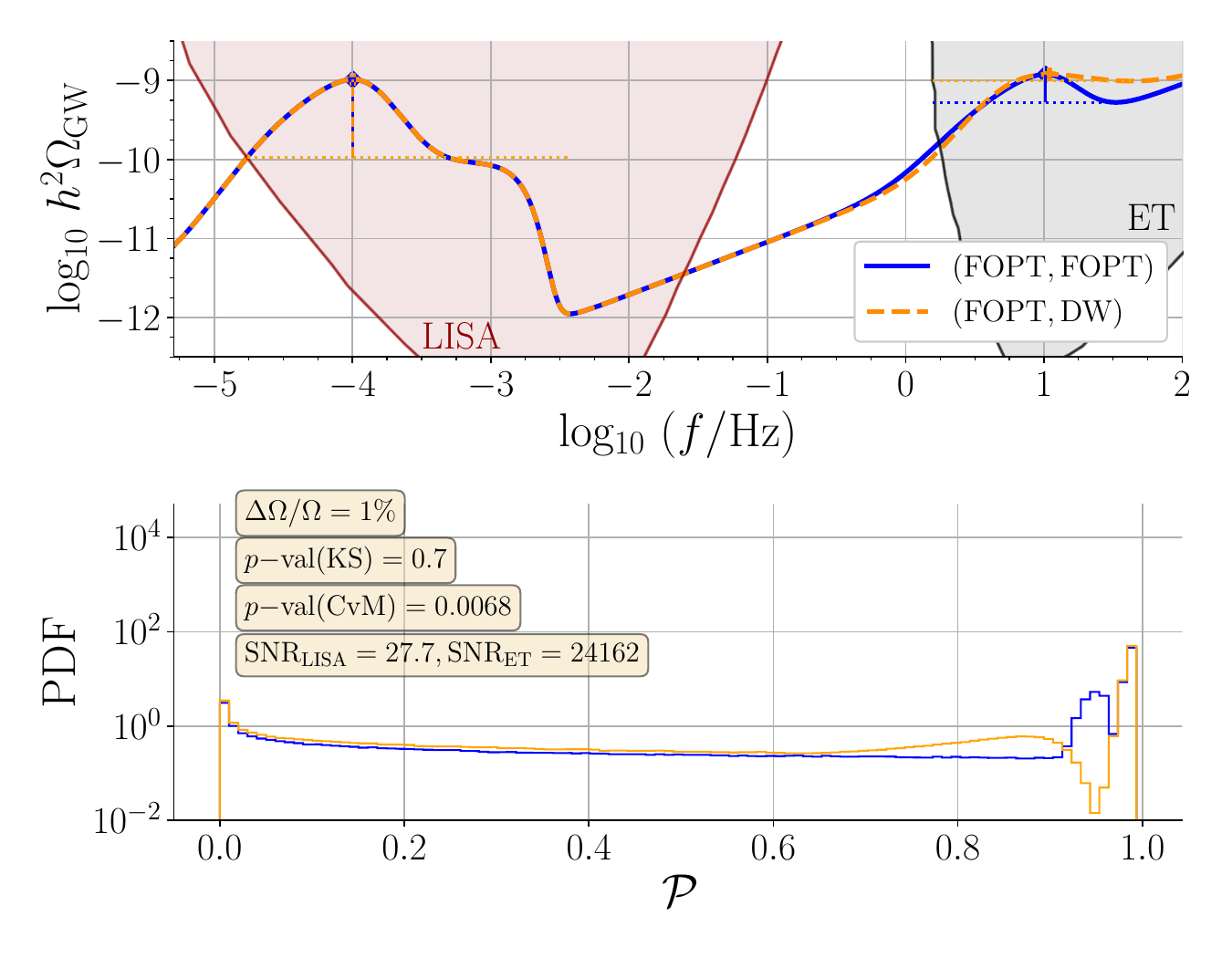}} 
    \subfloat[]{\includegraphics[width=0.5\textwidth]{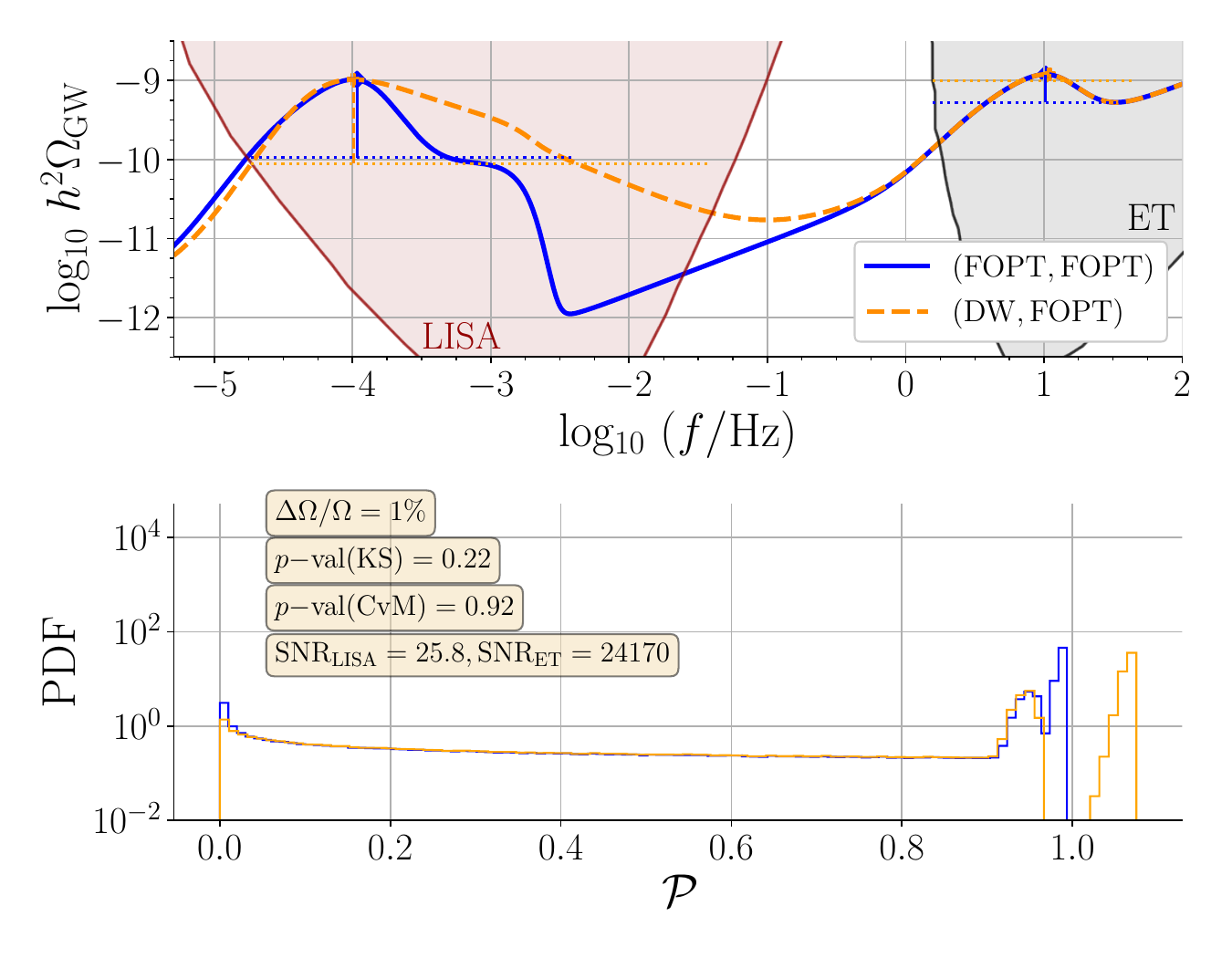}} \\
    \subfloat[]{\includegraphics[width=0.5\textwidth]{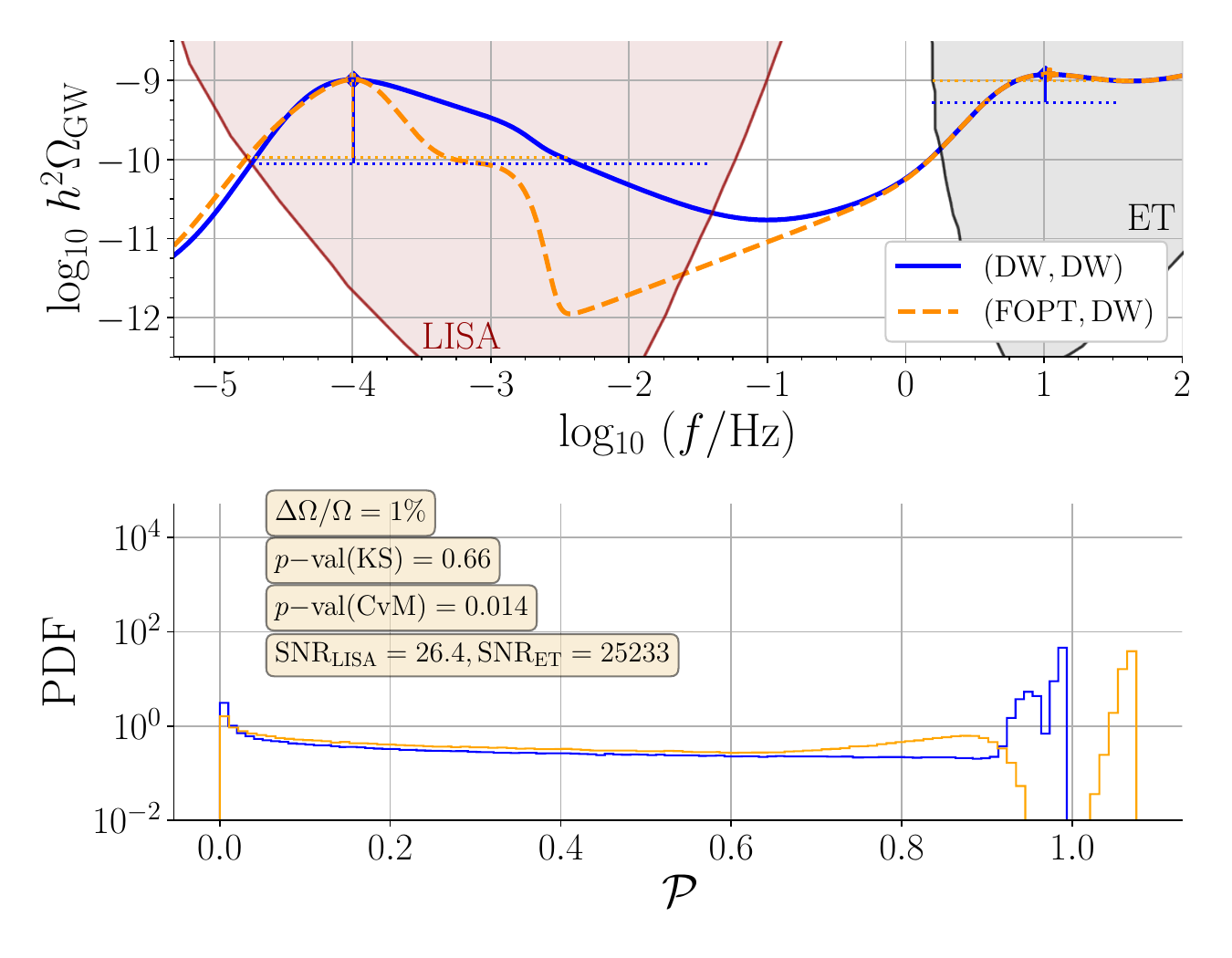}}
    \subfloat[]{\includegraphics[width=0.5\textwidth]{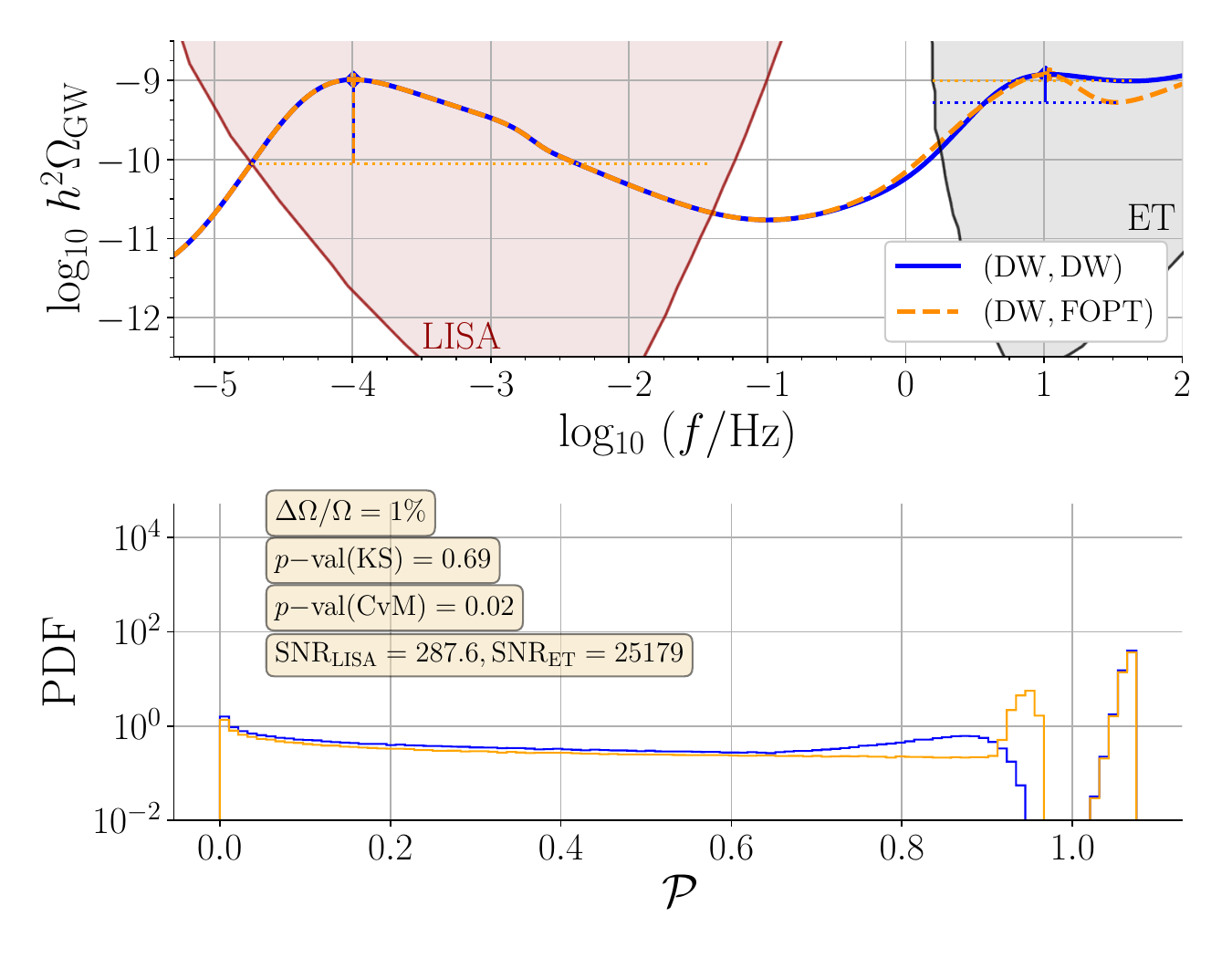}}
    \caption{ Similar to Fig.~\ref{fig:prom_twopeak} for two-peaked GW spectra with the higher frequency peak in the ET frequency band. The discriminating power of $\mathcal{P}$ is reduced if the second peak lies in the ET frequency band because the large extragalactic background obscures the signal.
    }
    \label{fig:prom_twopeak_ET}
\end{figure*}
For the ET signal, we set $\log_{10}[\mathcal{E}/\mathrm{GeV}^3] = 34.02$ and $\log_{10}[V_\mathrm{bias}/\mathrm{GeV}^4] = 32.46$ for DWs, and $\beta/H(T_*) = 50$, $T_* = 3.70\times 10^6~\mathrm{GeV}$, $\alpha = 1.80$ for the FOPT. We find that discrimination in the signals is generally poor. For the KS test, even in the best-case scenario (Case A), the two-peaked GW spectra cannot be distinguished with a CL exceeding 78\%. However, in several cases the CvM test yields superior discrimination power compared to the KS test. 
This lack of sensitivity of the KS test arises from the large extragalactic background in the ET band, which dominates the DW and FOPT signals. This is evident from the PDFs, which nearly overlap at small $\cal{P}$ values and differ only in their height, thus minimizing the differences in the CDFs and reducing the discriminating power of the KS test. However, when small differences persist over a wide range of $\cal{P}$, the CvM test becomes more sensitive, as in Fig.~\ref{fig:prom_twopeak_CDFs}(c) and Case~C.
 For larger $\cal{P}$ values, which occur in the LISA band, the ability to discriminate between the DW and FOPT sources improves.

\subsection{Impact of cosmic strings}
\begin{figure*}[t]
	\centering
    \subfloat[$G\mu = 10^{-14}$]{\includegraphics[width=0.5\textwidth]{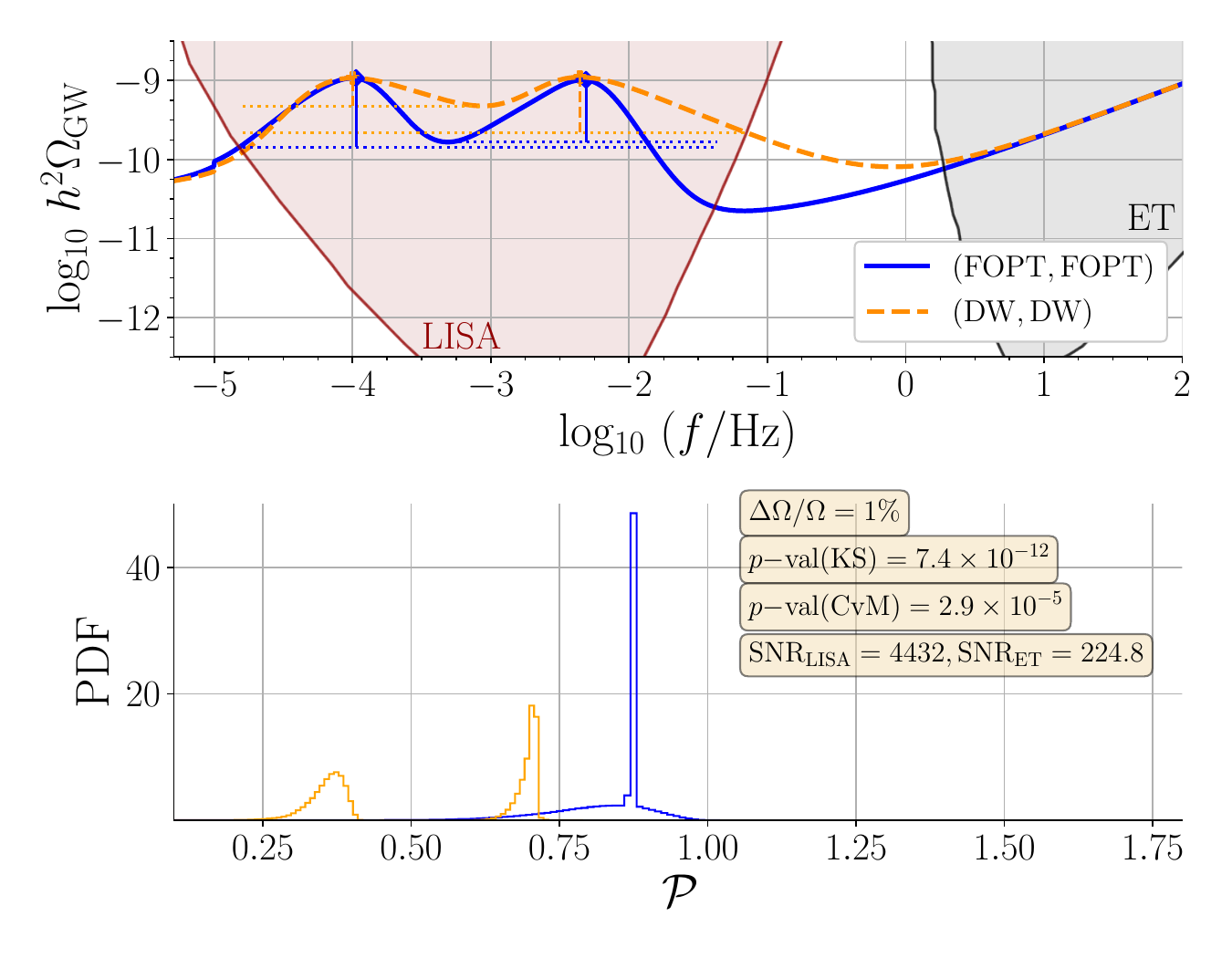}} 
    \subfloat[$G\mu = 10^{-14}$]{\includegraphics[width=0.5\textwidth]{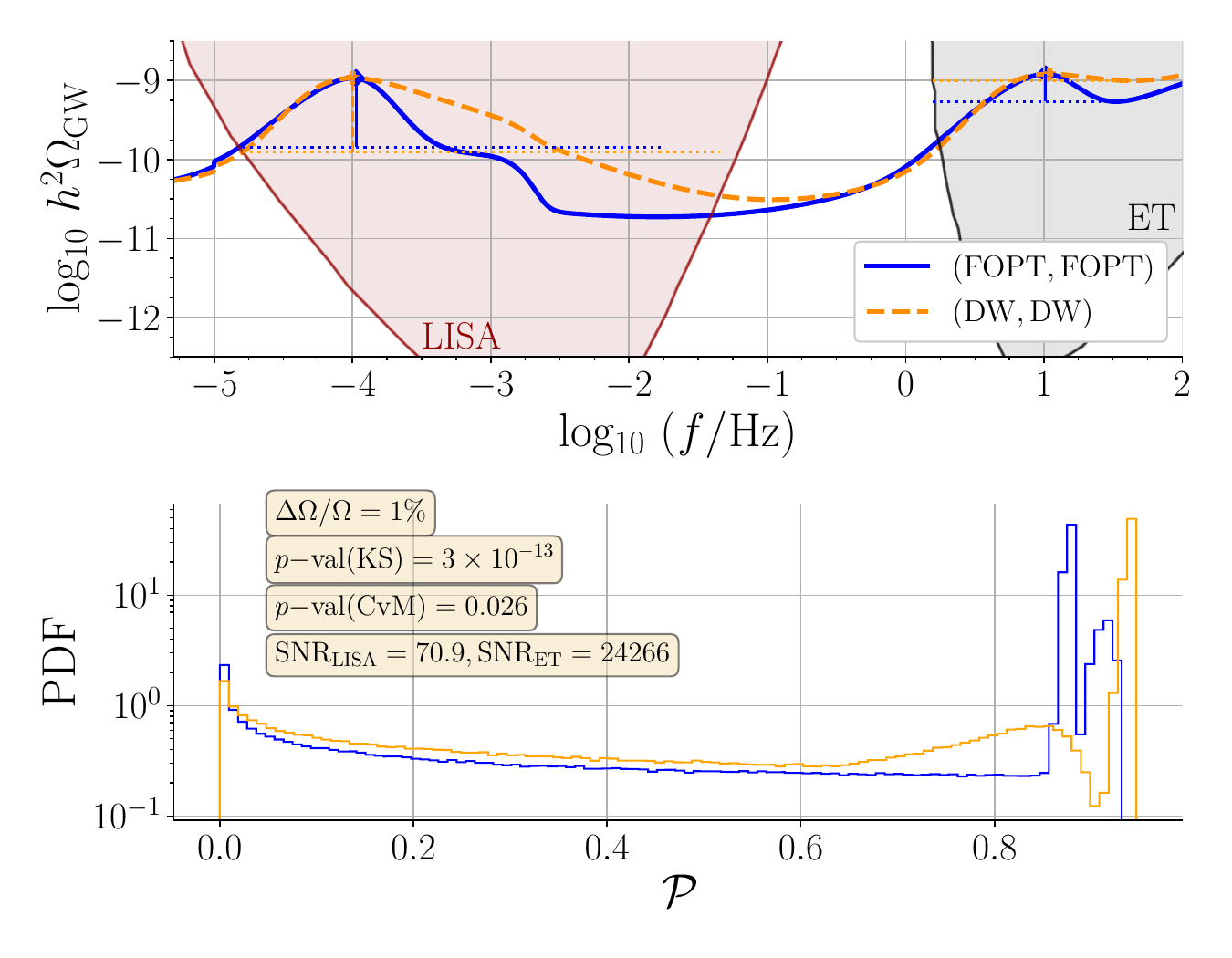}} \\
    \subfloat[$G\mu = 10^{-12}$]{\includegraphics[width=0.5\textwidth]{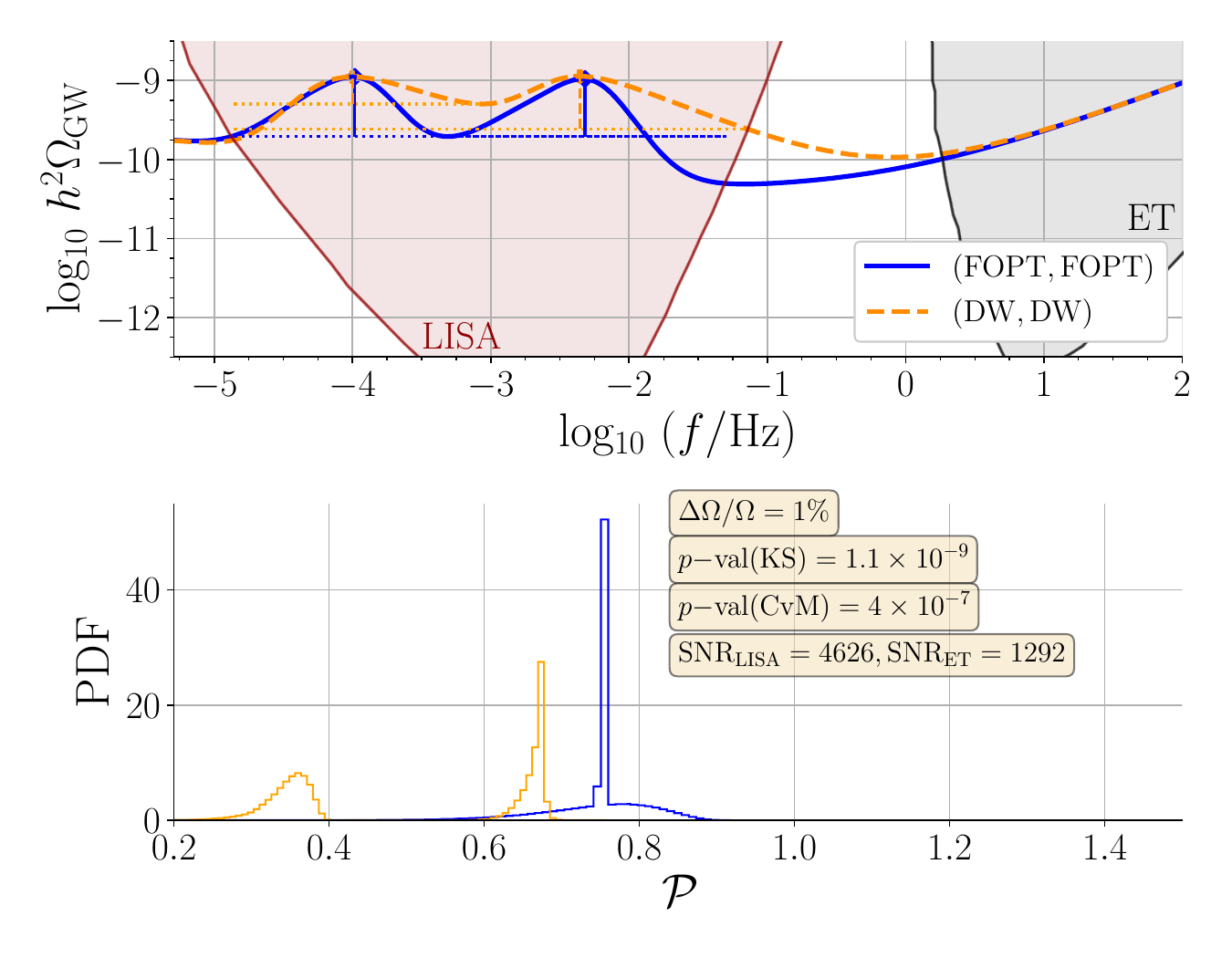}} 
    \subfloat[$G\mu = 10^{-12}$]{\includegraphics[width=0.5\textwidth]{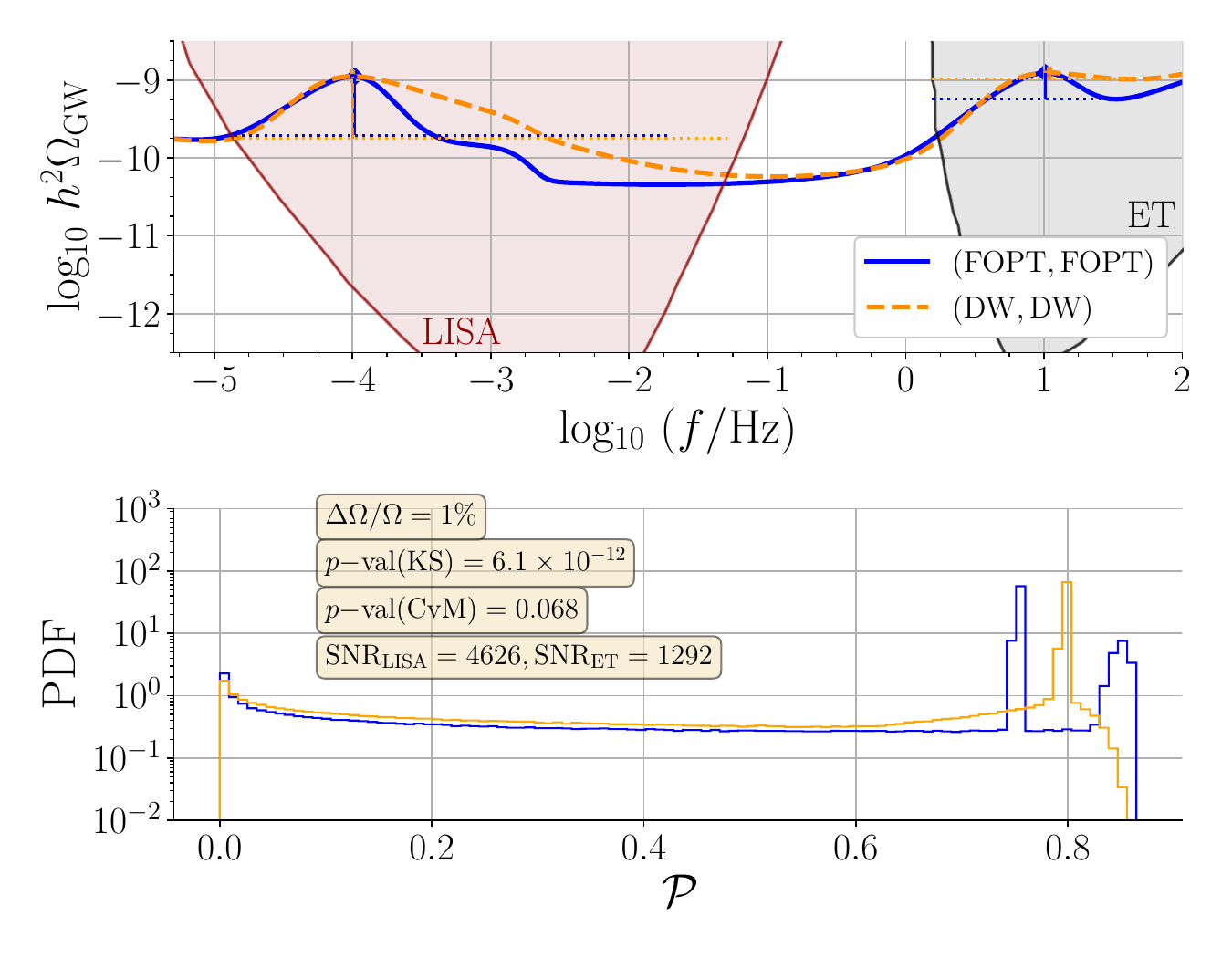}}    
    \caption{Similar to Fig.~\ref{fig:prom_twopeak} for a two-peaked GW spectrum plus a CS signal with $G \mu = 10^{-14}$ (upper panels), and $G \mu = 10^{-12}$ (lower panels). In the left panels both peaks are in the LISA band, and in the right panels the higher frequency peak is in the ET band. A CS signal can enhance the discriminating power of $\mathcal{P}$ by burying the background.}
    \label{fig:prom_twopeak_CS}
\end{figure*}

Unlike GW signals from FOPTs and DWs, the GW spectrum from CSs in the interferometer frequency band is quite flat and may be confused with the GW background, which motivates a closer treatment. Another aspect that bears consideration is that since CSs are produced from spontaneous breaking of a continuous symmetry, the SGWB may have contributions from CSs and an FOPT. Moreover, in many particle physics models, a discrete symmetry may also be broken, producing a GW signal from DWs as well. We therefore consider scenarios in which CSs, FOPTs, and DWs produce a SGWB together.

It is important to note that for Prominence to serve as an effective observable, the GW spectrum should have at least one peak. Therefore, we select $G\mu$ values such that the CS contribution is subdominant to the DW and FOPT peaks. 
We superimpose the CS signal on the two-peaked spectra in panel~(A) of \cref{fig:prom_twopeak,fig:prom_twopeak_ET}.
The result is shown in \cref{fig:prom_twopeak_CS} for $G\mu = 10^{-14}$ in the upper panels, and for $G\mu = 10^{-12}$ in the lower panels. We find that a CS signal can enhance the discrimination power of $\cal{P}$, even if the CS signal is weaker than the dominant backgrounds and FOPT/DW signals, as illustrated in \cref{fig:prom_twopeak_CS}(a). This effect is more pronounced for larger CS signals, as is evident from the lower panels, where the CvM test yields lower $p$-values compared to the case with $G\mu = 10^{-14}$. Recall that we are considering the most conservative case in which the DW and FOPT peaks overlap exactly. 

\begin{figure*}[t!]
	\centering
    \subfloat[]{\includegraphics[width=0.5\textwidth]{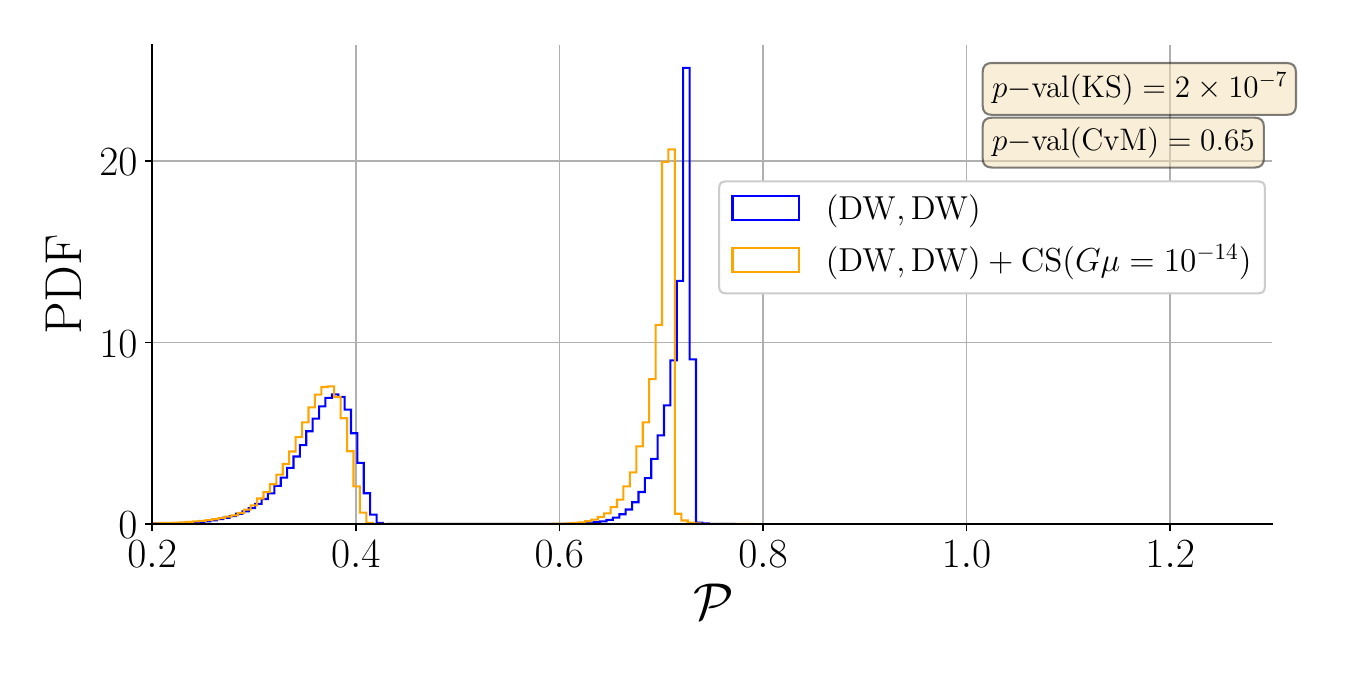}} 
    \subfloat[]{\includegraphics[width=0.5\textwidth]{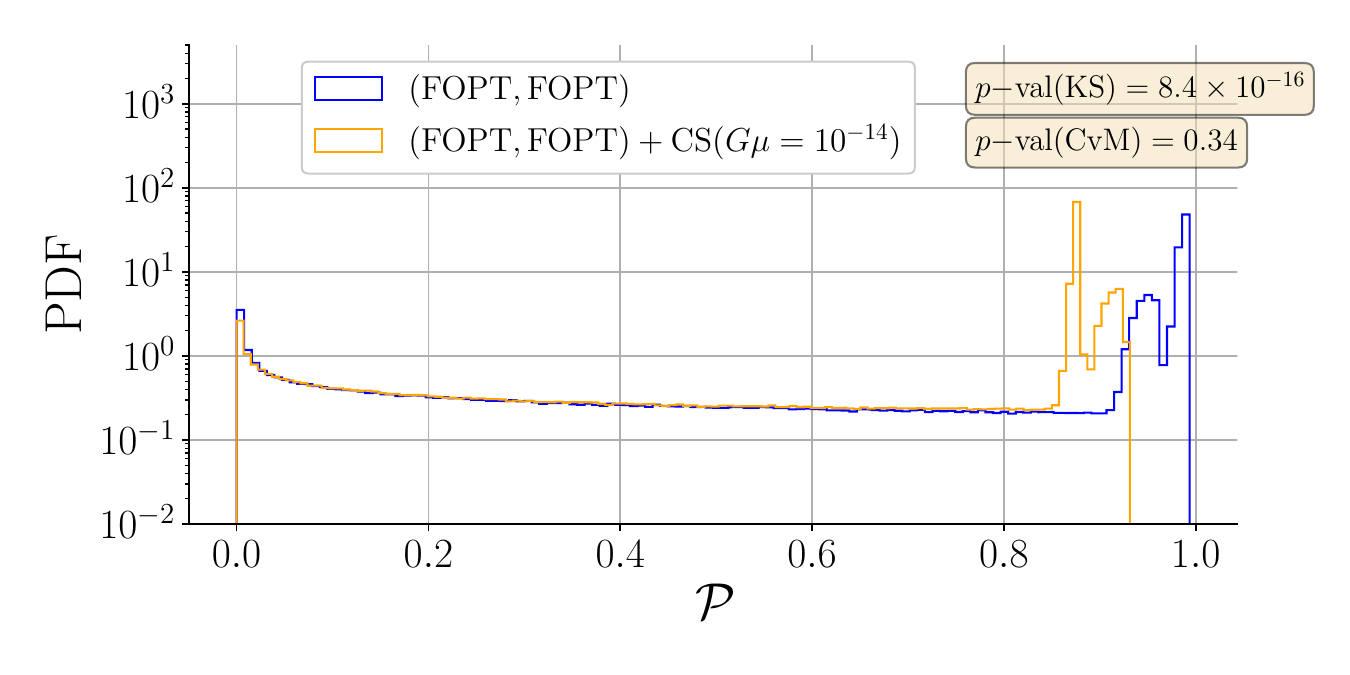}} \\
    \subfloat[]{\includegraphics[width=0.5\textwidth]{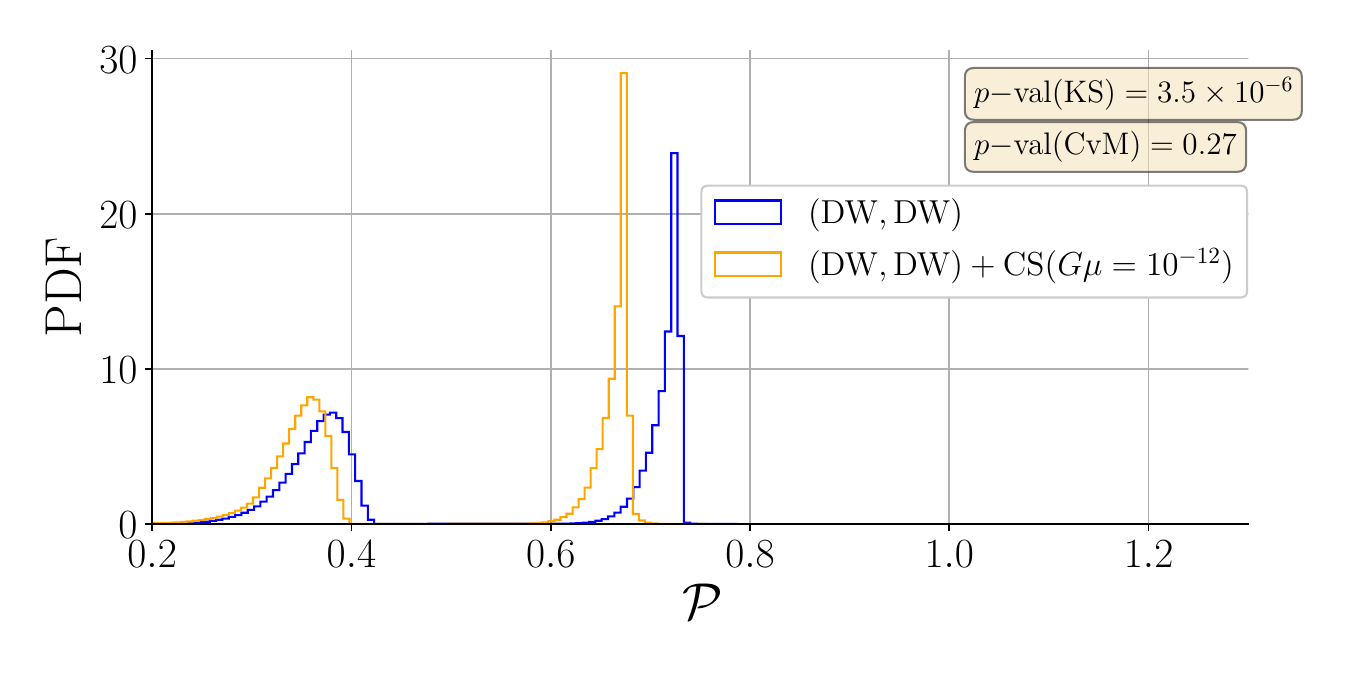}} 
    \subfloat[]{\includegraphics[width=0.5\textwidth]{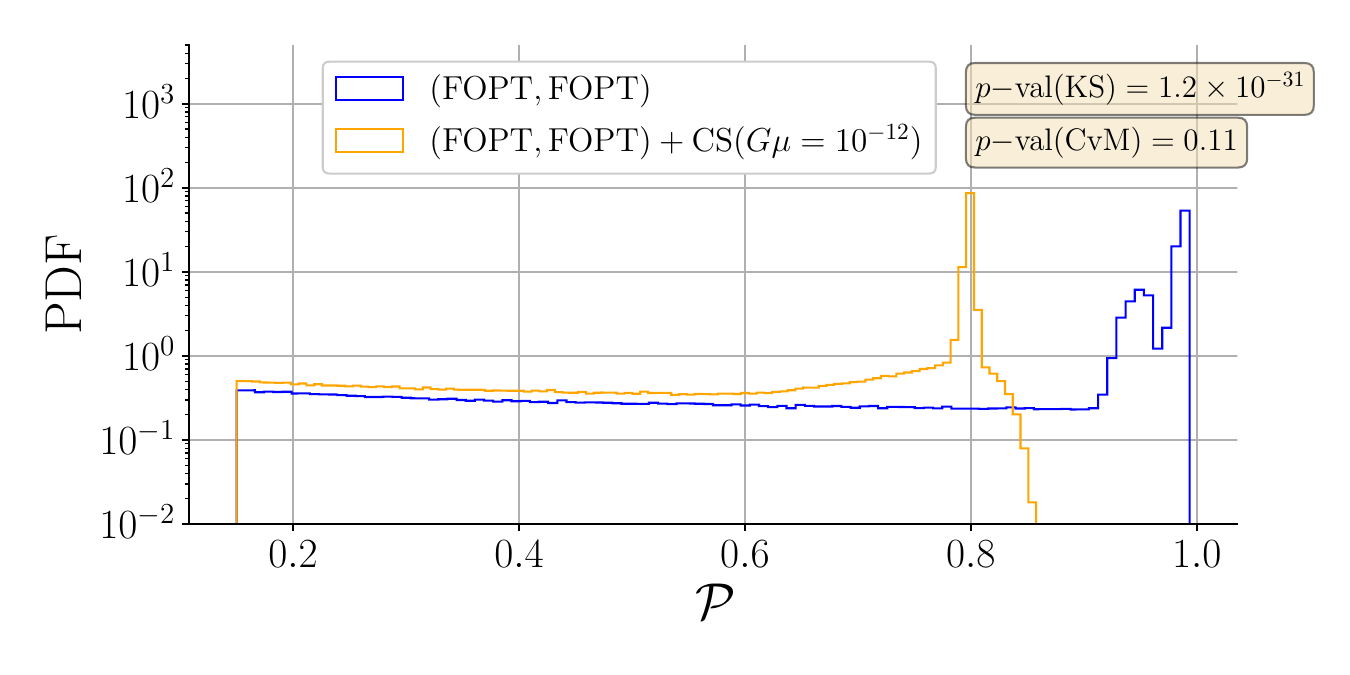}} \\   
    \caption{ PDFs of $\cal{P}$, comparing two-peak signals without a CS signal (in blue) to that including a CS signal (in orange). In the upper panels $G\mu = 10^{-14}$, and in the lower panels $G\mu = 10^{-12}$. The CS signal can bury the background and produce large local differences in the PDFs, thereby yielding small $p$-values for the KS test. The effect is larger for signals from FOPTs than from DWs because the spectrum from FOPTs is narrower. 
    }
    \label{fig:prom_CS_v_NoCS}
\end{figure*}

To understand this behavior, we present in \cref{fig:prom_CS_v_NoCS} the PDFs for two-peak signals, comparing the scenario without a CS signal (blue) to those including a CS signal (orange), with $G\mu = 10^{-14}$  (top panels) and $G\mu = 10^{-12}$ (bottom panels).
The left panels show the (DW,DW) case, and the right panels show the (FOPT,FOPT) case. For (DW,DW) with $G\mu = 10^{-14}$ the PDFs are essentially unchanged by the CS signal and  exhibit significant overlap, as indicated by the relatively large $p$-value of the CvM test (0.65). Similarly, for the FOPT scenario shown in panel~(b), the impact of the CS signal is small. The CS signal causes a shift towards lower $\cal{P}$ values, although the distributions overlap significantly for $\mathcal{P} < 0.8$. This manifests as a large $p$-value of 0.34 for the CvM test. These results demonstrate that while a subdominant CS signal does not affect the global behavior of the PDFs, it produces localized differences, as is evident from the small $p$-values for the KS test. This effect is more pronounced in the FOPT case compared to the DW case. The difference arises from the large width of the DW spectrum, so that the CS contribution has a small effect on the PDFs. For $G\mu = 10^{-12}$ the impact of CSs remains important, particularly for the FOPT case, yielding KS $p$-values of $3.5\times 10^{-6}$ for DWs and $1.2\times 10^{-31}$ for FOPTs. Note that the CS signal is flat and sizable in the interferometer band. Therefore, for certain choices of background parameters, the CS signal dominates the background. In combination with the fact that the Prominence window is determined by the frequency range of the sensitivity curves, this implies that $\mathcal{P}$ remains fixed, generating the spikes in the PDFs in panels~(c) and~(d). 

\subsection{Prominence in the PTA band}

We now examine the application of Prominence in the PTA frequency band ($f < 10^{-7}~\mathrm{Hz}$), with a focus on the NANOGrav 15-year dataset~\cite{NANOGrav:2023gor}. To determine the best-fit parameters for the DW, FOPT and SIGW signals that explain the NANOGrav data, we use \texttt{PTArcade}~\cite{Mitridate:2023oar} in its default \texttt{ceffyl} configuration~\cite{lamb2023rapid}. The expected contribution from SMBHBs is also included, with their parameters sampled according to \cref{eq:musigma}. In addition to the NANOGrav signal, we also assess the prospects for future experiments. 
We do not consider CS sources because
as shown in Ref.~\cite{NANOGrav:2023gor}, GWs from cusps do not fully explain the NANOGrav data. 

\begin{figure*}[t!]
	\centering
    \subfloat[]{\includegraphics[width=0.50\textwidth]{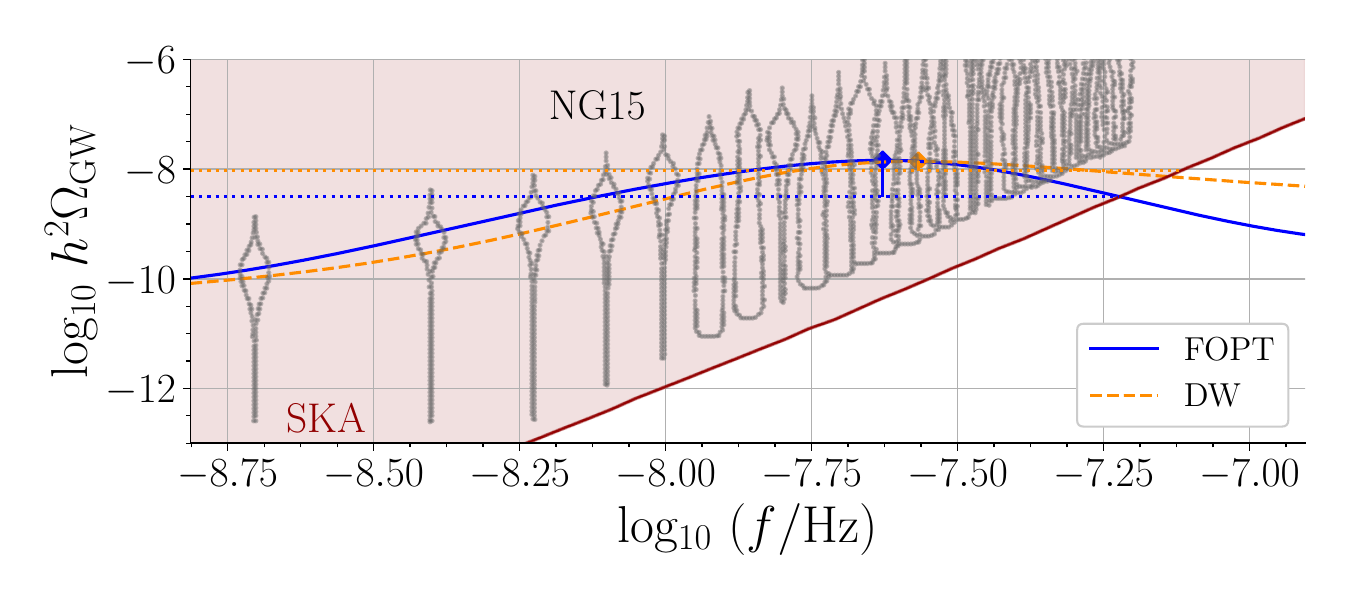}}
    \subfloat[]{\includegraphics[width=0.50\textwidth]{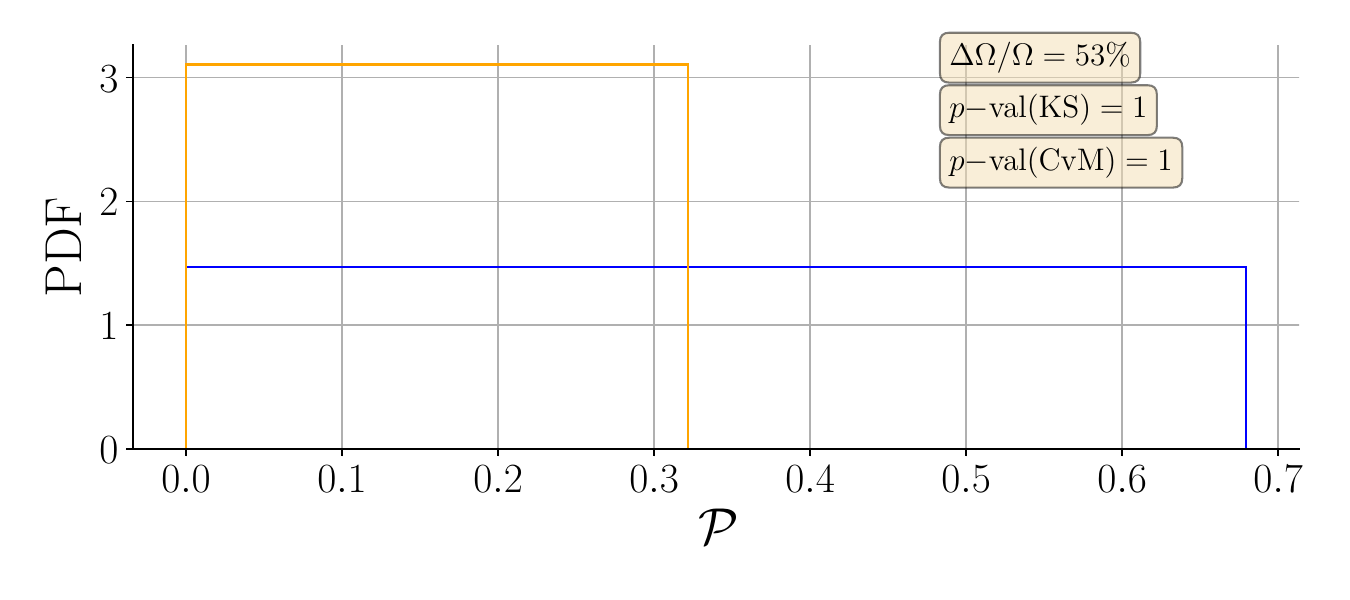}} \\
    \subfloat[]{\includegraphics[width=0.50\textwidth]{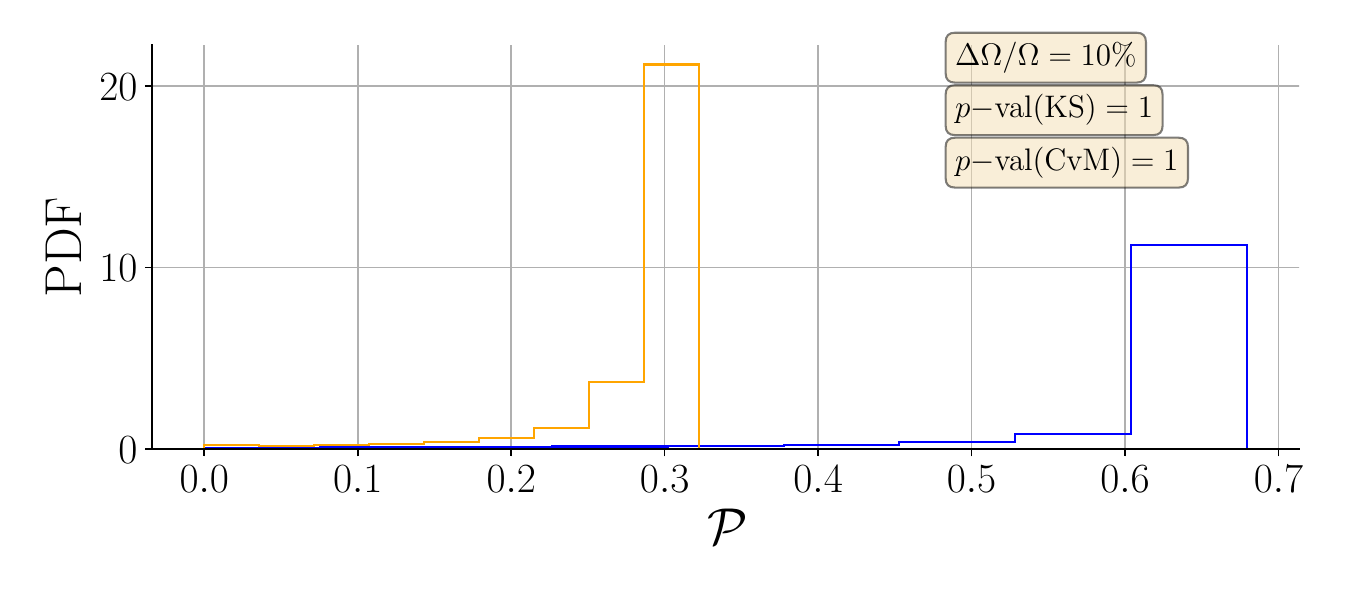}}
    \subfloat[]{\includegraphics[width=0.50\textwidth]{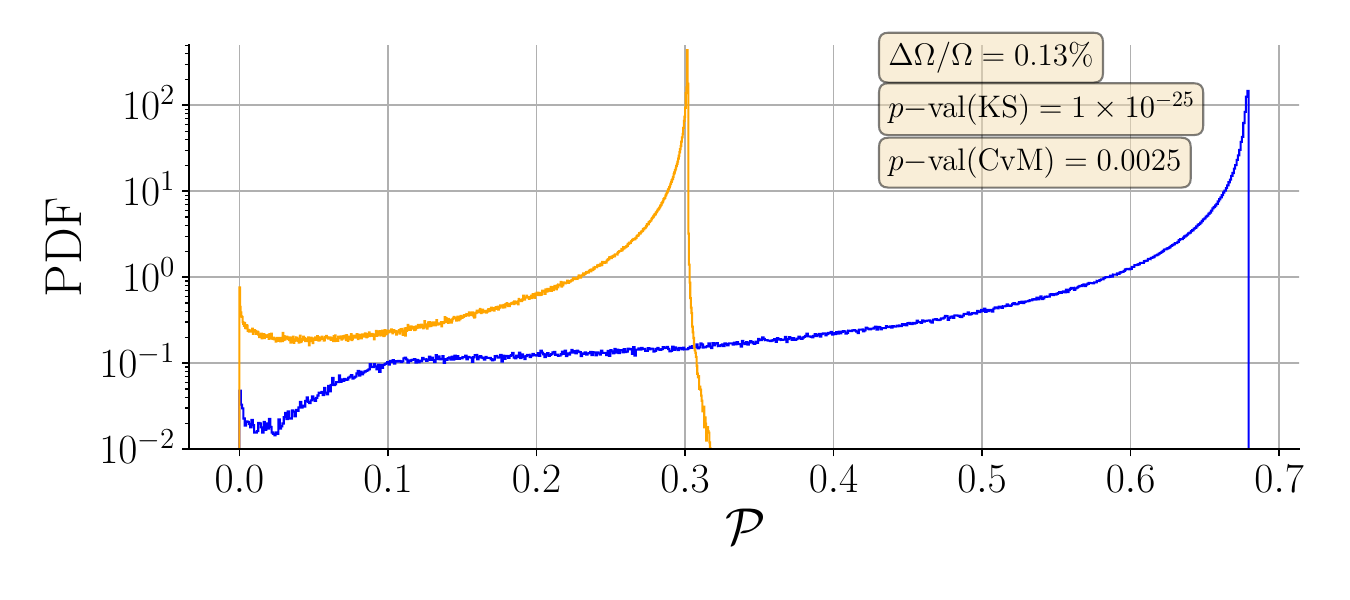}}
    \caption{Panel~(a) shows the SGWB spectrum for DW and FOPT signals in the PTA band.  Panel~(b) shows the $\cal{P}$ distribution based on the uncertainties reported by NANOGrav \cite{NANOGrav:2023gor}, panel~(c) shows results based on SKA forecasts, and panel~(d) shows the minimum uncertainty needed to discriminate between the two signals at $3\sigma$. The current (NANOGrav) and projected (SKA) uncertainties in the PTA frequency region are too large for $\mathcal{P}$ to be an effective discriminator. An uncertainty at the sub-percent level is needed for a 99\%~CL discrimination.}
    \label{fig:prom_PTAs}
\end{figure*}

In \cref{fig:prom_PTAs}, we present the result of applying the same methodology as in the interferometer band. In panel~(a) we show the SGWB spectrum for the DW and FOPT signals. In panel~(b), the bin width is the $1\sigma$ amplitude uncertainty reported by the NANOGrav collaboration~\cite{NANOGrav:2023gor} at $f=32~\mathrm{nHz}$. In panel~(c), the bin width is chosen to be the relative uncertainty expected from SKA observations of 200 pulsars over 20~years: $\Delta \Omega/\Omega \sim 10\%$~\cite{Babak:2024yhu}. 
We find that current NANOGrav data are not yet precise enough for Prominence to serve as an effective discriminator, yielding $p$-values of unity for both the KS and CvM tests. Similarly, in its default configuration SKA will also lack the precision required for Prominence to be a viable discriminator. As shown in panel~(d), 
the FOPT and DW signals can be distinguished at $3\sigma$ if $\Delta \Omega/\Omega = 0.13\%$. 

\begin{figure*}[htb!]
    \centering
    \subfloat[]{\includegraphics[width=0.50\textwidth]{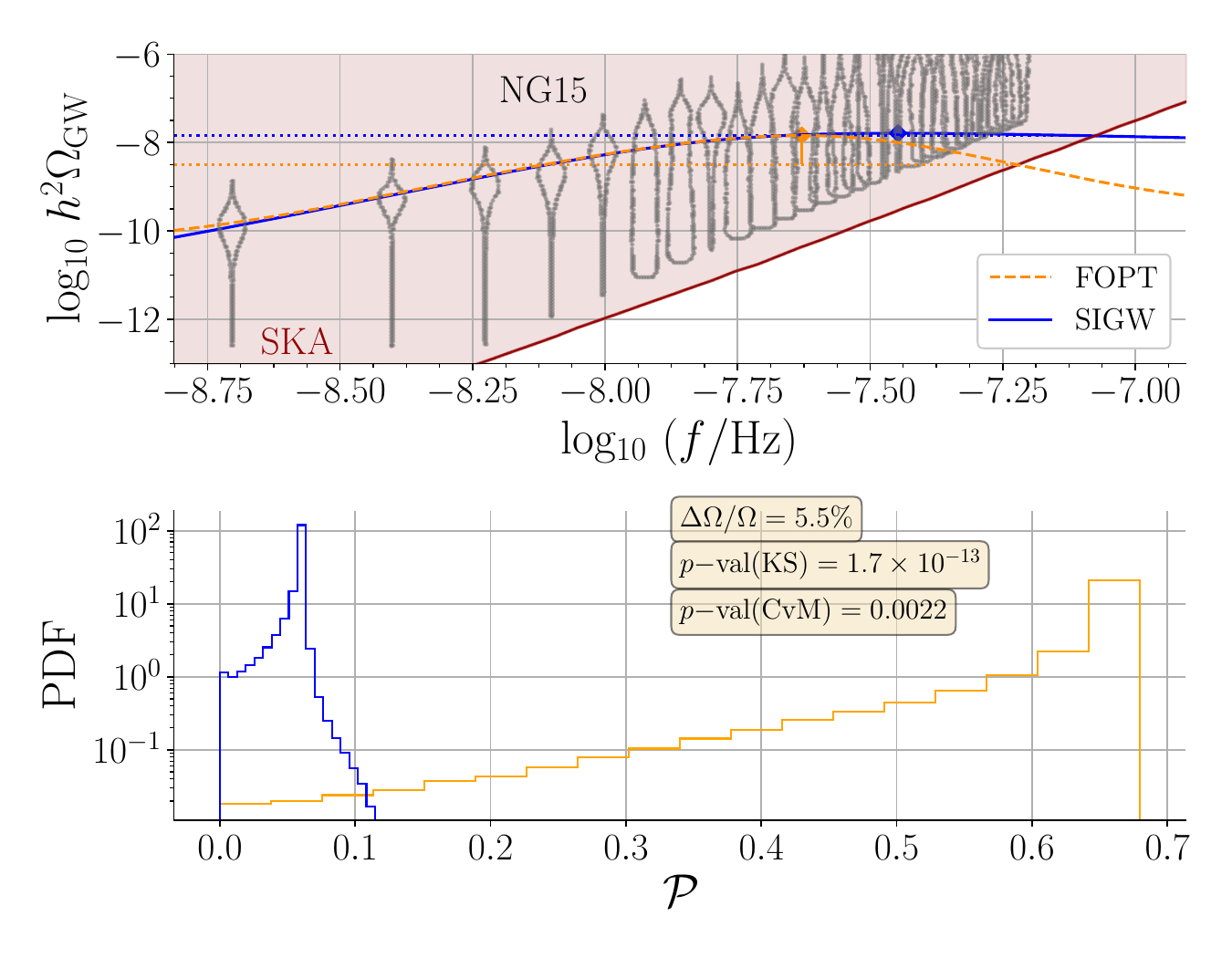}}
    \subfloat[]{\includegraphics[width=0.50\textwidth]{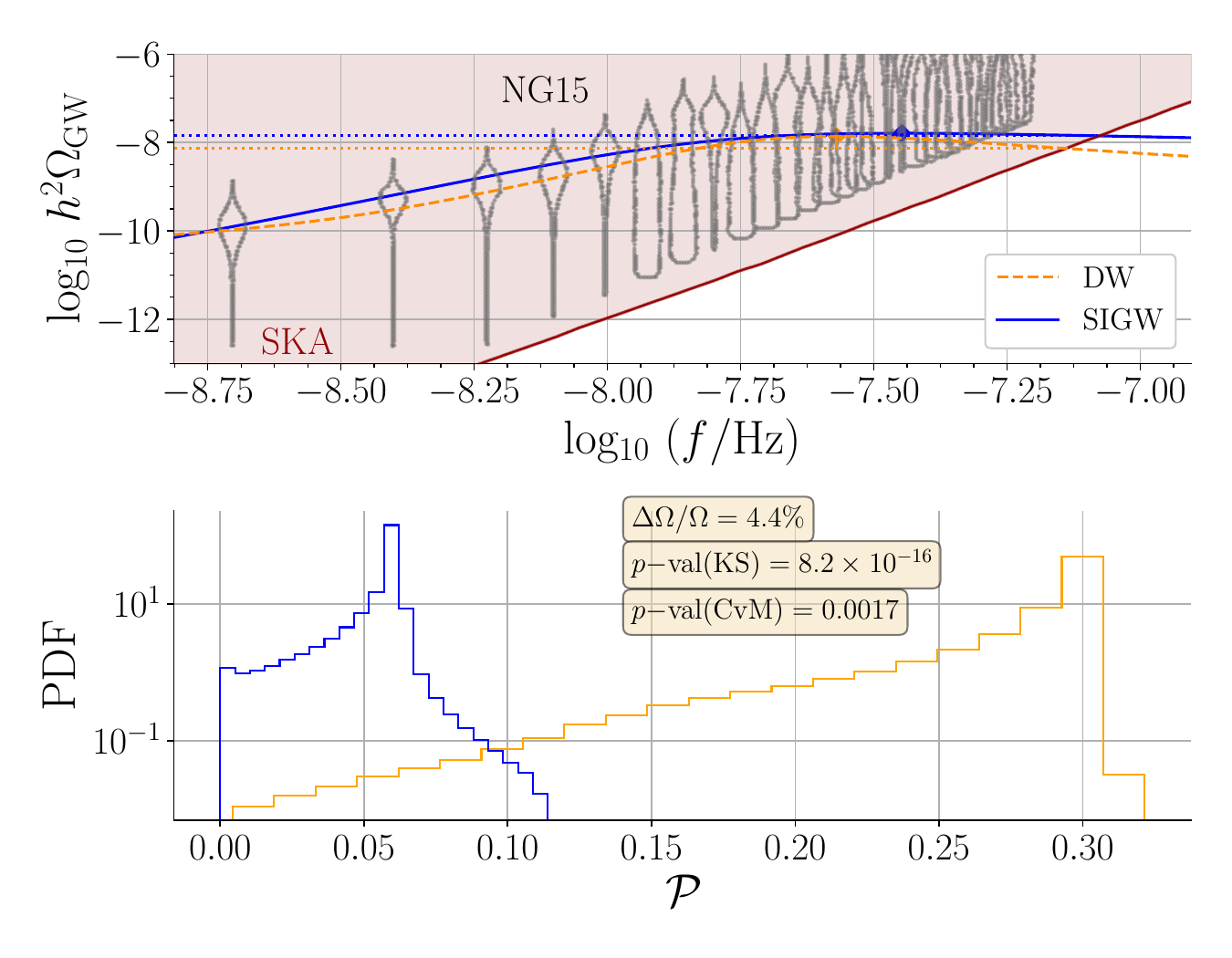}} 
    \caption{SGWB spectra in the PTA band and the corresponding PDFs. For the GW spectra, the background and signal parameters have been fixed to the best-fit values obtained using \texttt{PTArcade}. The blue solid curves correspond to the the SIGW source, and the orange dashed curves correspond to the FOPT source in panel~(a), and the DW source in panel~(b). We set $\Delta \Omega / \Omega$ to the minimum uncertainty needed to discriminate between the two signals at $3\sigma$ with the CvM test. The larger amplitude uncertainties indicate that a SIGW signal is easier to distinguish from FOPT and DW signals than from each other.}
    \label{fig:prom_PTAs_SI}
\end{figure*}

In \cref{fig:prom_PTAs_SI} we show the ability to discriminate an SIGW signal from the DW and FOPT signals in \cref{fig:prom_PTAs}, with the minimum experimental uncertainty required to achieve a $3\sigma$ discrimination with the CvM test. We find that SIGW and FOPT signals can be distinguished at greater than 3$\sigma$ with larger uncertainties, an improvement with respect to the FOPT and DW signals in \cref{fig:prom_PTAs}. Similarly, the DW and SIGW signals can be distinguished at $3\sigma$ if $\Delta \Omega/\Omega = 4.4\%$. SKA can achieve this sensitivity by observing $\sim 600$ pulsars for 20 years~\cite{Babak:2024yhu}. 

\subsection{Prominence versus SNR-based signal discrimination}\label{sec:prom_SNR}

SNR can be used to assess whether two sources/spectra are distinguishable by an experiment.
Suppose that we want to discriminate between the signal+background spectra, $\Omega_1=\Omega_\mathrm{SIG_1}+\Omega_\mathrm{Bkg}$ and  
$\Omega_2=\Omega_\mathrm{SIG_2}+\Omega_\mathrm{Bkg}$.
Inspired by waveform model analyses~\cite{Finn:1992wt,Cutler:1994ys}, we define the inner product of the two spectra,
\begin{equation}\label{eq:inner_prod}
    \expval{\Omega_{\mathrm{1}}|\Omega_{\mathrm{2}}} = T_\mathrm{obs} \int_{f_\mathrm{min}}^{f_\mathrm{max}} df\, \frac{\Omega_{\mathrm{1}}(f) \Omega_{\mathrm{2}}(f)}{\Omega^2_{\mathrm{Sens}}(f)}\,,
\end{equation}
which reduces to $\expval{\Omega_\mathrm{1}|\Omega_\mathrm{1}} = \mathrm{SNR}_{1}^2$ if the two signals are identical; see Eq.~\eqref{eq:SNR_calc}. 
To quantify the difference between the two signals we define the signal-space statistic,
\begin{equation}
    \chi^2 = \expval{\Omega_\mathrm{1} - \Omega_\mathrm{2}|\Omega_\mathrm{1}-\Omega_\mathrm{2}}\,.
\end{equation}
which in terms of the {\it mismatch},
\begin{equation}
    \mathcal{M}_{12} \equiv 1- \frac{\expval{\Omega_\mathrm{1}|\Omega_\mathrm{2}}}{\rm{SNR}_1 \,\rm{SNR}_2}\,,
\end{equation}
can be written as
\begin{equation}\label{eq:chi_squared}
    \chi^2 = (\rm{SNR}_1 - \rm{SNR}_2)^2 + 2\mathcal{M}_{12}\, \rm{SNR}_1 \,\rm{SNR}_2  \,.
\end{equation}
In the limit of identical signals, the mismatch vanishes and $\chi^2 = 0$. In the opposite limit of no signal overlap, $\mathcal{M}_{12}=1$ and $\chi^2 = \rm{SNR}_1^2 + \rm{SNR}_2^2$.

We apply this $\chi^2$ statistic to estimate the signal discrimination power of SKA and LISA, and demonstrate how Prominence can aid in signal discrimination. We fit a fixed $\Omega_1$ spectrum with $\Omega_2$ by varying $\Omega_\mathrm{SIG_2}$ and $\Omega_\mathrm{Bkg}$. We then employ $\cal P$ to distinguish between $\Omega_1$ and $\Omega_2$ if the application of Eq.~\eqref{eq:chi_squared} finds them to be compatible with each other at 2$\sigma$.

In the PTA band, we consider a fixed FOPT signal corresponding to the one shown in Figs.~\ref{fig:prom_PTAs} and \ref{fig:prom_PTAs_SI}, and a fixed SIGW signal as shown in Fig.~\ref{fig:prom_PTAs_SI}. In the LISA band, the signal is fixed to the FOPT case shown in 
Fig.~\ref{fig:prom_SNR_vary}. These are to be interpreted as the simulated signals detected. For each of these signals, the background is also fixed as in the corresponding figure. We use a Markov Chain Monte Carlo algorithm to scan over the astrophysical and cosmological parameters of the simulated source and a hypothetical source, as detailed in Table~\ref{tab:parameters}, and select points with $\Delta\chi^2$ (for each source) corresponding to the 95.4\% CL (2$\sigma$) threshold. Note that the best-fit $\chi^2$ is zero if the hypothesis is the same as the simulated source.
We require that the background and the DW parameters $b$ and $c$ fall within the ranges specified in Section~\ref{sec:GWs_backs}. For the FOPT signals, we require $\beta/H(T_*) > 1$. 

\begin{table}[t]
    \centering
    \resizebox{\textwidth}{!}{%
    \begin{tabular}{c c c c c c}
        \hline
        {\textbf{Simulated source}} & \textbf{Hypothesis} & \textbf{Signal parameters} & \textbf{Background parameters} & \bf{$\chi_\mathrm{min}^2$} & \bf{$\Delta \chi^2$} \\
        \hline
        FOPT (PTA) & DW & $V_\mathrm{bias},\, \mathcal{E},\, b,\, c$ & $A_\mathrm{BHB},\, \gamma_\mathrm{BHB}$ & $0.13$ & $9.70$ \\[0.1em]
        FOPT (PTA) & FOPT  & $\alpha,\, \beta/H(T_*),\, T_*$ & $A_\mathrm{BHB},\, \gamma_\mathrm{BHB}$ & $0$ & $8.02$ \\[0.1em]
        FOPT (LISA) & DW & $V_\mathrm{bias},\, \mathcal{E},\, b,\, c$ & $h^2\Omega_\mathrm{Gal},\,h^2\Omega_\mathrm{Ext},\, n_s,\,\alpha_\mathrm{Ext}$ & $0.90$ & $9.70$ \\[0.1em]
        FOPT (LISA) & FOPT & $\alpha,\, \beta/H(T_*),\, T_*$ &$h^2\Omega_\mathrm{Gal},\,h^2\Omega_\mathrm{Ext},\, n_s,\, \alpha_\mathrm{Ext}$ & $0$ & $8.02$  \\[0.1em]
        SIGW (PTA) & SIGW & $\alpha_\mathrm{SIG},\, \beta_\mathrm{SIG},\, A,\,f_c$ & $ A_\mathrm{BHB},\, \gamma_\mathrm{BHB}$ & $0$ & $9.70$ \\[0.1em]
        SIGW (PTA) & DW & $V_\mathrm{bias},\,\mathcal{E},\, b,\, c$ & $A_\mathrm{BHB},\, \gamma_\mathrm{BHB}$ & $0.90$ & $9.70$ \\[0.1em]
        SIGW (PTA) & FOPT & $\alpha,\, \beta/H(T_*),\, T_*$ & $A_\mathrm{BHB},\, \gamma_\mathrm{BHB}$ & $0.07$ & $8.02$  \\
        \hline
    \end{tabular}%
    }
    \caption{
    Signals detected in the PTA and LISA frequency bands and the parameters varied for each hypothetical source (and background), with the corresponding values of $\Delta \chi^2=\chi^2 - \chi^2_{\rm min}$ for compatibility at $2\sigma$. We marginalize over the background parameters. }  
    \label{tab:parameters}
\end{table}
\begin{figure*}[htb!]
	\centering
    \subfloat[]{\includegraphics[width=0.5\textwidth]{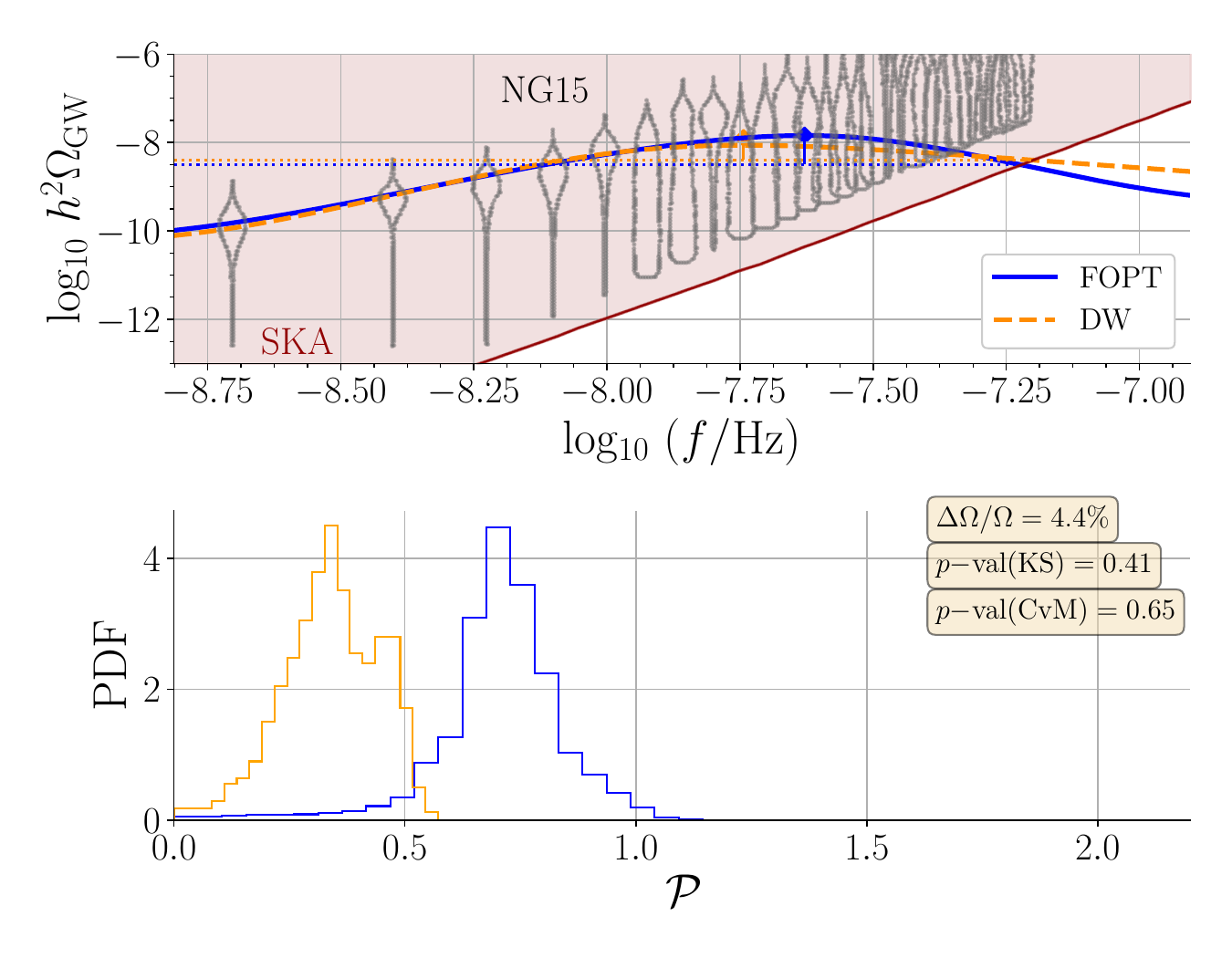}} 
    \subfloat[]{\includegraphics[width=0.5\textwidth]{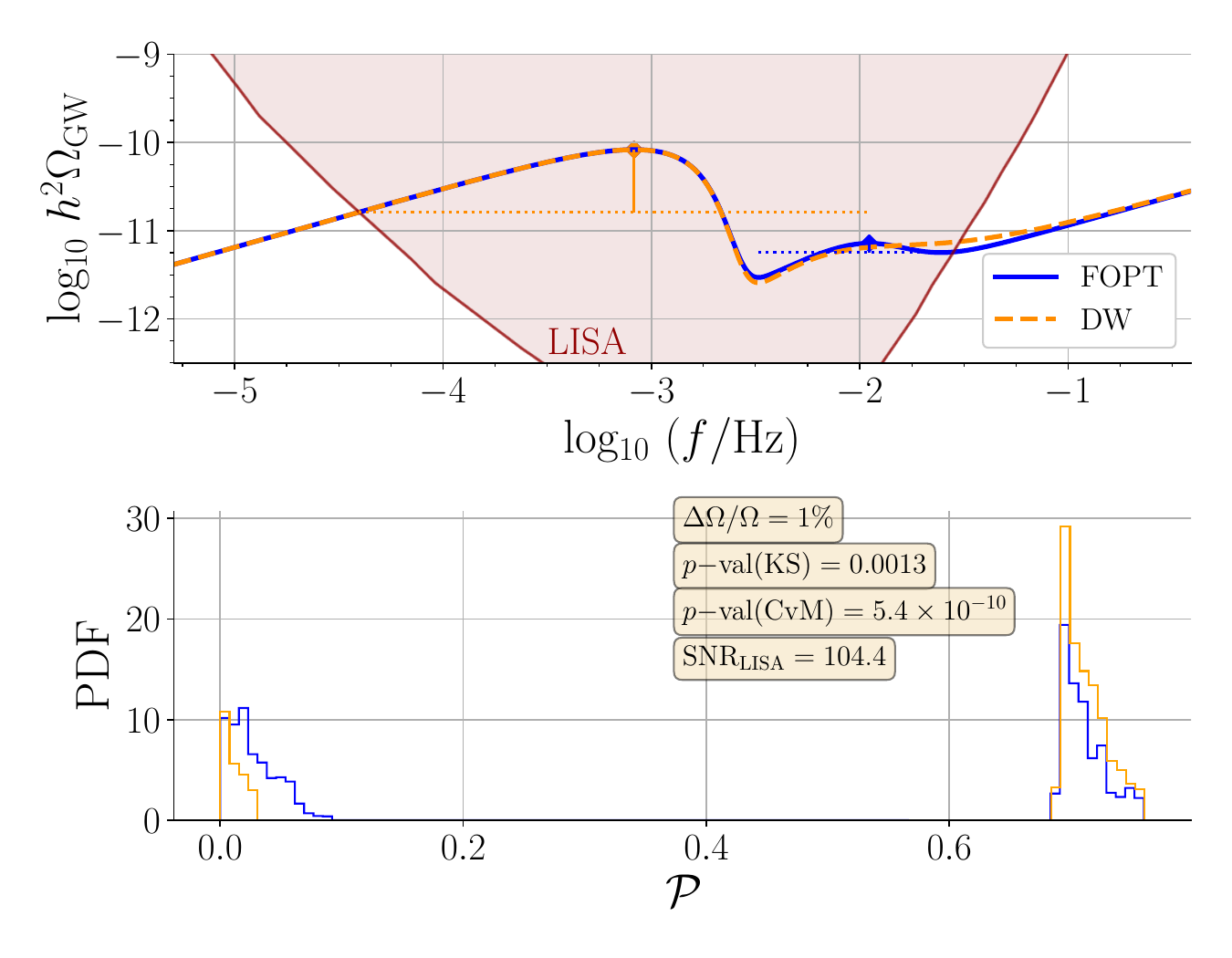}} \\
    \subfloat[]{\includegraphics[width=0.5\textwidth]{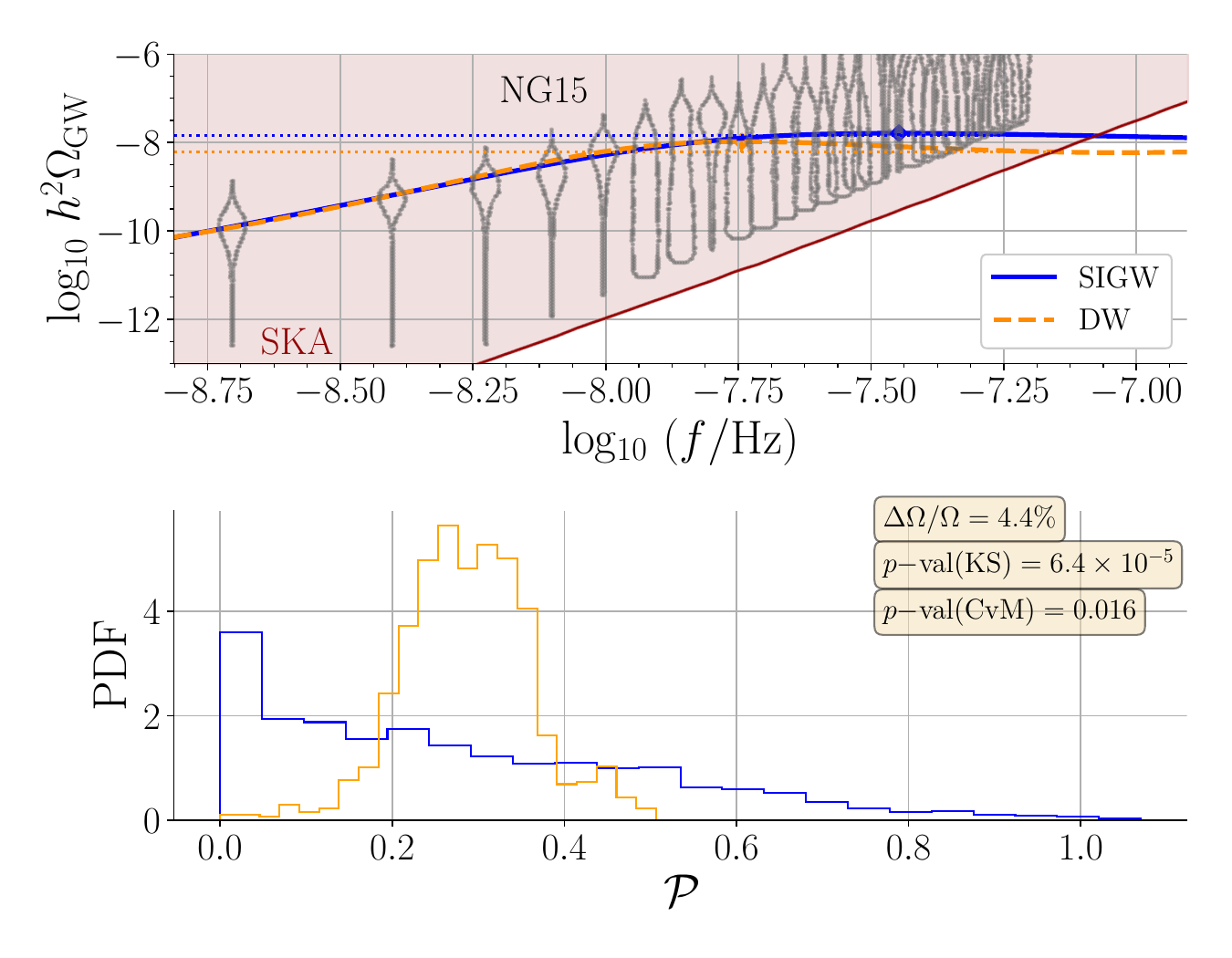}}
    \subfloat[]{\includegraphics[width=0.5\textwidth]{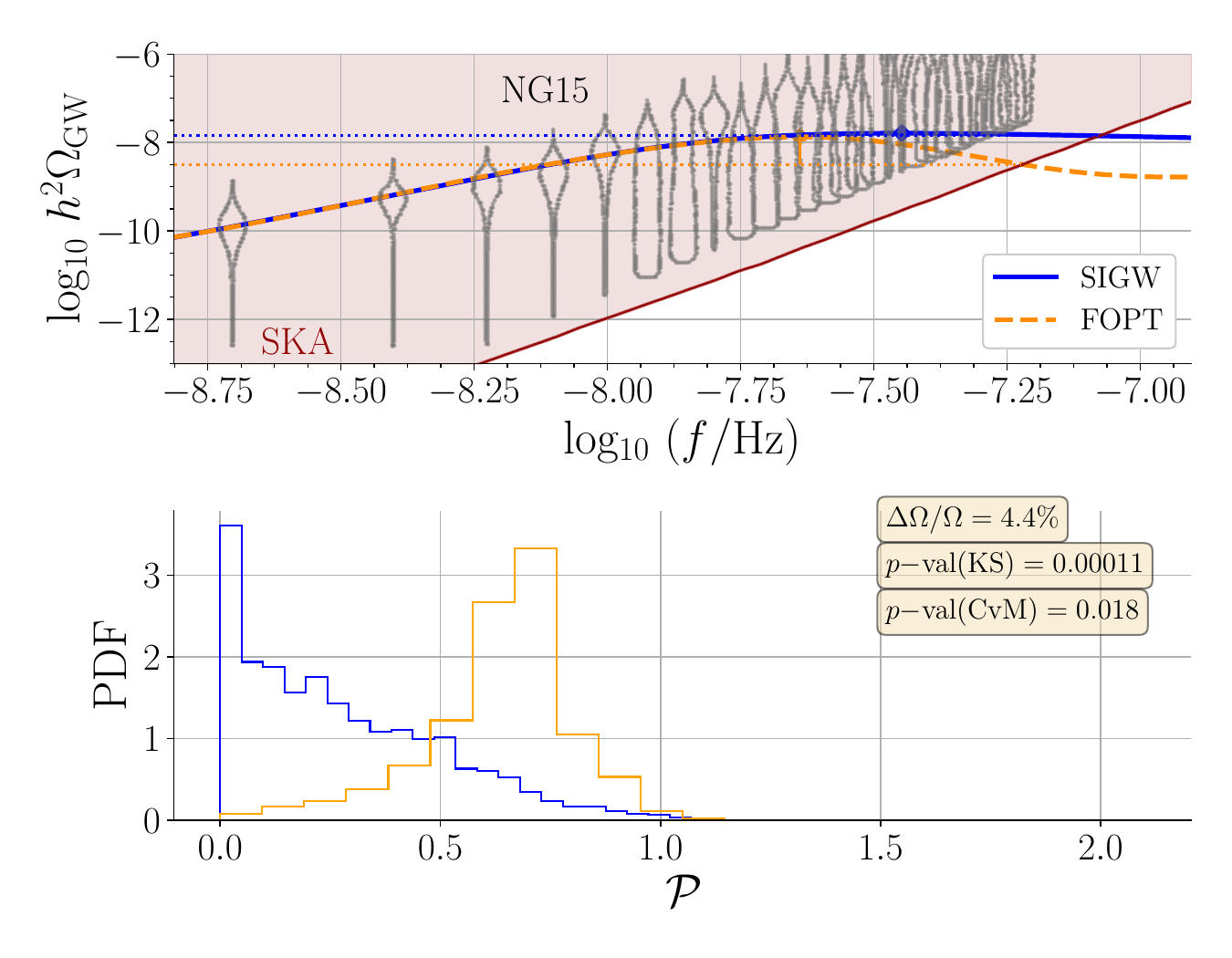}}
    \caption{SGWB spectra and the corresponding PDFs of $\mathcal{P}$. The blue curves correspond to the fixed spectrum and the dashed orange curves to the best-fit spectra. In panel~(a) the background+FOPT signal is fixed to the spectrum in Fig.~\ref{fig:prom_PTAs}. In panel~(b) the background+FOPT signal is fixed to the spectrum in Fig.~\ref{fig:prom_SNR_vary}, and in panels~(c) and~(d) the background+SIGW signal is fixed to the spectrum in Fig.~\ref{fig:prom_PTAs_SI}. Signals that cannot be distinguished at $2\sigma$ using SNR can 
be distinguished using $\mathcal{P}$ at much more than $3\sigma$ in some cases.} 
    \label{fig:promSNR_PTAs}
\end{figure*}

We present our results in Fig.~\ref{fig:promSNR_PTAs}. In panels~(a) and~(b), the solid blue curves show the fixed FOPT plus background spectrum, and the dashed orange curves are the best-fit DW spectra. In the PTA and LISA bands we use $\Delta \Omega/\Omega = 4.4\%$ and $\Delta \Omega/\Omega = 1\%$, respectively. 
We find that in the LISA frequency band $\cal P$ exhibits strong discriminating power even if distinct SGWB signals have similar SNRs within $2\sigma$, while in the PTA band, the discrimination power is poor. In the LISA band, the signals can be distinguished with >99.5\%~CL, while the CvM test yields an even stronger result. In particular, signals can be distinguished with the KS test with a $p$-value of 0.0013, which corresponds to a 3.2$\sigma$~CL, while the more conservative CvM test yields a $p$-value of $5.4\times 10^{-10}$, corresponding to a $6.2\sigma$ discrimination. 
Panels~(c) and~(d) show the case where a fixed SIGW signal in the PTA band is fitted with an FOPT and DW signal, respectively. We find strong discriminating power, with $p$-values from the KS test as low as $6.4\times 10^{-5}$ for SIGW versus DW (corresponding to 4$\sigma$) and 0.00011 for SIGW versus FOPT (corresponding to 3.87$\sigma$). The CvM test can distinguish signals with a significance greater than 2.9$\sigma$.

The comparison presented here is intended to demonstrate that Prominence can provide spectral-shape information lost in the SNR. By encoding information about both the overall signal strength and the spectral shape, Prominence can help lift degeneracies between hypotheses with comparable significance according to SNR.

\subsection{Direct hypothesis testing with Prominence}

 Only a single value of Prominence is obtainable from the GW spectrum reconstructed from an experimental dataset. Another way to assess its discriminating power is to compare this extracted value with the PDF distributions corresponding to different signal hypotheses.

As an illustration, consider the PTA spectra in Fig.~\ref{fig:promSNR_PTAs}. We suppose that the reconstructed signal matches the simulated source. The reference hypothesis  $\mathcal{H}_0$ is FOPT in panel (a) and SIGW in panels (c) and (d). The other source in each panel is the signal hypothesis $\mathcal{H}_1$. We evaluate the $p$-value for 
the $\cal{P}$  PDF for $\mathcal{H}_1$ (by varying the signal and background parameters) given the measured value of $\cal{P}$:
\begin{equation}\label{eq:pvals_single}
\begin{aligned}
& \mathrm{Fig.}~\ref{fig:promSNR_PTAs}\mathrm{(a)}:\quad p\text{-}\mathrm{value}(\mathcal{H}_1=\mathrm{DW}~|~\mathcal{H}_0=\mathrm{FOPT}) = 0.027\,, \\
& \mathrm{Fig.}~\ref{fig:promSNR_PTAs}\mathrm{(c)}:\quad p\text{-}\mathrm{value}(\mathcal{H}_1=\mathrm{DW}~|~\mathcal{H}_0=\mathrm{SIGW}) = 0.41\,, \\
& \mathrm{Fig.}~\ref{fig:promSNR_PTAs}\mathrm{(d)}:\quad p\text{-}\mathrm{value}(\mathcal{H}_1=\mathrm{FOPT}~|~\mathcal{H}_0=\mathrm{SIGW}) = 0.091\,.
\end{aligned}
\end{equation}
The discrimination for the case in Fig.~\ref{fig:promSNR_PTAs}(a) improves and the discrimination for the cases in Figs.~\ref{fig:promSNR_PTAs}(c) and \ref{fig:promSNR_PTAs}(d) worsens. The $p$-value depends on how well within the PDF of $\mathcal{H}_1$ the value of $\cal P$ for $\mathcal{H}_0$ is contained. If it lies in the tails of the PDF, the $p$-value is smaller than if it is more central.
 Only the sources in 
panel~(a) can be distinguished at more than 95\%~CL.


\section{Summary}\label{sec:summary}

We have introduced and demonstrated the effectiveness of \textit{Prominence}, a concept originally developed in topography and signal processing, as a powerful and model-independent discriminator of SGWB sources. By quantifying the relative height of spectral peaks with respect to their surrounding features, Prominence is sensitive not only to the absolute peak amplitude but also to the overall shape of the signal. Our analysis focused on SGWB sources such as FOPTs, DWs, CSs and SIGWs, although the methodology can be readily extended to other sources.

Employing statistical CvM and KS tests on Prominence CDFs, we showed that this observable can effectively distinguish between GW signal arising from different new physics, even under the most conservative assumption that their peak amplitude and frequency are identical. This result holds for both single-peaked and multi-peaked spectra.

As a relative measure of GW amplitude, the effectiveness of Prominence depends on the precision with which $h^2\Omega_\mathrm{GW}$ can be determined. We find that better than 3$\sigma$ discrimination is possible if the fractional uncertainty $\Delta \Omega/\Omega < 1\%$, a level expected for LISA~\cite{Caprini:2024hue}. In contrast, SGWB signals in the PTA frequency band ($f < 10^{-7}~\mathrm{Hz}$) are subject to much larger uncertainties ($\Delta \Omega_\mathrm{PTA}/\Omega_\mathrm{PTA} \sim 10\%$--$50\%$), which significantly limits the discriminating power of Prominence. We estimate that $\Delta \Omega_\mathrm{PTA}/\Omega_\mathrm{PTA} \sim 0.1\%\mbox{-}6\%$ is needed for Prominence to distinguish signals at $3\sigma$. A few percent uncertainty may be achieved by SKA with observations of several hundred millisecond pulsars over a 20-year period. It is also worth noting that the large extragalactic backgrounds in the ET band obscure some of the spectral features that distinguish FOPT and DW signals, making them appear more similar and weakening the discrimination power of Prominence. 

 We also find that GWs from CSs can, in some cases, enhance the ability to discriminate between FOPT and DW signals. While the global behavior of the PDFs remain similar, large localized differences enhance the discrimination power. This effect is generally more pronounced for FOPTs, as they tend to produce larger Prominences due to their narrow peaks, resulting in CDFs that are more sensitive to the presence of CSs. 

To explore scenarios in which signal discrimination with SNR is challenging,  we studied signals with small SNRs and substantial overlap. This is motivated by the fact that for signals with sufficiently large SNRs (as in Fig.~\ref{fig:prom_twopeak}) and small reconstruction uncertainties of  $\sim 1\%$, SNR alone may serve as a robust discriminator. In contrast, if the uncertainties are large enough that the SNRs expected from two sources are nearly equal, $\cal P$ becomes particularly valuable, effectively distinguishing between them even when their overlap would render a traditional SNR-based approach inconclusive. Thus, Prominence provides its greatest discriminating power precisely when signals are significantly degenerate, with similar spectra and relatively small SNRs.  

The implementation of our analysis is available for download on GitHub~\cite{Prominence_code}.

\section*{Acknowledgments}

We thank M.~Finetti for a discussion. J.G.~is directly funded by the Portuguese Foundation for Science and Technology (FCT - Funda\c{c}\~{a}o para a Ci\^{e}ncia e a Tecnologia) through the doctoral program grant with the reference 2021.04527.BD (\url{https://doi.org/10.54499/2021.04527.BD}).
D.M.~is supported in part by the U.S. Department of Energy under Grant No.~DE-SC0010504.
A.P.M.~is supported by FCT through the project with reference 2024.05617.CERN (\url{https://doi.org/10.54499/2024.05617.CERN}).
J.G.~and A.P.M.~are also supported by LIP and FCT, reference LA/P/0016/2020 (\url{https://doi.org/10.54499/LA/P/0016/2020}) and by the ERC-PT A-Projects 'Unveiling', financed by PRR, NextGenerationEU.

\bibliographystyle{JHEP}
\bibliography{references}
\end{document}